\documentclass[epj,nopacs]{svjour}
\usepackage{epsfig,array}
\usepackage{multirow,color}
\usepackage{epsf,graphicx}
\usepackage{times,latexsym,euscript,amstext,amssymb,amsbsy,amsmath,amsfonts}
\usepackage{mdwlist}\usepackage{epsfig}
\usepackage{slashbox,multirow}
\newcommand{\be}{\begin{eqnarray}}
\newcommand{\ee}{\end{eqnarray}}
\newcommand{\bc}{\begin{center}}
\newcommand{\ec}{\end{center}}
\newcommand{\bea}{\begin{eqnarray}}
\newcommand{\eea}{\end{eqnarray}}

\newcommand{\beq}{\begin{equation}}
\newcommand{\eeq}{\end{equation}}

\def\fun#1#2{\lower3.6pt\vbox{\baselineskip0pt\lineskip.9pt
\ialign{$\mathsurround=0pt#1\hfil##\hfil$\crcr#2\crcr\sim\crcr}}}

\begin{document}
\title{\boldmath Study of ambiguities in $\pi^-p\to \Lambda K^0$ scattering amplitudes }
\titlerunning{Study of ambiguities in $\pi^-p\to \Lambda K^0$ scattering amplitudes}
\author{
A.V.~Anisovich$\,^{1,2}$, R. Beck$\,^1$, E.~Klempt$\,^1$,
V.A.~Nikonov$\,^{1,2}$, A.V.~Sarantsev$\,^{1,2}$, U.~Thoma$\,^{1}$,
and Y. Wunderlich$\,^{1}$}
\authorrunning{A.V.~Anisovich \it et al.}
\institute{$^1\,$Helmholtz-Institut f\"ur Strahlen- und Kernphysik,
Universit\"at Bonn, Germany\\
$^2\,$Petersburg Nuclear Physics Institute, Gatchina, Russia}

\date{\today}
\abstract{Amplitudes for the reaction $\pi^-p\to \Lambda K^0$ are
reconstructed from data on the differential cross section
$d\sigma/d\Omega$, the recoil polarization $P$, and on the spin
rotation parameter $\beta$. At low energies, no data on $\beta$
exist, resulting in ambiguities. An approximation using $S$ and $P$
waves leads only to a fair description of the data on
$d\sigma/d\Omega$ and $P$; in this case, there are two sets of
amplitudes. Including $D$ waves, the data on $d\sigma/d\Omega$ and
$P$ are well reproduced by the fit but now, there are several
distinct solutions which describe the data with identical precision.
In the range where the spin rotation parameter $\beta$ was measured,
a full and unambiguous reconstruction of the partial wave amplitudes
is possible. The energy-independent (single-energy) amplitudes are
compared to the energy dependent amplitudes which resulted from a
coupled channel fit (BnGa2011-02) to a large data set including both
pion and photo-induced reactions. Significant deviations are
observed. Consistency between energy dependent and energy
independent solutions is obtained by choosing the energy independent
solution which is closest to the energy dependent solution. In a
second step, the {\it known} energy dependent solution for low (or
high) partial waves is imposed and only the high (or low) partial
waves are fitted leading to smaller uncertainties.
 \vspace{1mm}   }

\maketitle

\section{Introduction}
The excitation spectrum of the nucleon has been studied in
experiments on $\pi N$ elastic scattering, including experiments in
which the target nucleon is polarized or in which the recoil
polarization of the scattered nucleon is measured in a secondary
reaction. Data on the $\pi^-p\to n\pi^+$ charge~ex\-change are
required to separate the two isospin contributions. Elastic
scattering yields differential cross sections $d\sigma/d\Omega$. A
transversely polarized target - or the decay asymmetry of hyperons
in the final state - can be used to determine the analyzing power
$P$, the spin transfer from a nucleon polarized longitudinally
(along the pion beam line) to the final state baryon yields the spin
rotation parameters $A$ and $R$ or the spin rotation angle $\beta
=\arctan{(-R/A)}$. $d\sigma/d\Omega$ and two polarization
observables need to be known to reconstruct the scattering amplitude
without using further constraints; the third polarization variable
can be calculated up to a sign ambiguity from the relation
\be
\label{norm}
P^2+A^2+R^2 = 1.
\ee
In practice, experimental information on the spin rotation
parameters $A$ and $R$ is mostly missing. In this case, the
scattering amplitude is not defined unambiguously
\cite{Arenhovel:1998vj}. A unique solution
can be constructed using constraints from dispersion relations
linking the real and the imaginary part of the scattering amplitude
\cite{Hohler:1979yr,Hohler:1993xq,Cutkosky:1980rh}. One can get a
limited number of different solutions without the use of theoretical
input, by fitting the data with Legendre polynomials (with a finite
number of coefficients) which provide a link between data at
different angles.

Alternatively, the ambiguity problem can be solved by starting from
an energy-dependent fit to the data \cite{Arndt:2006bf}. In an
energy-dependent fit, the amplitudes are constrained by analytic
functions in energy and angle. The ambiguity problem can be solved
by selecting a solution which is compatible with the amplitude
determined from the energy-dependent fit. The amplitudes from the
energy-independent solution can then be used as input for an
iterative procedure \cite{Arndt:2006bf}. The energy independent
solution thus obtained is then the correct one, provided that the
energy dependent amplitudes were close to the correct values. In a
recent article, these methods were applied to the reactions
$\pi^-p\to p\eta$ and $\pi^-p\to \Lambda K^0$. Energy independent
solutions for the $S_{11}$, $P_{11}$, $P_{13}$, $D_{15}$, $F_{15}$,
and $F_{17}$ amplitudes were given \cite{Shrestha:2012va} and used
to extract parameters of contributing resonances in a multichannel
analysis \cite{Shrestha:2012ep}.

In this article we restrict ourselves to the reaction $\pi^-p\to
\Lambda K^0$ using the data from
\cite{Knasel:1975rr,Baker:1978qm,Saxon:1979xu,Bell:1983dm}. Our aim
here is however not to determine the $\pi^-p\to \Lambda K^0$
scattering amplitudes but rather to study the ambiguities which are
intrinsic parts of the method when the data are incomplete and of
limited accuracy.

\section{Pion induced reaction}
\subsection{Amplitudes, partial waves and  observables}

Scattering processes of a pseudoscalar meson off a nucleon to a
final state with a pseudoscalar meson plus a spin-1/2 baryon are
conventionally described in terms of a scattering matrix $M$ with
the following structure in the reaction center-of-mass system

\be
&&M=f(W,z)+g(W,z)i(\overrightarrow{\sigma} \overrightarrow{n})
\ee
where $f(W,z)$ is the non-spin-flip amplitude, $g(W,z)$ is the
spin-flip amplitude, $\overrightarrow n$ is the normal vector of the
production plane and $\overrightarrow\sigma$ are the Pauli spin
matrices. The amplitudes $f(W,z)$ and $g(W,z)$ depend on total
energy $W$ and on $z=\cos \Theta$ where $\Theta$ is the scattering
angle of the outgoing meson in the center-of-mass system (cms). The
normal vector of the production plane is defined as
\be
\overrightarrow n= \frac{\overrightarrow q \times \overrightarrow
k}{|\overrightarrow q \times \overrightarrow k|},
\ee
where  $\overrightarrow q$ is the initial cms momentum of the meson,
$q$ its modulus, $\overrightarrow k$ is the final meson cms
momentum, $k$ its modulus. The amplitudes $f$ and $g$ can be
expanded into partial waves

\be
\nonumber
 &&f(W,z)=\frac{1}{\sqrt{qk}}\sum\limits_{l=0}^{L} \big [(l\!+\!1)A_l^+(W)+ l A_l^-(W)\big ]
P_l(z) \;, \nonumber \\
 &&g(W,z)=\frac{1}{\sqrt{qk}} \sin \Theta \sum\limits_{l=1}^L \big [A_l^+(W)- A_l^-(W)\big ]
P'_l(z) \;.
\label{piN_expan}
\ee

The partial amplitudes $A^\pm_l$ depend only on the total energy $W$
of the reaction, the $A^+_l$ functions describe the $1/2^-\!$,
$3/2^+\!$, $5/2^-\!$, $\ldots$ states and the $A^-_l$ functions the
$1/2^+\!$, $3/2^-\!$, $5/2^+\!$, $\ldots$ states. $P_l$ are Legendre
polynomials in $z$ and $P_l'$ are their derivatives. The initial
$\pi N$ system has isospin $I=1/2$ and $I=3/2$ and the amplitudes
$A^\pm_l$ can be decomposed into the isospin amplitudes as follows:
\be
A^\pm_l = C_{\frac12} A^\pm_{l\frac12} + C_{\frac32}
A^\pm_{l\frac32}\,.
\ee
For reaction $\pi^-p\to \Lambda K^0$ $C_{\frac12} =
-\sqrt{\frac23}$ and $C_{\frac32} = 0$.

 The amplitudes $f$ and $g$ are complex functions. Except for an
arbitrary phase, $f$ and $g$ can be calculated up to one discrete
ambiguity when three observables are known. These can be chosen to
be the differential cross section $d\sigma/d\Omega$, the
polarization $P$ of the outgoing baryon in the final state, and
spin-rotation angle $\beta$. The differential cross section is given
by
\be
\frac{d\sigma}{d\Omega} = \frac{k}{q}(|f|^2 + |g|^2)\,,
\ee
and the total cross section is
\be
\sigma = \frac{2\pi}{q^2}(2J+1)\sum\limits_{l=0}^L \big [|A_l^+(W)|^2 + |A_l^-(W)|^2\big ]
\ee
where $J$ is the total spin of the state (remember that $J = |l \pm
1/2|$ for $\pm$ states).

The  polarization in the final state is given by
\be
P = \frac{-2 Im(f^* g)}{|f|^2 + |g|^2}\,.
\ee
The third observable is the spin-rotation angle:
\be
\beta = arg\Big(\frac{f-i g}{f+i g}\Big) =tan^{-1}\Big(\frac{-2 Re(f^* g)}{|f|^2 - |g|^2}\Big)\,.
\ee
In some pion-induced experiments not the $\beta$ angle was measured
but one or both spin-rotation parameters, $R$ and $A$. They are
defined as
\be
R = \frac{2 Re(f^* g)}{|f|^2 + |g|^2}\;,\;\;\; A = \frac{|f|^2 -
|g|^2}{|f|^2 + |g|^2}\,.
\ee
$P$, $R$ and $A$ are not the independent observables. The
polarization variables are constrained by the relation \ref{norm}.

\subsection{\boldmath Amplitude ambiguities when only $d\sigma/d\Omega$ and $P$ are known}

Often, only the observables $d\sigma/d\Omega$ and $P$ are measured.
We face this situation even in the simplest case of $\pi N$ elastic
scattering. The target nucleon can be polarized transversely giving
access to $P$. Experimentally more difficult are measurements of the
spin rotation parameters $A$ and $R$. Their determination requires
the measurement of the proton polarization in a secondary scattering
process. Therefore, $A$ and $R$ have been determined only in a
rather limited range of energies and angle. In the case of
pion-induced hyperon production ($K\Lambda$ or $K\Sigma$), the
analyzing power $P$ of the final-state hyperon can be inferred from
its decay, and then no secondary scattering process is required for
a complete experiment. The spin rotation variables can then be
determined using a target which is polarized longitudinally. Also
here, the data covers only a limited range in energy and solid
angle.

The problem of ambiguities in the case of so-called ``incomplete"
experiments (with lack of spin-rotation information) has been
discussed many decades ago, see
\cite{Gersten:1969ae,Barrelet:1971pw,Baker:1976xp}. Let us briefly
recall the origin of such ambiguities. The cross section with
polarization information (assuming that the target nucleon is fully
polarized) can be defined as
\be
(1\pm P)\frac{d\sigma}{d\Omega} =  |f \pm i g|^2.
\ee
The idea is to replace the two functions $(1\pm
P)\frac{d\sigma}{d\Omega}$ by a single function. To do this we
expand the physical region of the scattering angle $\Theta$ from
$[0, \pi]$ to $[0, 2\pi]$. Using a new variable $w= e^{i \Theta}$
one can see that the $f$ amplitude is even in power of $w$ and
$w^{-1}$ since it depends on $z= \frac12(w+w^{-1})$. The $g$
amplitude contains $\sin \Theta = \frac{1}{2i}(w-w^{-1})$ and so
behaves as $g(w^{-1}) = - g(w)$. Let us define the function
\be
F(w) = f(w) + i g(w)\,.
\ee
For $\Theta\,\epsilon\,[0, \pi]$
\be
F(w) = f(z) + i g(z) \;,\;\;\;   |F(w)|^2 = (1 +
P)\frac{d\sigma}{d\Omega}\,
\label{Fp}
\ee
holds. For $\Theta\,\epsilon\,[\pi, 2\pi]$,  $\sin\Theta < 0 $ and
 $g(w) = - g(z)$, hence
\be
F(w) = f(z) - i g(z) \;,\;\;\;   |F(w)|^2 = (1 -
P)\frac{d\sigma}{d\Omega}\,.
\label{Fm}
\ee
Instead of a real and positive cross section in the region $ -1 \leq
z \leq 1$ (the case of scalar particle rescattering), we now have a
real and positive cross section in the region $ 0 \leq \Theta \leq 2
\pi$ or on the unit circle of the $w$ plane.

Let us rewrite $|F(w)|^2$ as a power series in $w$:
\be
\label{sumn}
|F(w)|^2 = \sum_{n=-N}^N a_n w^n
\ee
where $N$ depends on maximal orbital momentum $L$ in
eq.~(\ref{piN_expan}). Since $|F(w)|^2$ is real on the unit circle
in the $w$-plane, we can write this function as a product of roots
in the following form
\be
|F(w)|^2 = C \prod_{i=1}^{N} (w - w_i)(w^{-1} - w_i^{*})\,.
\ee
Remember that $w^* = w^{-1}$ on the unit circle. Finally we have
\be
F(w) = C^{\frac12}e^{i\phi}  \prod_{i=1}^{N} (w - w_i)\,.
\label{Fexpr}
\ee
Equation~(\ref{Fexpr}) is not the only possible solution, one can as
well take $(w - w_i)$ or $(w^{-1} - w_i^{*})$ as a root, and this
gives $2^N$ different solutions. But not all of these solution are
physically sensible. In the next section we discuss the ambiguities
in the case when only a limited number of amplitudes are taken into
account.

\subsection{Amplitude near the threshold}

We might expect that near threshold only $S$ and $P$ waves are
important. As a simple and educative example let us consider the
case of two waves with $J^P=1/2^-$ and $1/2^+$ only. Let us denote
the magnitude and phase of the scattering amplitude as $r_{l\pm}$
and $\phi_{l\pm}$, respectively. Since each amplitude can be
multiplied by a factor $e^{i\Phi}$ without changing the observables,
we set $\phi_{0+} =0$ for simplicity. The cross section and
polarization in the final state can be calculated as
\be
\frac{d\sigma}{d\Omega} = \frac{k}{q} \;I_0\qquad{\rm and}
\ee
\be
P\;I_0 = 2 \sin\Theta\; r_{0+} r_{1-} \sin(\phi_{1-})
\ee
where
\be
I_0= r_{0+}^2 + r_{1-}^2 + 2 z \cos(\phi_{1-}) r_{0+} r_{1-}\,.
\ee
The interference of the $1/2^-$ and $1/2^+$ waves leads to a linear
dependence of the differential cross section in $z$ while the recoil
asymmetry multiplied with the differential cross section and divided
by $\sin\Theta$ should be flat. Using eqs.(\ref{Fp} - \ref{sumn}),
we can write the amplitude magnitude square $|F(w)|^2$ as
\be
|F(w)|^2 = \sum_{i=-1}^{1} a^{\pm}_i\; w^i
\ee
where
\be
a^\pm_{-1} = r_{0+} r_{1-} (\cos(\phi_{1-}) \pm i
\sin(\phi_{1-}))\,,
\ee
\be
a^\pm_0 = r_{0+}^2 + r_{1-}^2\;,\;\;\; a^\pm_{1} = (a^\pm_{-1})^*\,.
\ee
The coefficients $a^+_i$ define the amplitude $F(w)$ in the region $
0 \leq \Theta \leq \pi$, while the $a^-_i$ define $F(w)$ in the
region $ \pi \leq \Theta \leq 2 \pi$. The expressions above clearly
demonstrate the ambiguity in the determination of $1/2^-$ and
$1/2^+$ waves: they cannot be distinguished. Any PWA solution with
dominant $1/2^-$ and $1/2^+$ waves has an alternative solution where
the magnitudes of $S$ and $P$ waves replace each other.

The more general case includes another $P$ wave, $J^P=3/2^+$. The
cross section and polarization in the final state can be calculated
as
\be
\frac{d\sigma}{d\Omega} = \frac{k}{q} \;I_0 \label{spwave0}
\ee
\be
P\;I_0 = 2 \sin\Theta\; (r_{1-} r_{0+} \sin(\phi_{1-}) + \label{spwave1}\\
r_{1+} (3 z r_{1-} \sin(\phi_{1-}-\phi_{1+}) - r_{0+}
\sin(\phi_{1+}))) \nonumber
\ee
where
\be
I_0 = r_{1-}^2 + 2 z \cos(\phi_{1-}) r_{1-} r_{0+} + r_{0+}^2 +
\nonumber\\ 2 (-1+3 z^2) \cos(\phi_{1-}-\phi_{1+}) r_{1-} r_{1+} +
\label{spwave2}
\\4 z \cos(\phi_{1+}) r_{0+} r_{1+} + r_{1+}^2+3 z^2 r_{1+}^2\,.
\nonumber
\ee
Here, the differential cross section has a $z^2$ term while the
polarization multiplied by the differential cross section and
divided by $\sin\Theta$ has a $z$ term. The amplitude magnitude
square $|F(w)|^2$ can be calculated as
\be
|F(w)|^2 = \sum_{i=-2}^{2} a^{\pm}_i\; w^i
\ee
where
\be
a^\pm_{-2} = 3/4 r_{1+} (2 \cos(\phi_{1-}-\phi_{1+}) r_{1-} +
\label{aa1}
\\  r_{1+} \pm 2 i\; r_{1-}
\sin(\phi_{1-}-\phi_{1+})) \nonumber
\ee
\be
a^\pm_{-1} = r_{0+} (\cos(\phi_{1-}) r_{1-} + 2 \cos(\phi_{1+})
r_{1+} \pm \label{aa2}
\\ i (r_{1-}
\sin(\phi_{1-}) - r_{1+} \sin(\phi_{1+})))  \nonumber
\ee
\be
a^\pm_0 = r_{1-}^2 + r_{0+}^2+ \cos(\phi_{1-}-\phi_{1+}) r_{1-}
r_{1+}+ \frac52 r_{1+}^2
\label{aa3}
\ee
\be
a^\pm_{1} = (a^\pm_{-1})^*\;,\;\;\; a^\pm_{2} = (a^\pm_{-2})^*
\ee
Equations~(\ref{aa1} - \ref{aa3}) can be used to search for
ambiguities. For any found solution we can write a system of
equations \\$a^\pm_i(r_{l\pm}, \phi_{l\pm}) = C_i$. In this
particular case we have five equations for the real and imaginary
part of $a_i^\pm$ which define additional allowed solutions for the
$r_{l\pm}$ and $\phi_{l\pm}$.

\begin{figure*}
\begin{center}
\begin{tabular}{cccc}
\hspace{-3mm}\includegraphics[width=0.21\textwidth]{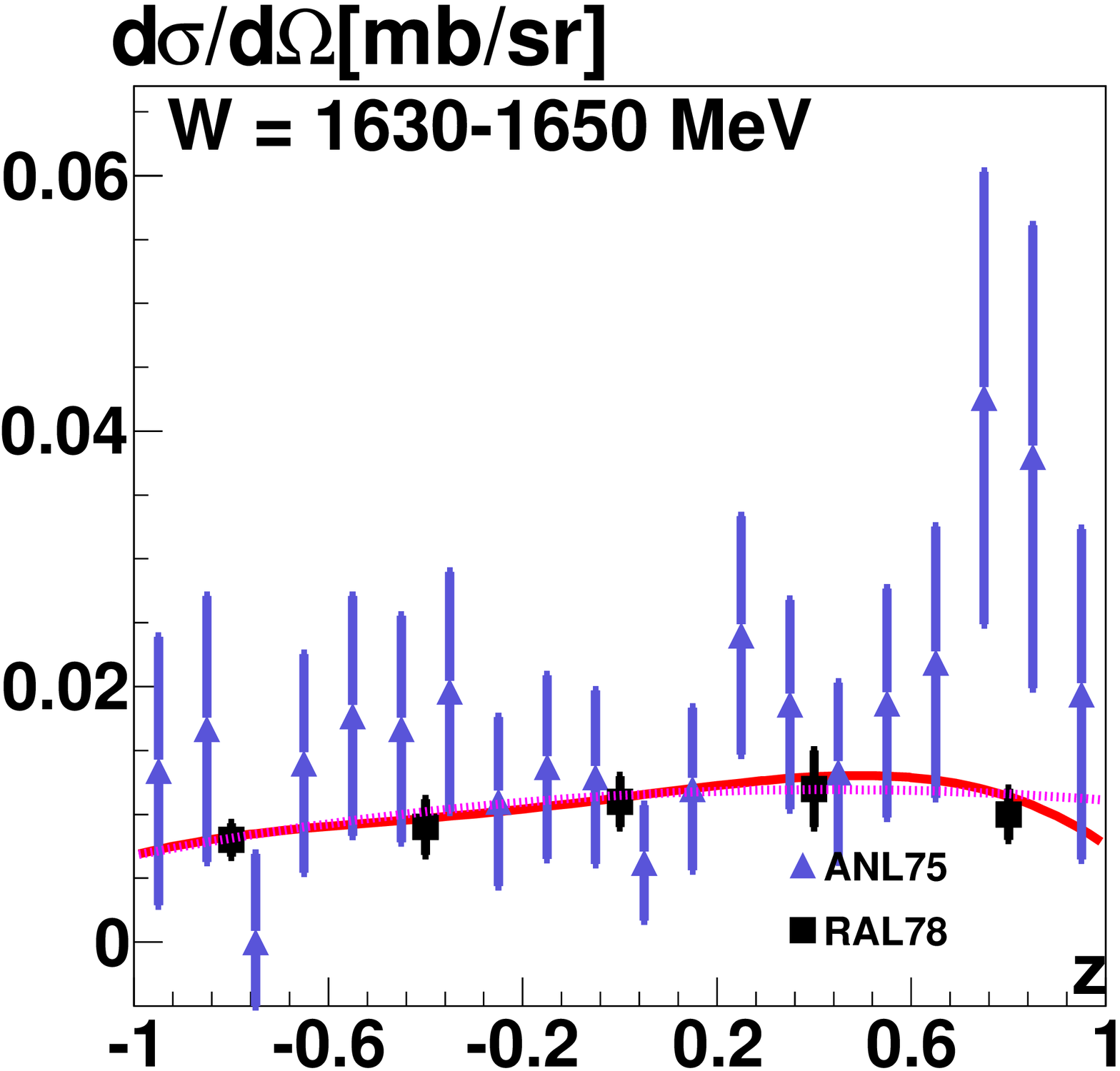}&
\hspace{-3mm}\includegraphics[width=0.21\textwidth]{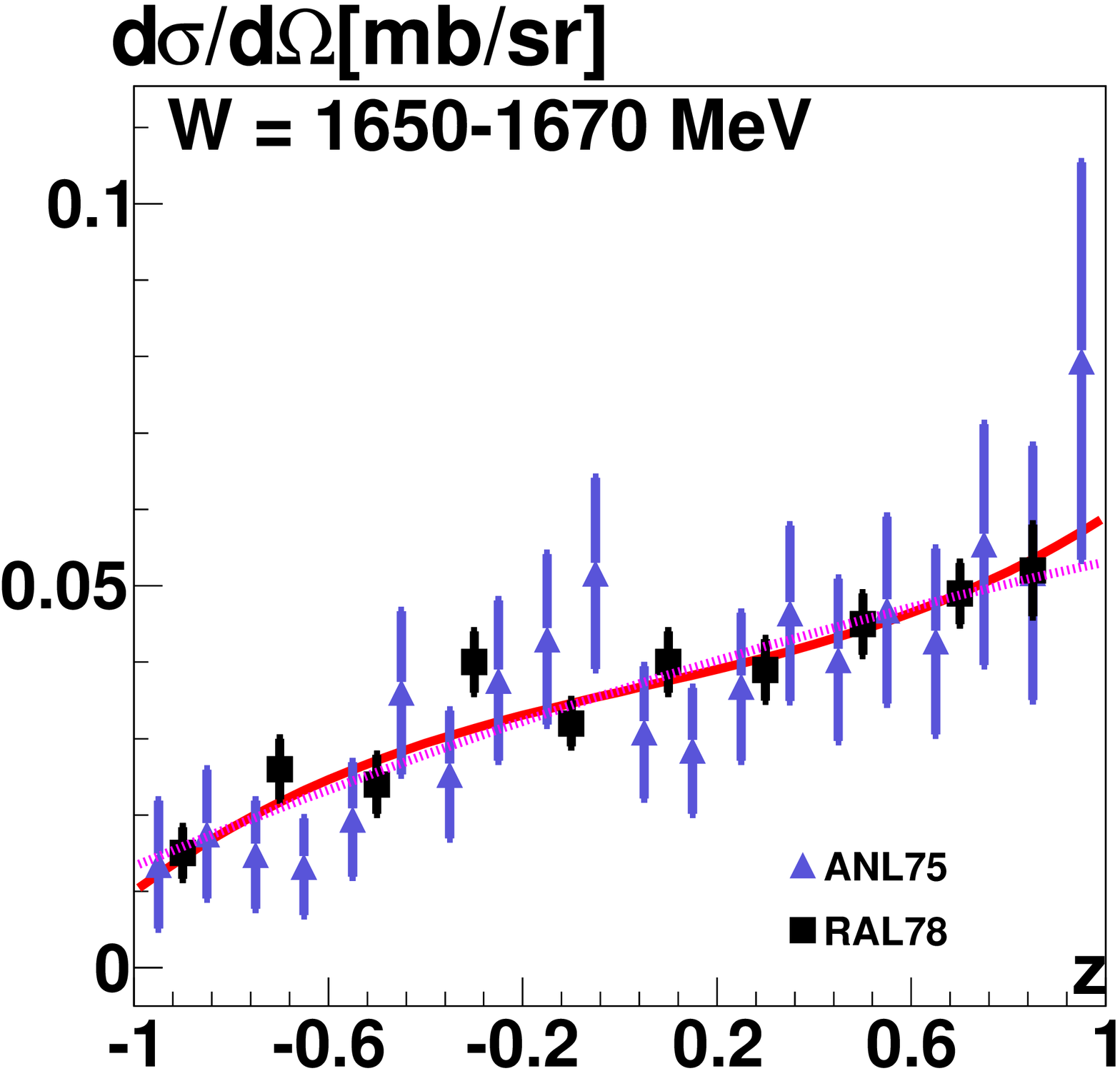}&
\hspace{-3mm}\includegraphics[width=0.21\textwidth]{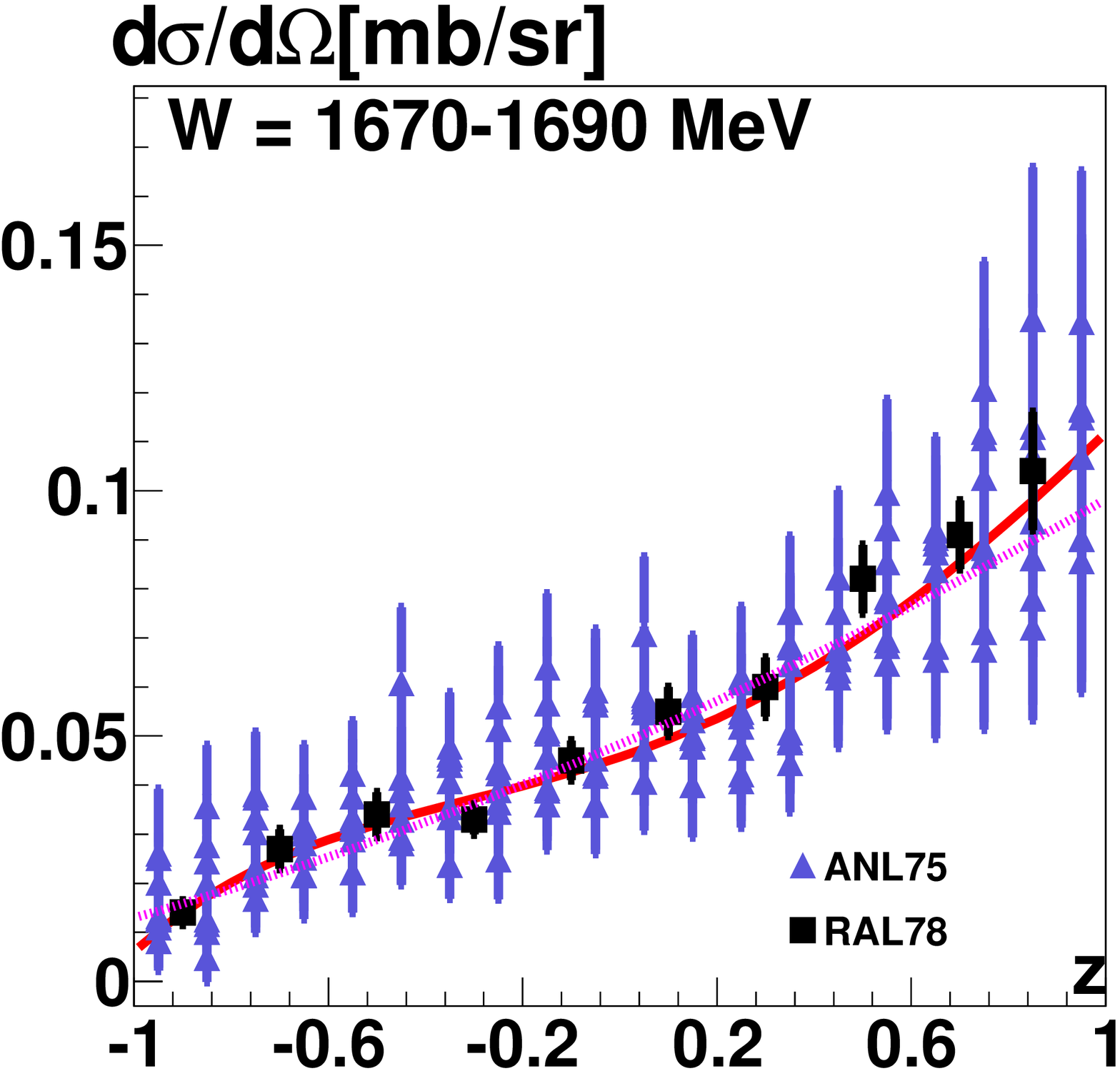}&
\hspace{-3mm}\includegraphics[width=0.21\textwidth]{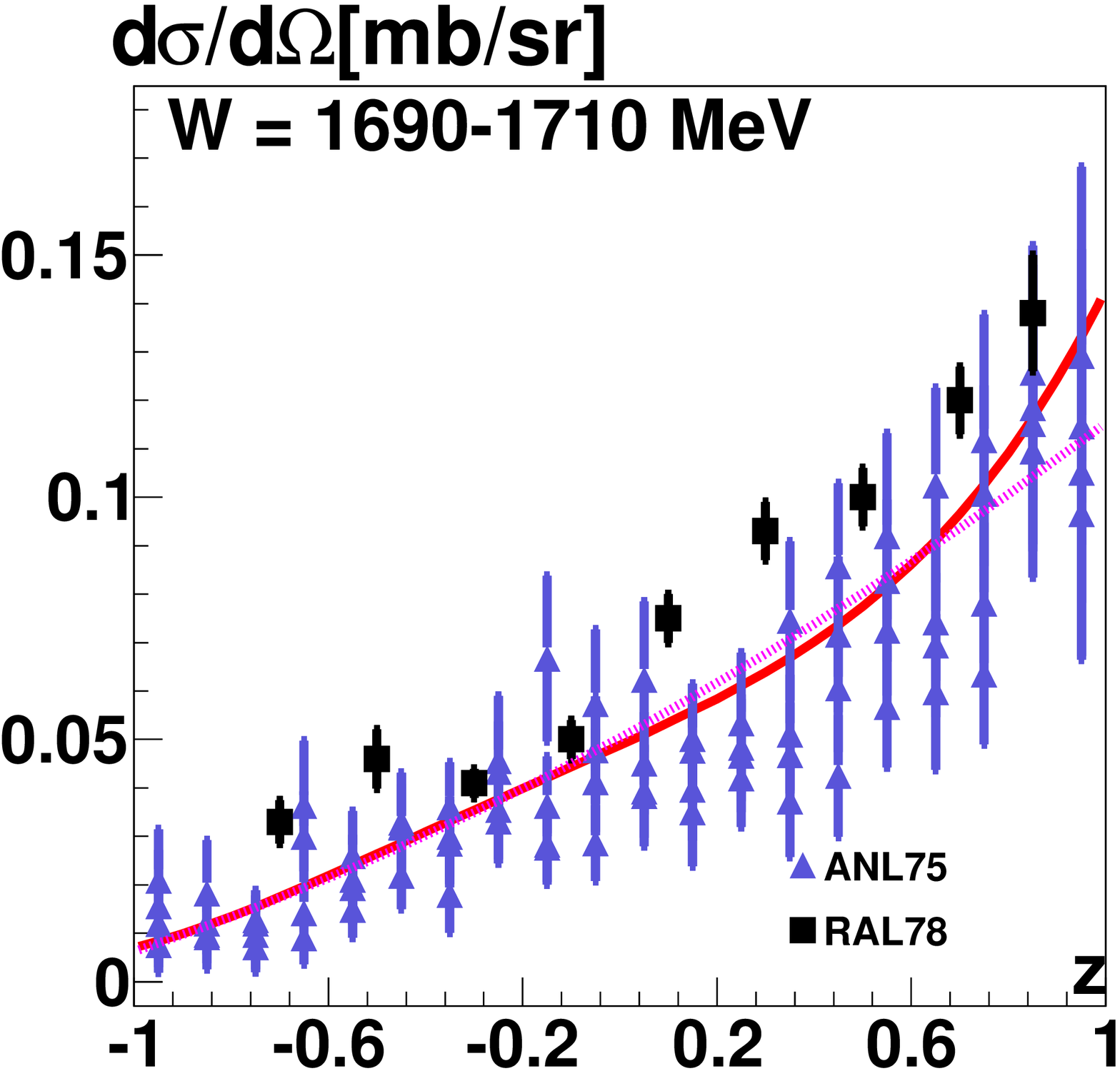}\\
\hspace{-3mm}\includegraphics[width=0.21\textwidth]{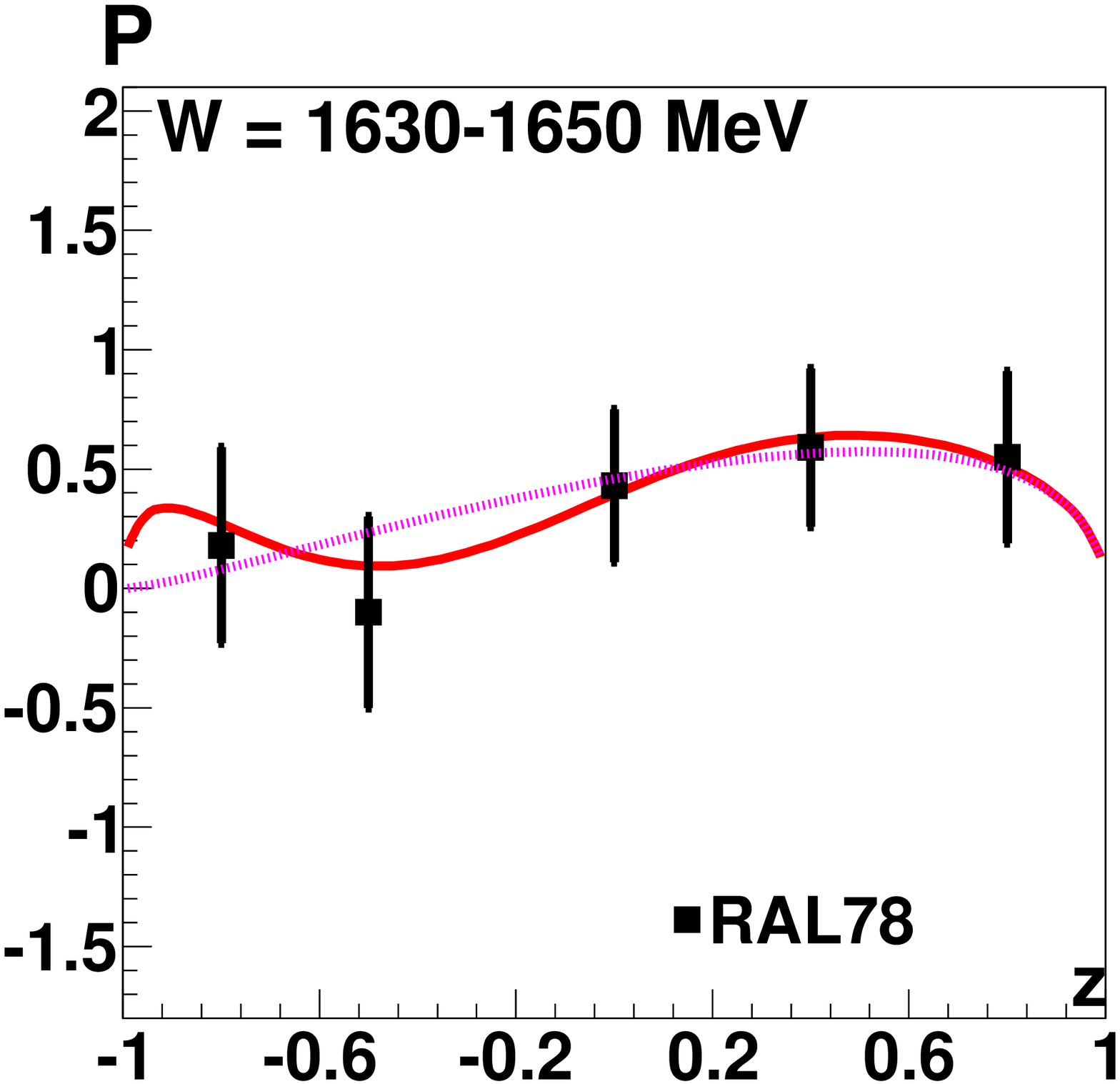}&
\hspace{-3mm}\includegraphics[width=0.21\textwidth]{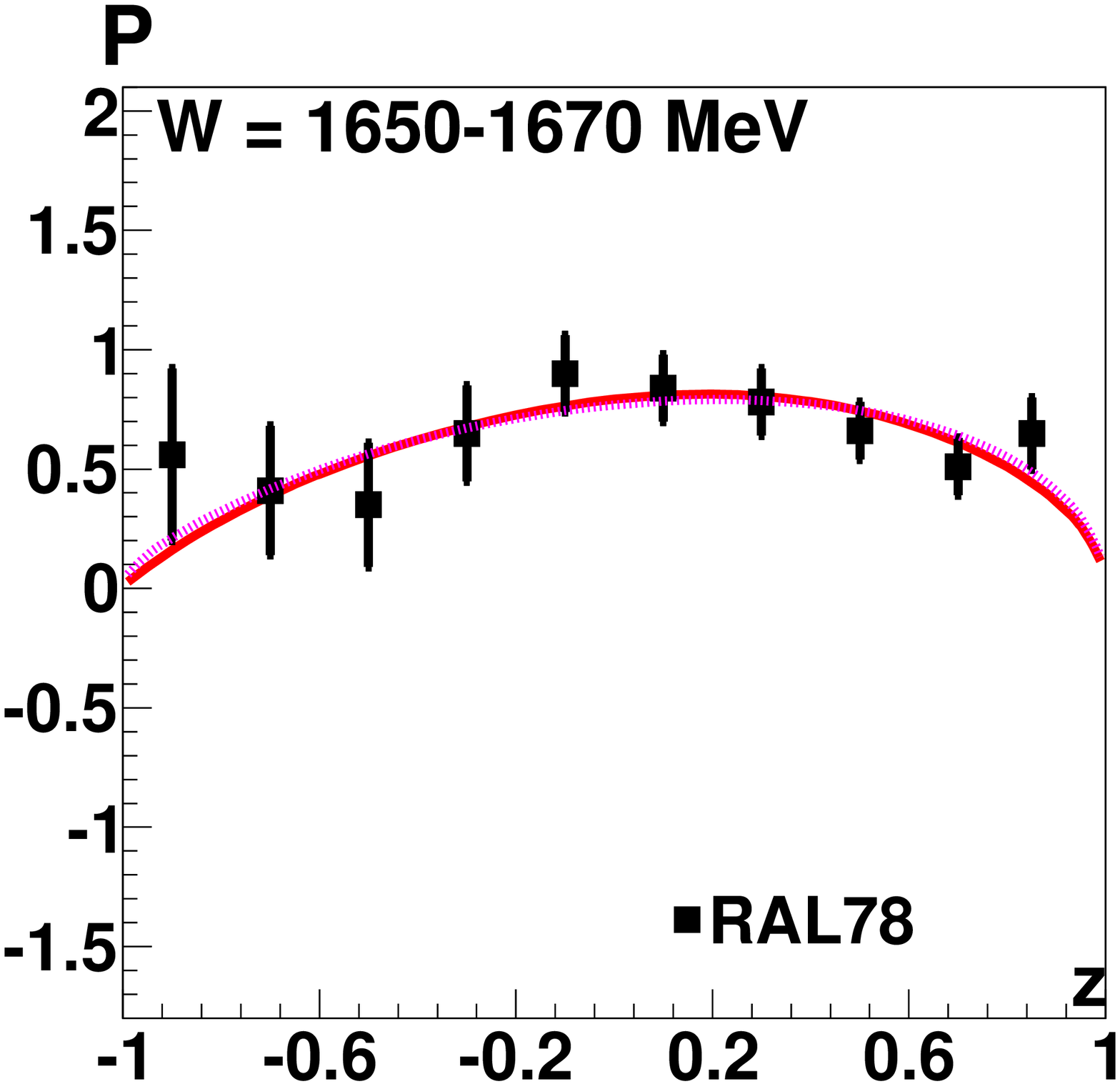}&
\hspace{-3mm}\includegraphics[width=0.21\textwidth]{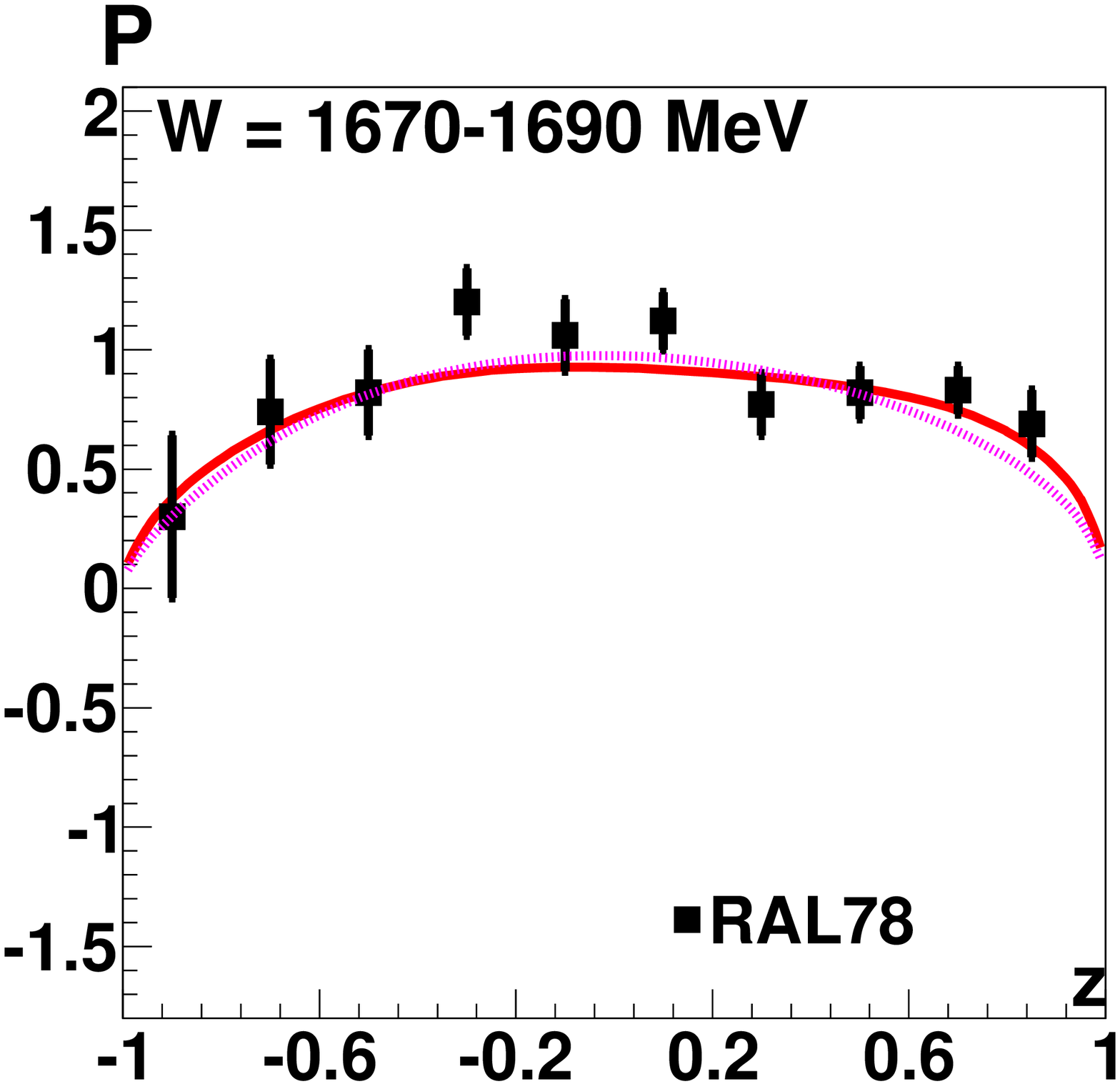}&
\hspace{-3mm}\includegraphics[width=0.21\textwidth]{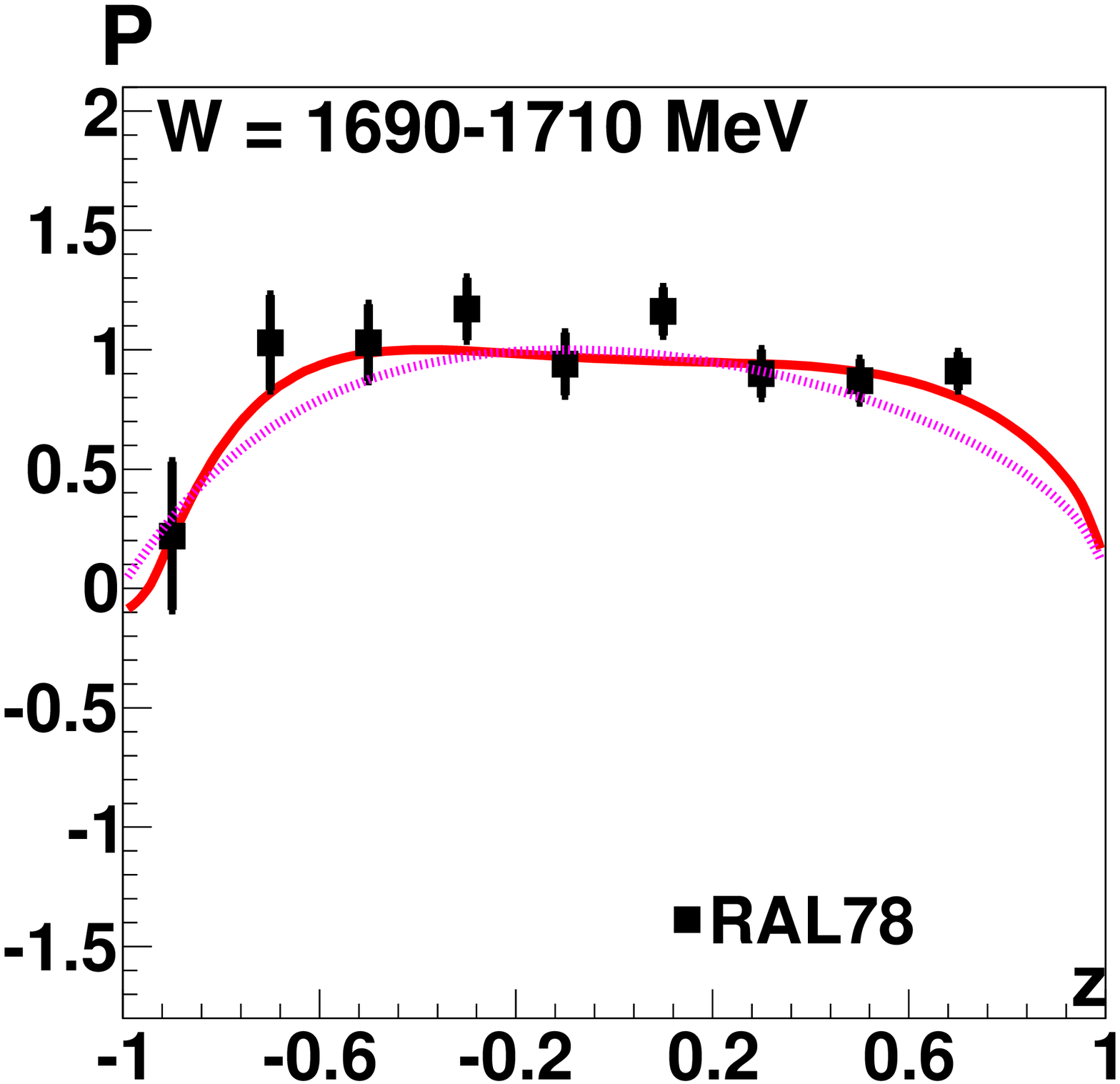}\\
\hspace{-3mm}\includegraphics[width=0.21\textwidth]{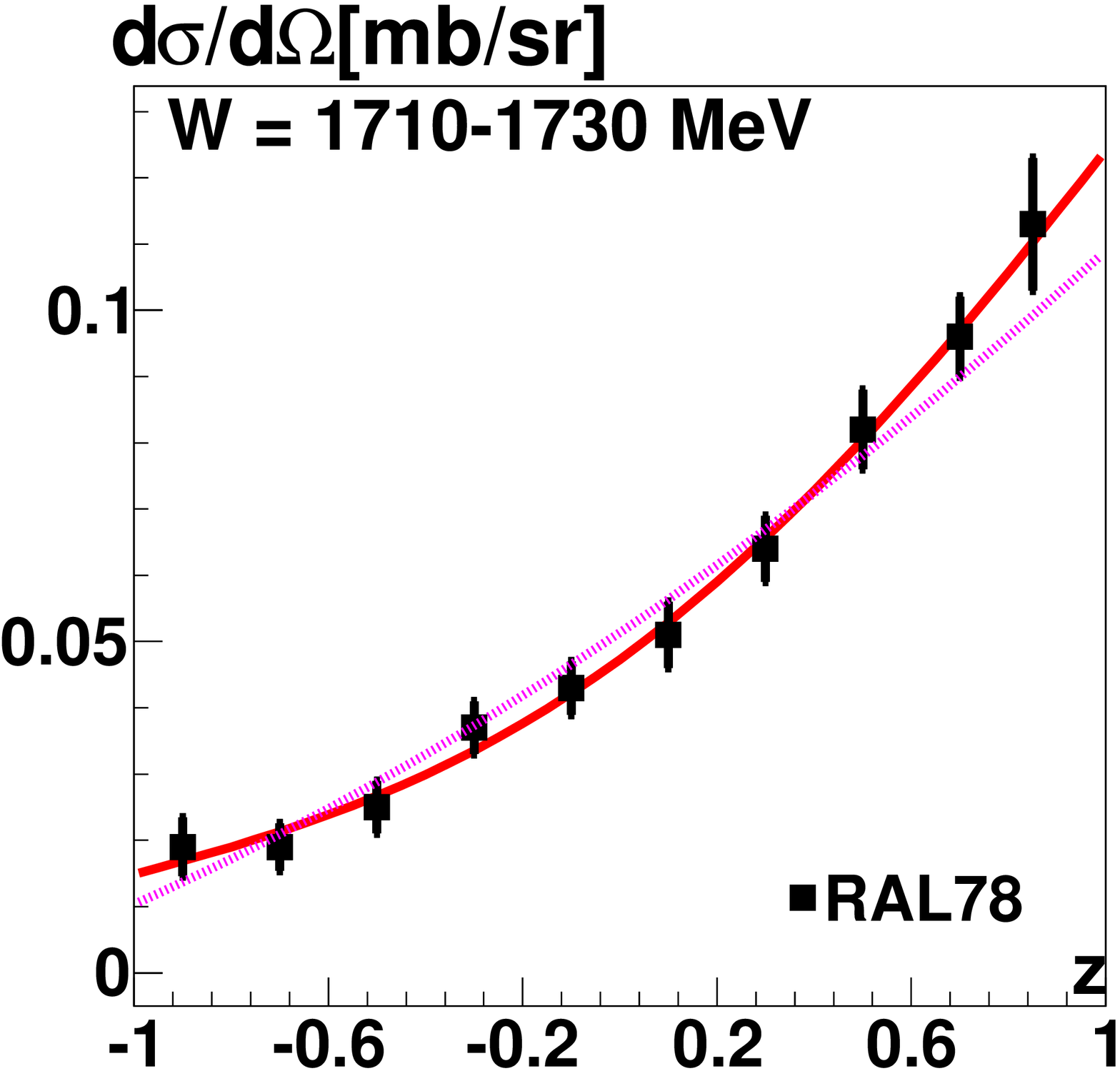}&
\hspace{-3mm}\includegraphics[width=0.21\textwidth]{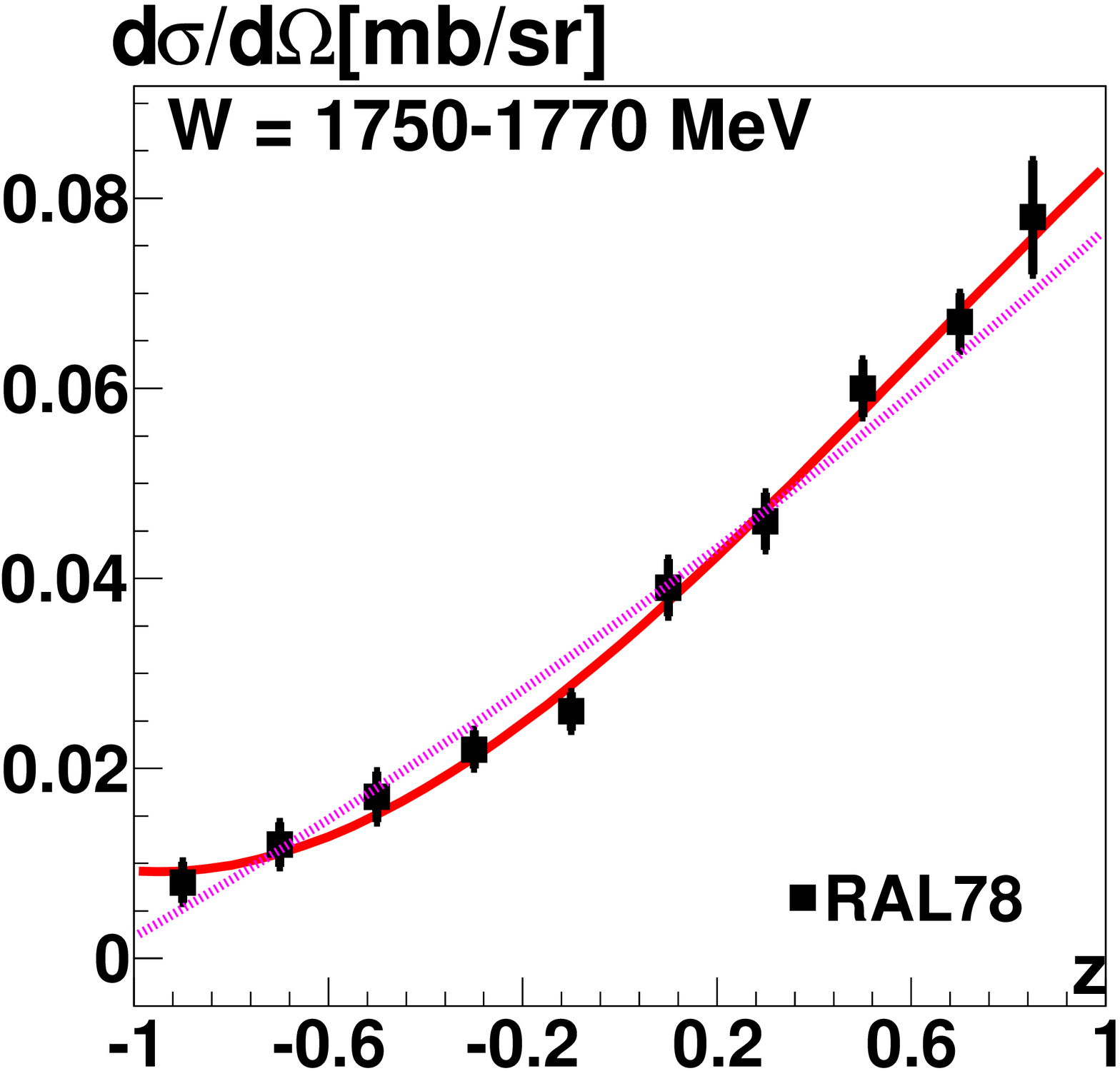}&
\hspace{-3mm}\includegraphics[width=0.21\textwidth]{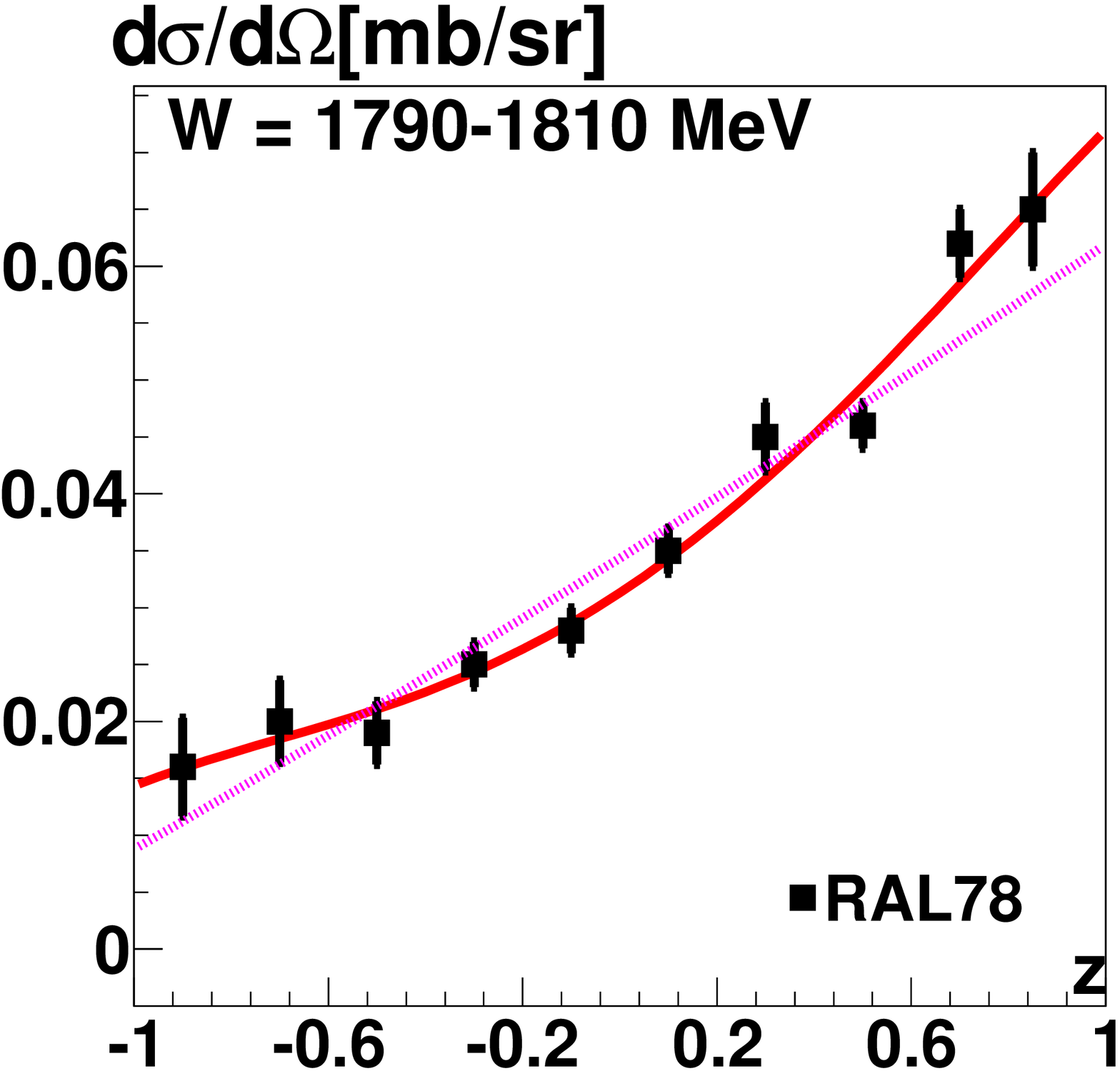}&
\hspace{-3mm}\includegraphics[width=0.21\textwidth]{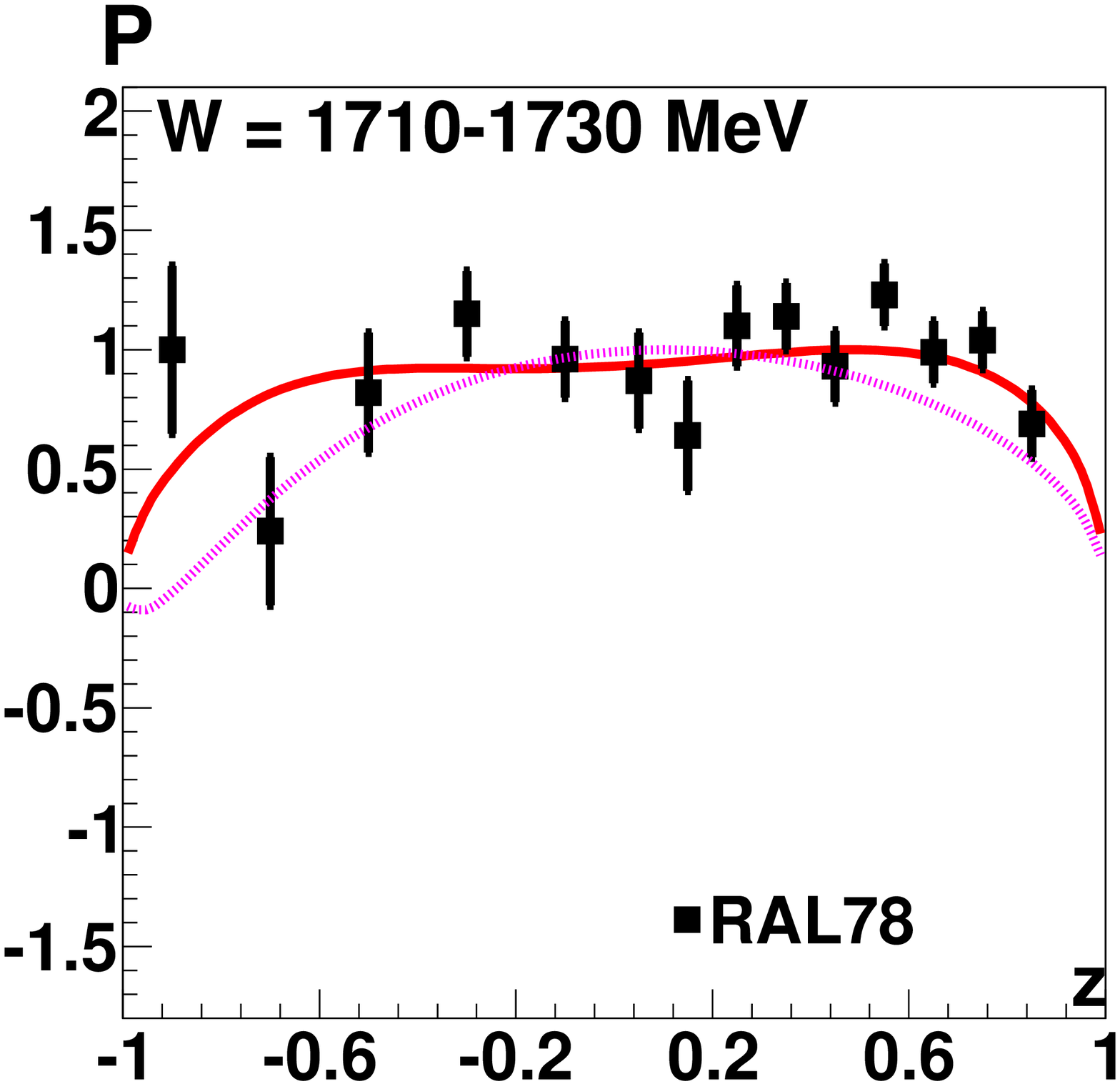}\\
\hspace{-3mm}\includegraphics[width=0.21\textwidth]{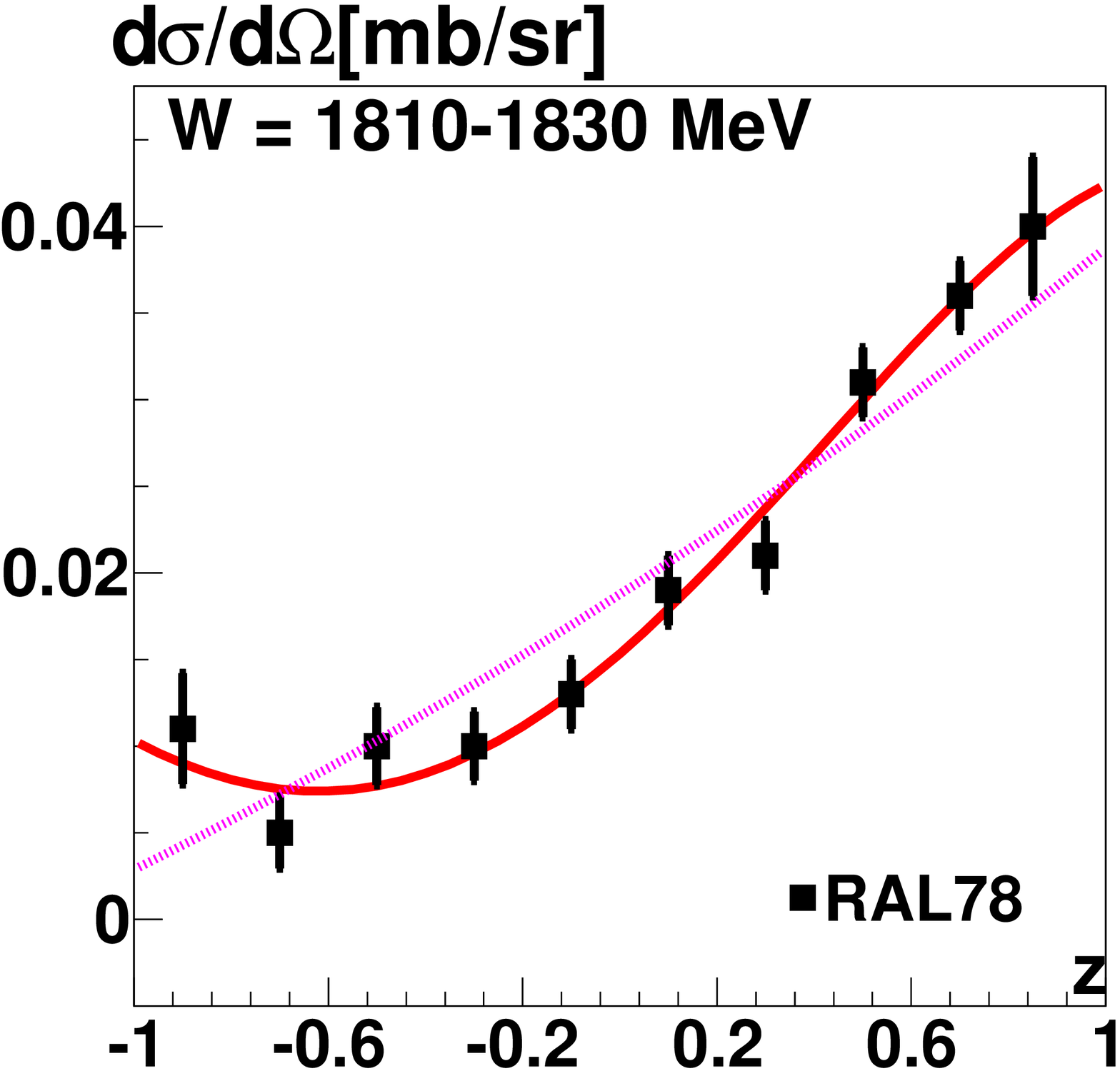}&
\hspace{-3mm}\includegraphics[width=0.21\textwidth]{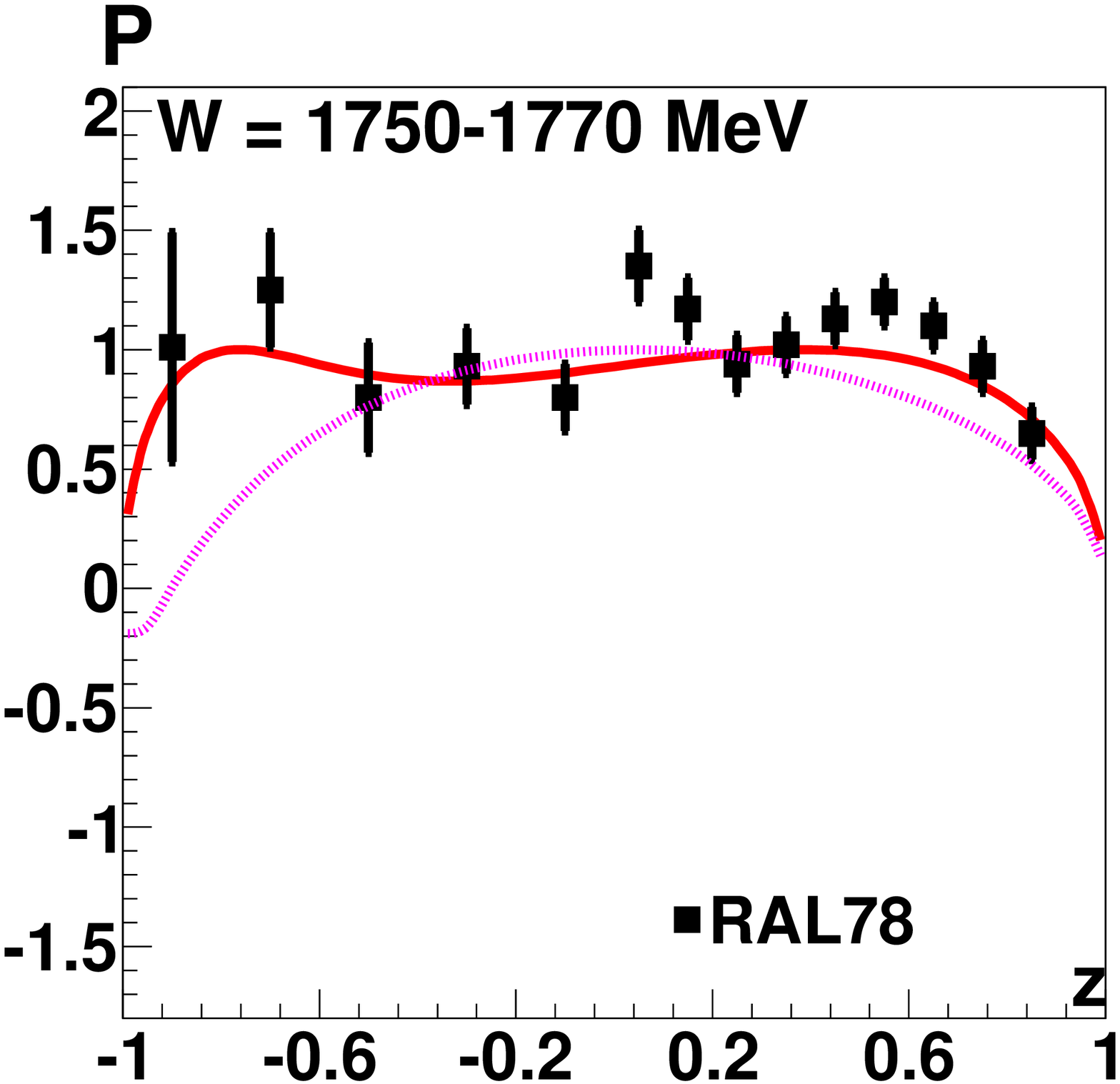}&
\hspace{-3mm}\includegraphics[width=0.21\textwidth]{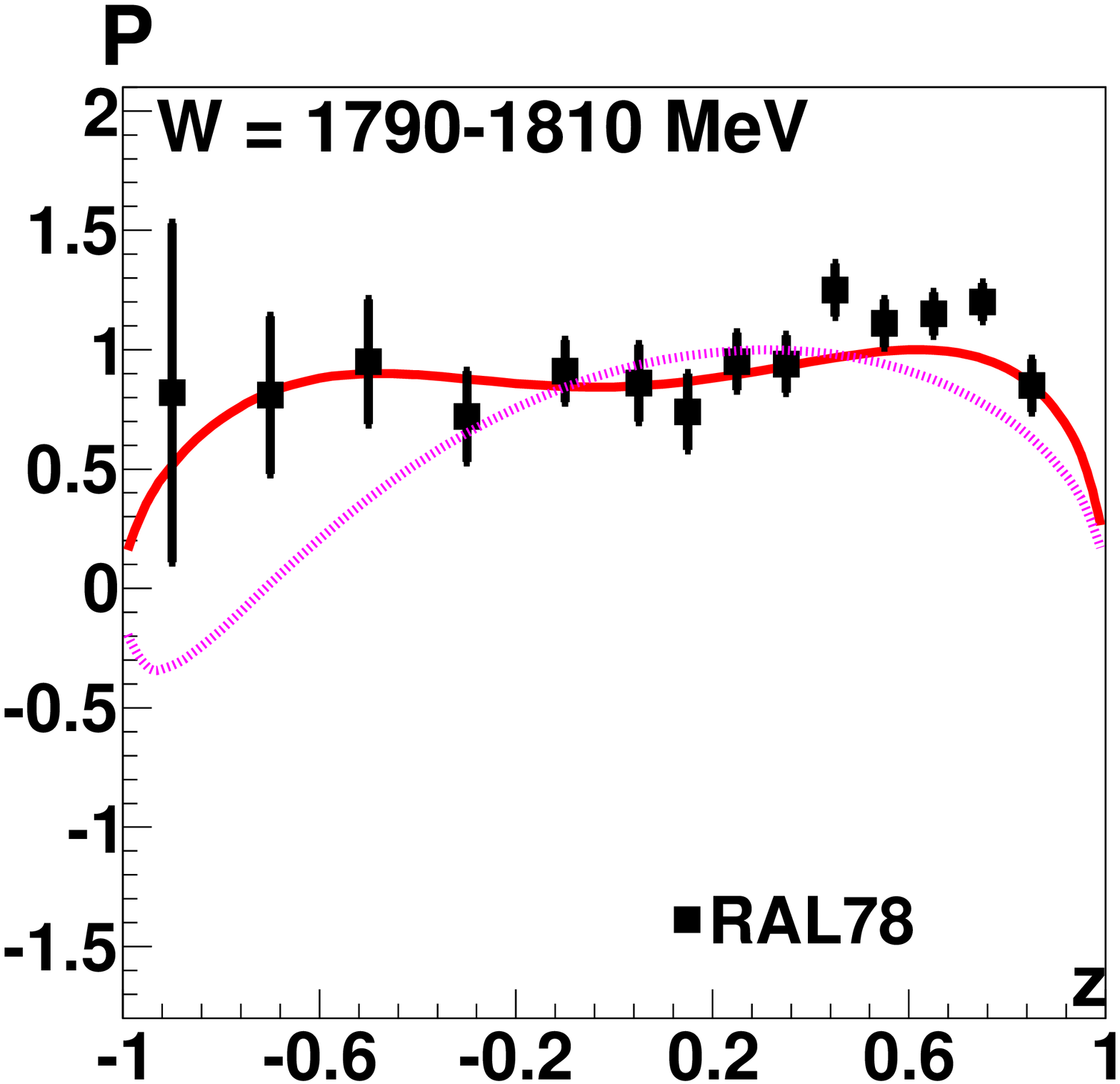}&
\hspace{-3mm}\includegraphics[width=0.21\textwidth]{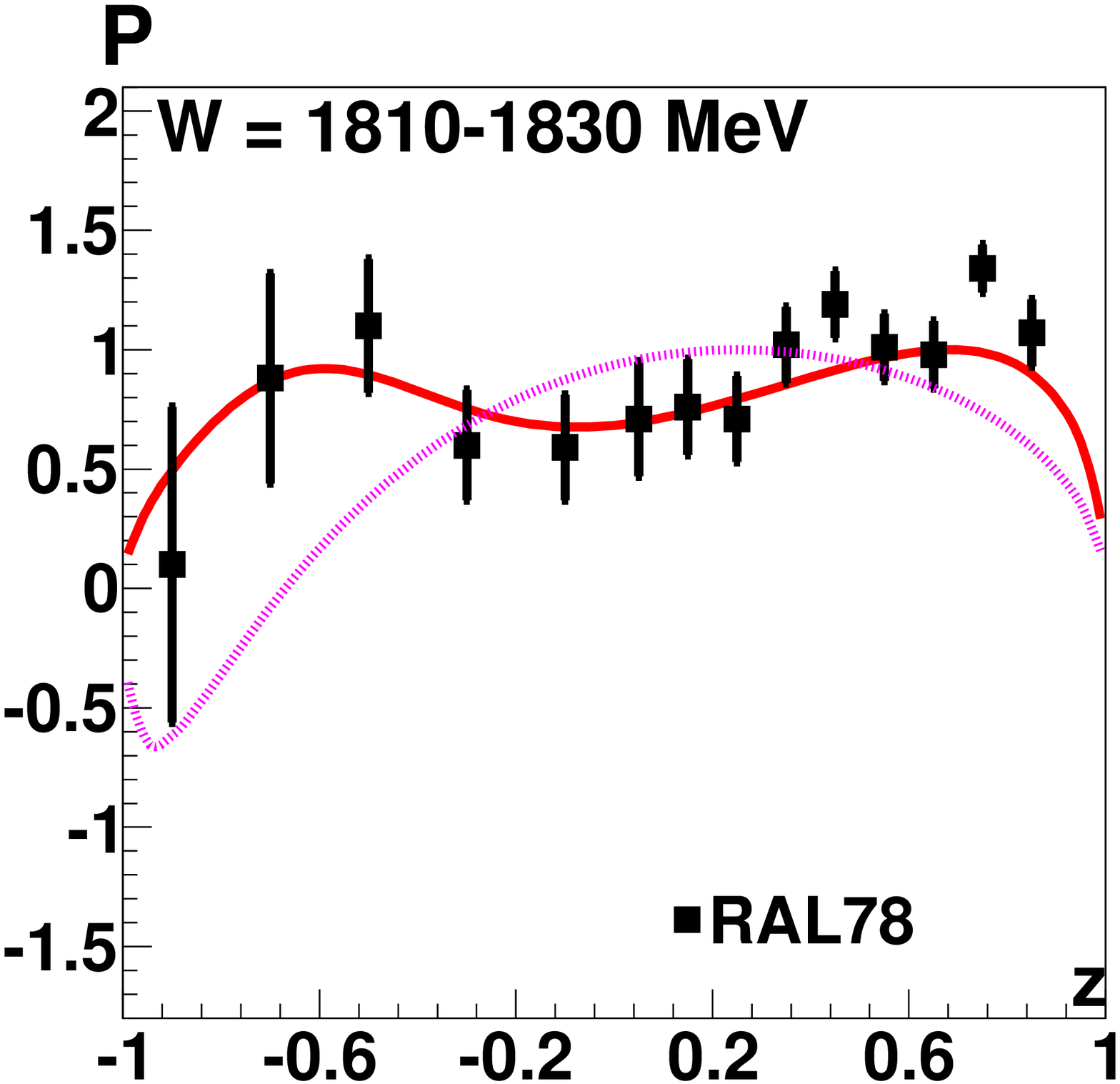}\\
\end{tabular}
\end{center}
\caption{\label{piLambdaK_obeser2}(Color online)Differential cross
sections and $\Lambda$ polarization for the reaction $\pi^- p
\rightarrow K^0 \Lambda$ from ANL75 (blue, grey)
\cite{Knasel:1975rr} and RAL78 (black) \cite{Baker:1978qm}.  Note
that a few differential cross sections from \cite{Knasel:1975rr}
fall into a single energy window.  The data are compared to two fits
using $S$ and $P$ waves (dotted line, red) and with $S$, $P$ and $D$
waves (solid line, red). The BnGa2010-02 fit to the data is shown in
figs. 1 - 4 of \cite{Anisovich:2010an}. \vspace{-0mm}}
\end{figure*}

\section{Energy-independent PWA near threshold}

We first consider an energy-independent (single-energy) fit to data
from the reaction $\pi^- p \rightarrow \Lambda K^0$ in the energy
region below $W=1830$ MeV. Experimental data on this reaction are
available for $d\sigma/d\Omega$, $Pd\sigma/d\Omega$ and $P$
\cite{Knasel:1975rr,Baker:1978qm}. The experimental data, divided
into $20$ MeV bins, are shown on Fig.~\ref{piLambdaK_obeser2}. In
this region the differential cross section has some small $z^2$
dependence, hence we expect that at least three partial waves need
to be included: $S_{11}$, $P_{11}$ and $P_{13}$. We then compare our
energy independent solutions with an energy dependent solution. The
latter was obtained from a multichannel fit to a large body of
photo- and pion-induced reactions \cite{Anisovich:2011fc}.


\subsection{\boldmath Energy independent PWA near threshold with $S$ and $P$ waves}

\begin{table}[pt] \caption{\label{Table} Quality of the
$S$ and $P$ waves energy independent fit: $\chi^2/N_{\rm data}$ and
number of data points (in brackets). The overall $\chi^2/N_{\rm
data}$ is 1.41}
\renewcommand{\arraystretch}{1.08}
\begin{center}
\begin{tabular}{|c|c|c|c|c|}
  \hline
  Energy bin& ANL75 & ANL75 & RAL78& RAL78 \\
            & $d\sigma/d\Omega$& $Pd\sigma/d\Omega$&$d\sigma/d\Omega$& P\\
  \hline
  1630 - 1650 & 0.8 (20)    & 1.3 (10) & 0.2 (5)  & 0.16 (5) \\
  1650 - 1670 & 0.5 (20)    & 0.6(10)& 0.9 (10)& 0.5 (10) \\
  1670 - 1690 & 0.7 (160)   & 0.9(80)& 1.0 (10) & 1.2 (10) \\
  1690 - 1710 & 1.4 (80)    & 0.5(40) &   10 (9) & 2.4 (9) \\
  1710 - 1730 & - &     -      & 0.8 (10)& 2.8 (14) \\
  1750 - 1770 & - &     -      & 1.8 (10)& 4.3 (14) \\
  1790 - 1810 & - &     -      & 2.0 (10)& 4.7 (14) \\
  1810 - 1830 & - &     -      & 2.2 (10)& 5.1 (14) \\
  \hline
\end{tabular}
\end{center}\vspace{-6mm}
\renewcommand{\arraystretch}{1.0}
\end{table}

The data were fitted with this hypothesis, the fit is shown by a
dotted line in Fig.~\ref{piLambdaK_obeser2}; the quality of the fit
for each energy bin is given in Table \ref{Table}.

\begin{table}[pb]
\caption{\label{Table_D} Quality of the $S$, $P$ and $D$ waves
energy independent fit: $\chi^2/N_{\rm data}$ and number of data
points (in brackets). The overall $\chi^2/N_{\rm data}$ is now 0.98}
\renewcommand{\arraystretch}{1.08}\begin{center}
\begin{tabular}{|c|c|c|c|c|}
  \hline
  Energy bin& ANL75 & ANL75 & RAL78& RAL78 \\
            & $d\sigma/d\Omega$& $Pd\sigma/d\Omega$&$d\sigma/d\Omega$& P\\
  \hline
  1630-1650 & 0.8 (20) &1.3(10)& 0.15 (5)  & 0.06 (5) \\
  1650-1670 & 0.6 (20) &0.6(10)& 0.8 (10)& 0.6 (10) \\
  1670-1690 & 0.65 (160) &0.8(80)& 0.75 (10) & 1.0 (10) \\
  1690-1710 & 1.2 (80) &0.5(40)& 10 (9) & 1.0 (9) \\
  1710-1730 & - &- &0.2 (10) & 1.15 (14) \\
  1750-1770 & - &- &0.4 (10) & 1.6 (14) \\
  1790-1810 & - &- &0.7 (10) & 1.5 (14) \\
  1810-1830 & - &- &0.5 (10) & 1.5 (14) \\
  \hline
\end{tabular}\vspace{-4mm}
\end{center}
\renewcommand{\arraystretch}{1.0}
\end{table}

In the lowest energy region, below 1700\,MeV, the fit agrees
reasonably well with the data but significant deviations between
data and fit are observed above this energy. Hence we expect that
higher partial waves are needed, at least above 1700\,MeV.
Nevertheless, we determined the amplitudes from this fit.

The amplitudes for the $S_{11}$, $P_{11}$, and $P_{13}$ waves are
determined from the data fit using
eqs.~(\ref{spwave0}-\ref{spwave2}). One overall phase remains
undetermined, hence the phases relative to the $S_{11}$ phase are
plotted. The latter phase is taken from the energy dependent
BnGa2011-02. The amplitude and phase errors correspond to an
increase of the full $\chi^2$ by 1 when the plotted parameter
(magnitude or phase) is changed and all other parameters are
refitted. From the best fit we use eqs.~(\ref{aa1} -\ref{aa3}) for a
numerical search for further solutions.  Even in this simplest case
there is no unambiguous solution. For each energy bin, we found two
different physical solutions.

\begin{figure}[pt]
\begin{center}
\begin{tabular}{cc}
\hspace{-2mm}\includegraphics[width=0.22\textwidth]{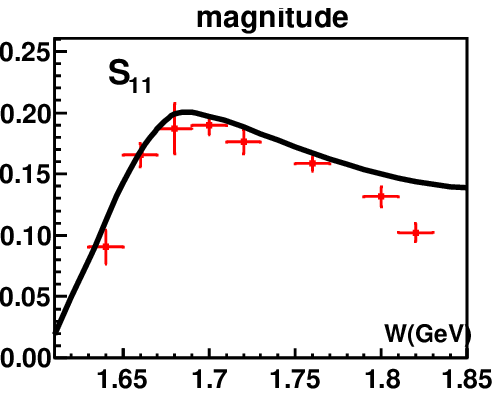}&
\hspace{-2mm}\includegraphics[width=0.22\textwidth]{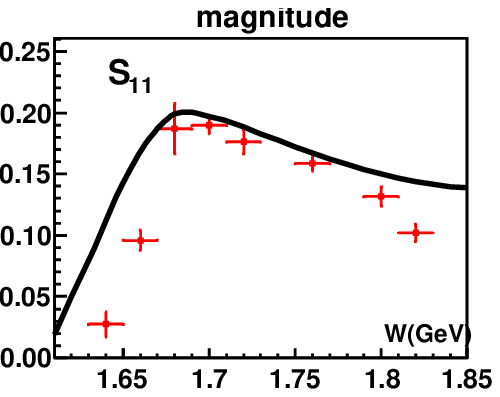}\\
\hspace{-2mm}\includegraphics[width=0.22\textwidth]{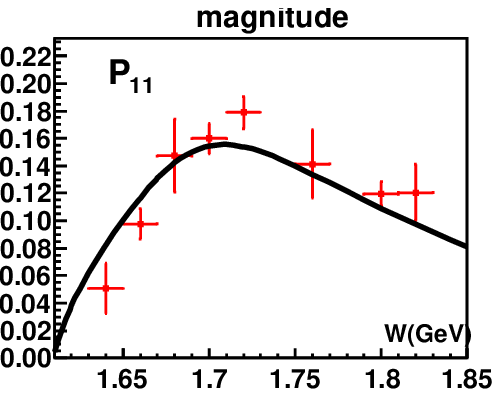}&
\hspace{-2mm}\includegraphics[width=0.22\textwidth]{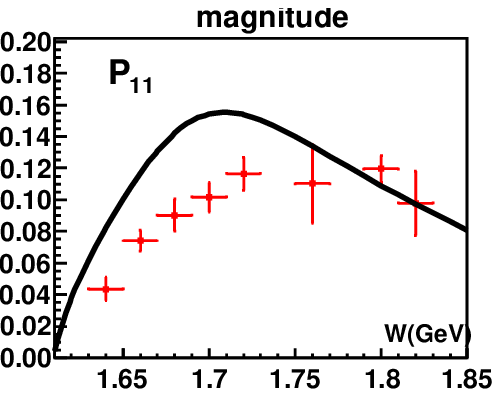}\\
\hspace{-2mm}\includegraphics[width=0.22\textwidth]{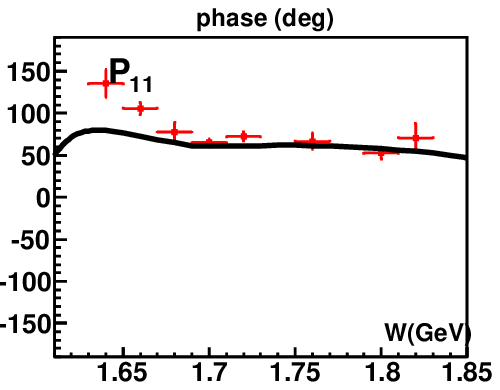}&
\hspace{-2mm}\includegraphics[width=0.22\textwidth]{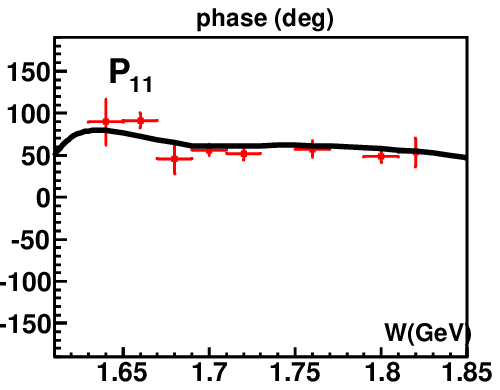}\\
\hspace{-2mm}\includegraphics[width=0.22\textwidth]{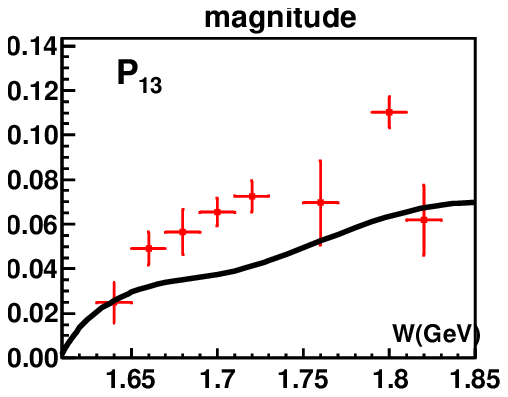}&
\hspace{-2mm}\includegraphics[width=0.22\textwidth]{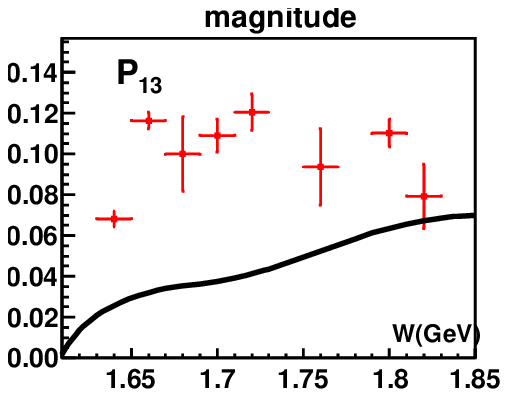}\\
\hspace{-2mm}\includegraphics[width=0.22\textwidth]{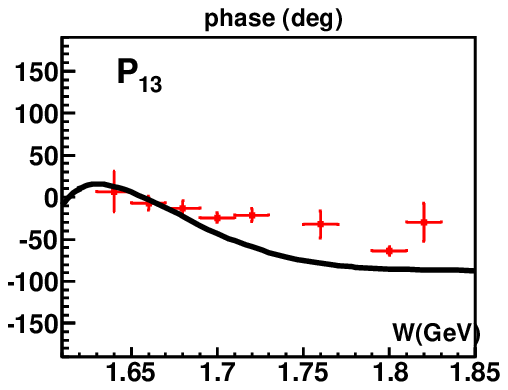}&
\hspace{-2mm}\includegraphics[width=0.22\textwidth]{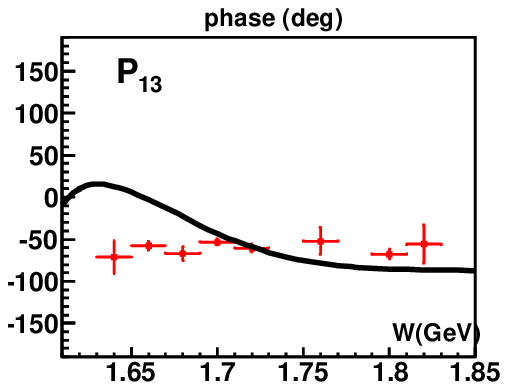}
\end{tabular}
\end{center}
\caption{\label{piLambdaK1}Two solutions (left and right)
for the decomposition of the $\pi N\to \Lambda K$  scattering
amplitude with $S$ and $P$ waves. A large number of further solutions can be drawn
by arbitrary choices of points from the left or right sub-figure.
The solid line is the energy
dependent solution BnGa2011-02.
}\vspace{-4mm}
\end{figure}

For every energy bin, one of the two solutions can be chosen, giving
a multitude of different energy dependencies of the three
amplitudes. In Fig.~\ref{piLambdaK1} the two solutions are sorted
according their proximity (in terms of $\chi^2$) to the energy
dependent solution BnGa2011-02. The ``best" solution agrees moderately
well with BnGa2011-02 with $\chi^2/N=262/40$.


 The left column in Fig.~\ref{piLambdaK1} shows the
energy independent solution (represented by ``data" points with
error bars) where the three magnitudes and the two phases are better
compatible with the energy dependent solution (represented by the
curves). The $S_{11}$ and $P_{11}$ amplitudes are reasonably
consistent with the energy dependent fit even though the phase of
the $P_{11}$ wave shows some discrepancy. The $P_{13}$ magnitude is
overestimated over a wide energy range, this could be due to the
neglect of higher waves.

The two solutions have similar $S_{11}$ amplitudes even though the
threshold behavior is different. Sizable differences are seen in
the magnitudes of the $P_{11}$ and $P_{13}$ amplitudes. Solution~1
is close to the energy-dependent solution but the analysis suggest
that the energy depend fit might underestimate the $P_{13}$
amplitude.

In the 1690 to 1710\,MeV mass slice, the differential cross section
$d\sigma/d\Omega$ from ANL75 \cite{Knasel:1975rr}  and RAL78
\cite{Baker:1978qm} are not consistent. To study the importance of
this effect, we left out the ANL75 data \cite{Knasel:1975rr}. The
reconstructed amplitudes changed a bit, the error bars increased but
the conclusions remained unchanged.

\subsection{\boldmath Energy independent PWA near threshold with $S$, $P$ and $D$ waves}

\begin{figure*}
\begin{center}
\begin{tabular}{ccccc}
\hspace{0mm}\includegraphics[width=0.185\textwidth]{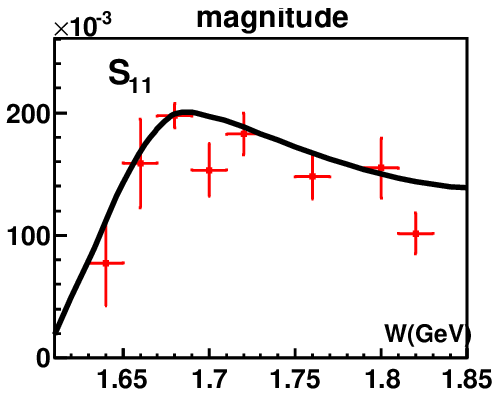}&
\hspace{-3mm}\includegraphics[width=0.185\textwidth]{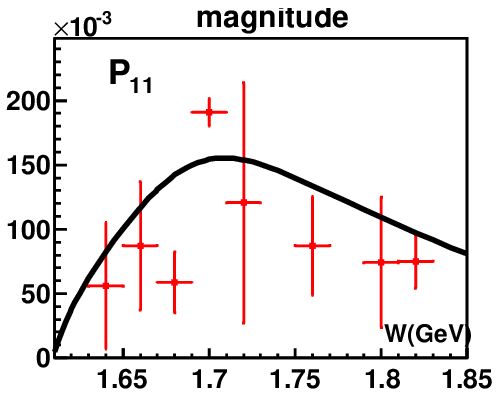}&
\hspace{-3mm}\includegraphics[width=0.185\textwidth]{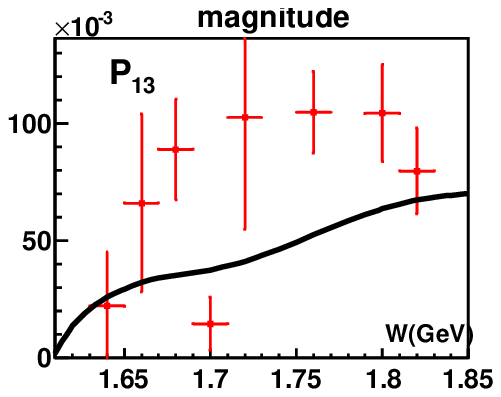}&
\hspace{-3mm}\includegraphics[width=0.185\textwidth]{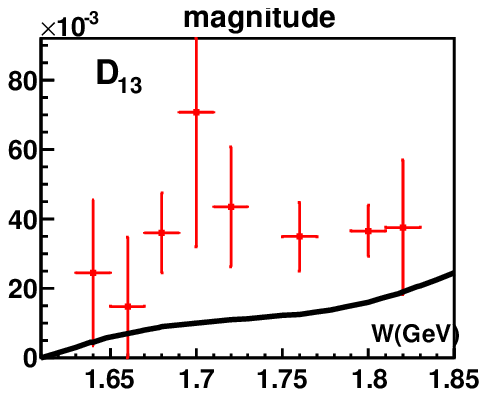}&
\hspace{-3mm}\includegraphics[width=0.185\textwidth]{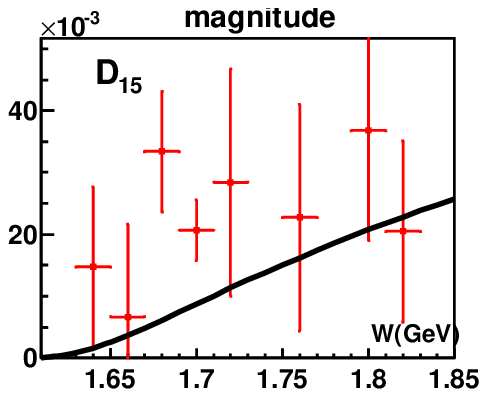}
\\
&
\hspace{-3mm}\includegraphics[width=0.185\textwidth]{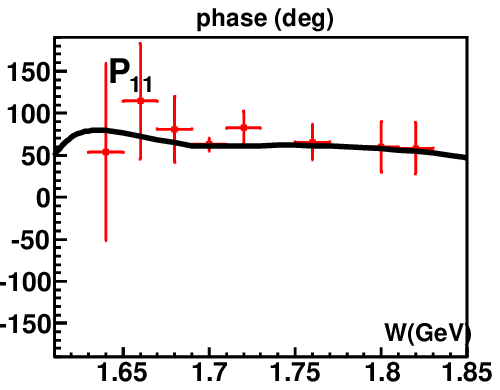}&
\hspace{-3mm}\includegraphics[width=0.185\textwidth]{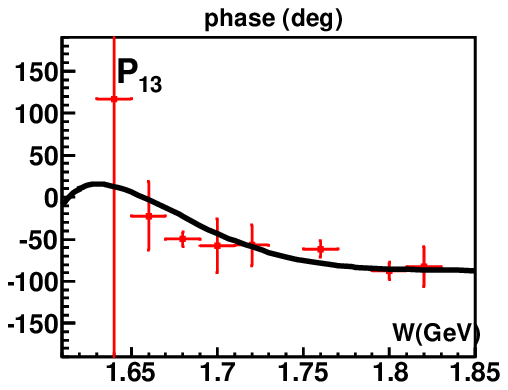}&
\hspace{-3mm}\includegraphics[width=0.185\textwidth]{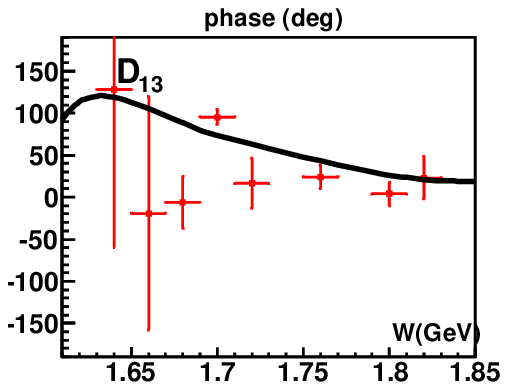}&
\hspace{-3mm}\includegraphics[width=0.185\textwidth]{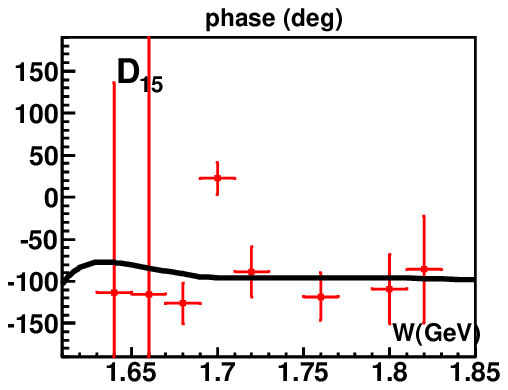}
\\
\hspace{0mm}\includegraphics[width=0.185\textwidth]{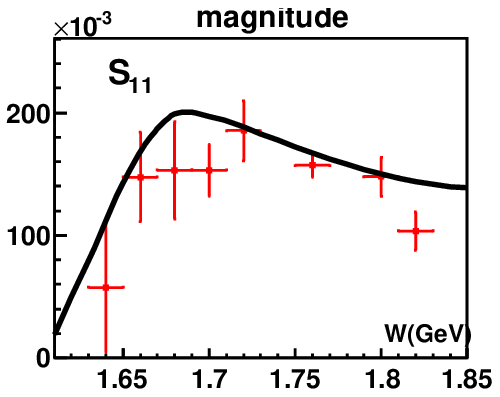}&
\hspace{-3mm}\includegraphics[width=0.185\textwidth]{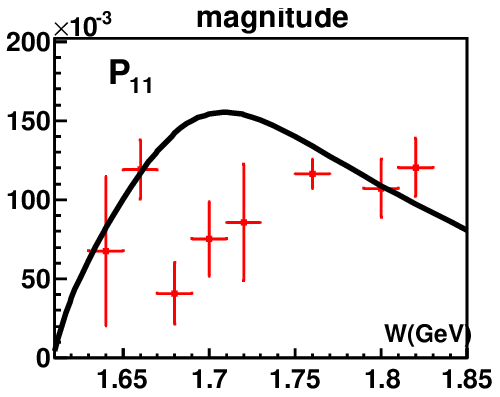}&
\hspace{-3mm}\includegraphics[width=0.185\textwidth]{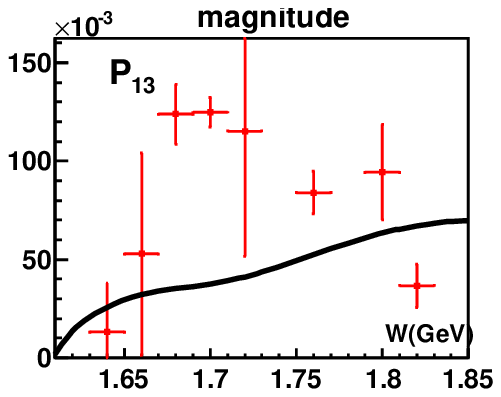}&
\hspace{-3mm}\includegraphics[width=0.185\textwidth]{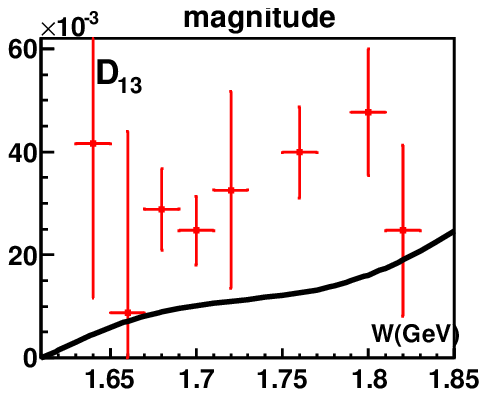}&
\hspace{-3mm}\includegraphics[width=0.185\textwidth]{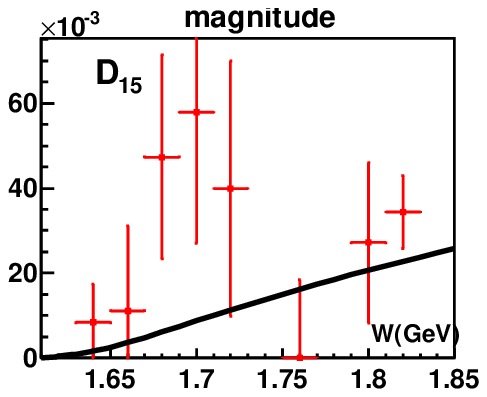}
\\
&
\hspace{-3mm}\includegraphics[width=0.185\textwidth]{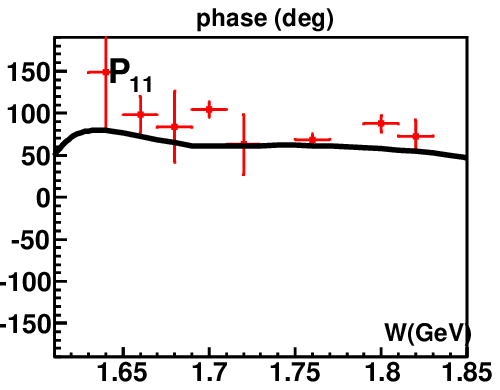}&
\hspace{-3mm}\includegraphics[width=0.185\textwidth]{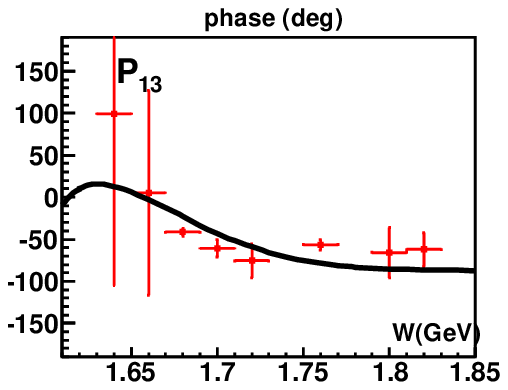}&
\hspace{-3mm}\includegraphics[width=0.185\textwidth]{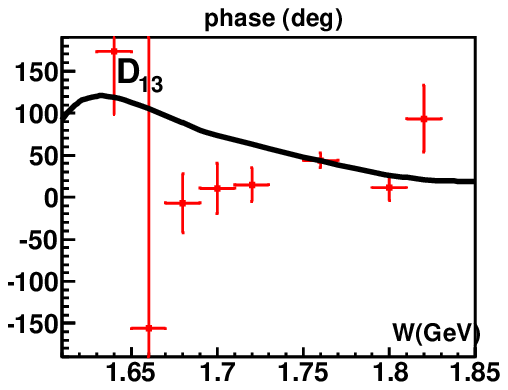}&
\hspace{-3mm}\includegraphics[width=0.185\textwidth]{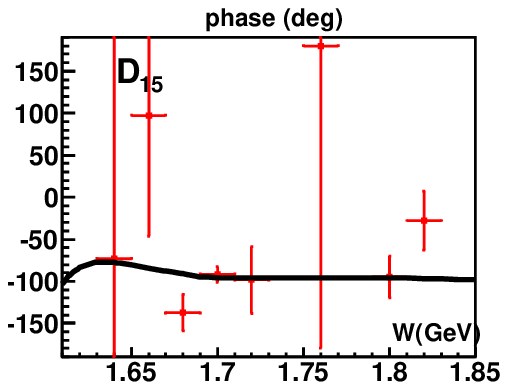}
\\
\hspace{0mm}\includegraphics[width=0.185\textwidth]{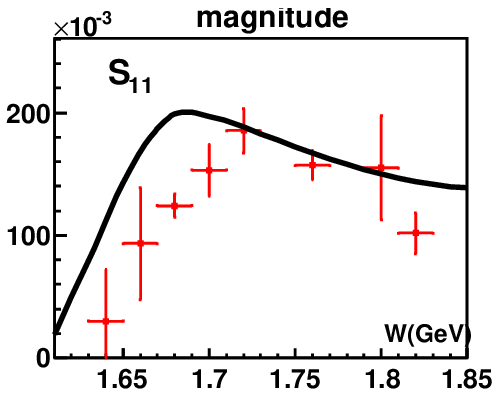}&
\hspace{-3mm}\includegraphics[width=0.185\textwidth]{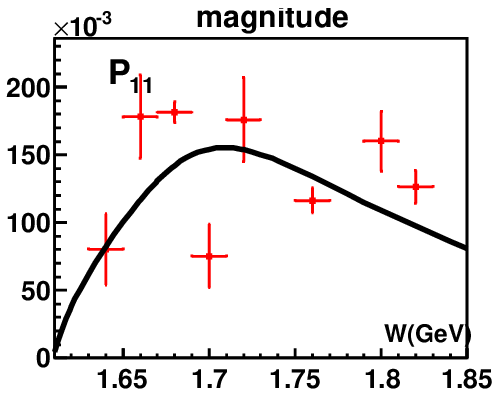}&
\hspace{-3mm}\includegraphics[width=0.185\textwidth]{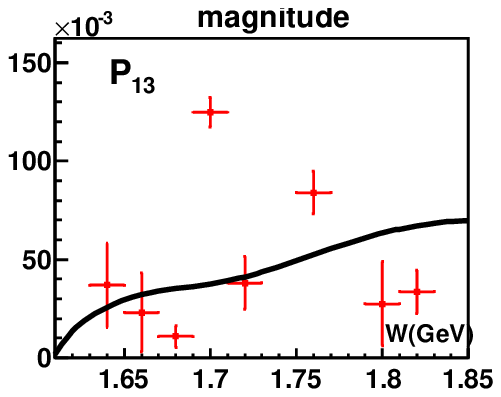}&
\hspace{-3mm}\includegraphics[width=0.185\textwidth]{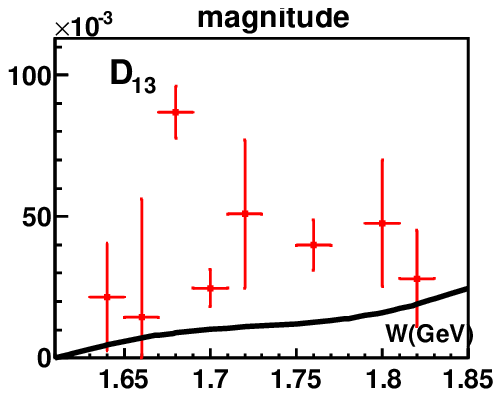}&
\hspace{-3mm}\includegraphics[width=0.185\textwidth]{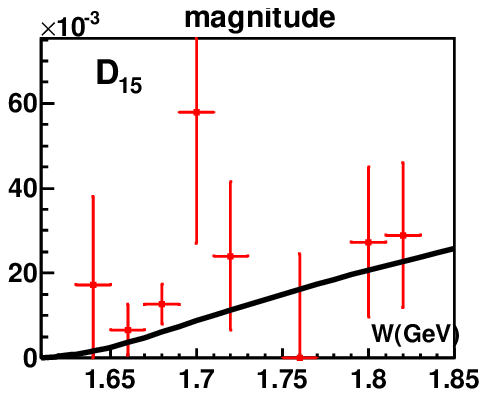}
\\
&
\hspace{-3mm}\includegraphics[width=0.185\textwidth]{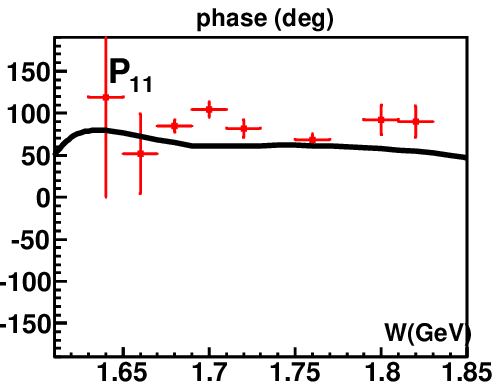}&
\hspace{-3mm}\includegraphics[width=0.185\textwidth]{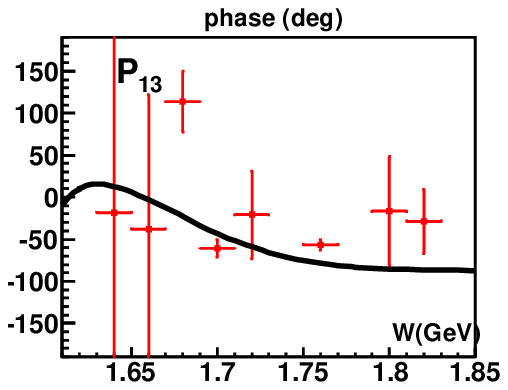}&
\hspace{-3mm}\includegraphics[width=0.185\textwidth]{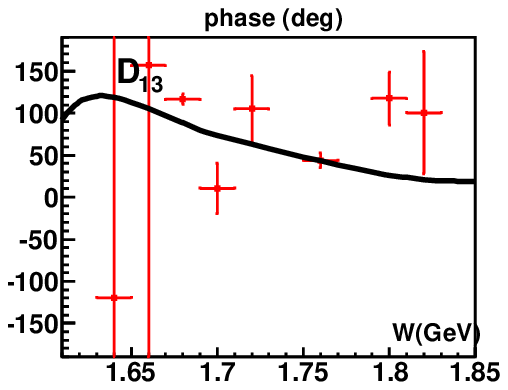}&
\hspace{-3mm}\includegraphics[width=0.185\textwidth]{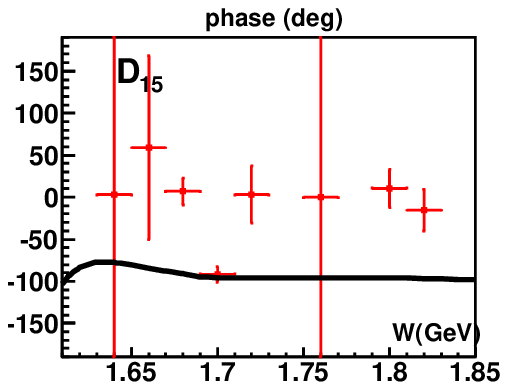}
\\
\hspace{0mm}\includegraphics[width=0.185\textwidth]{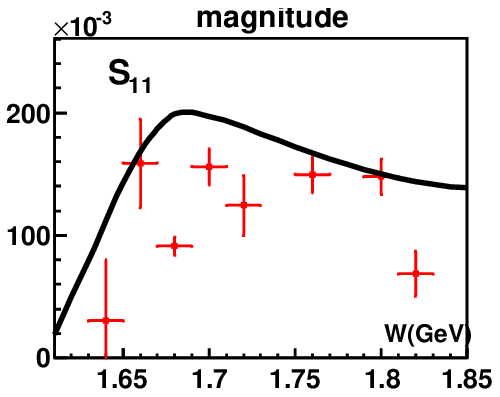}&
\hspace{-3mm}\includegraphics[width=0.185\textwidth]{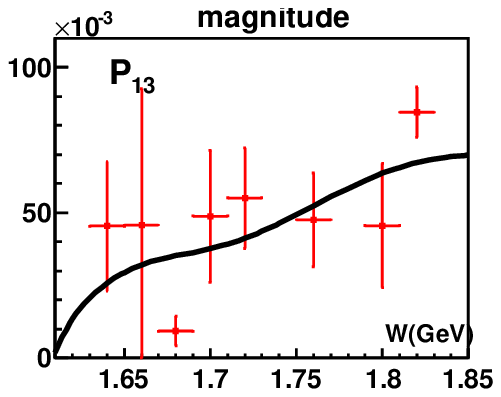}&
\hspace{-3mm}\includegraphics[width=0.185\textwidth]{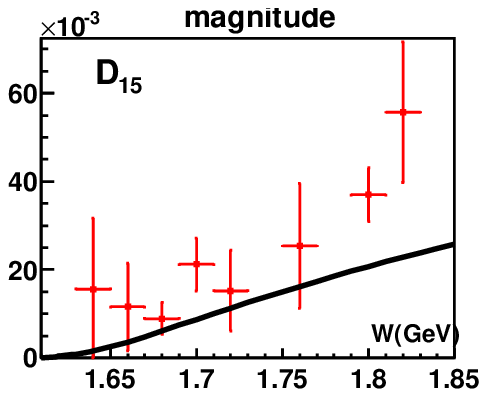}&
\hspace{-3mm}\includegraphics[width=0.185\textwidth]{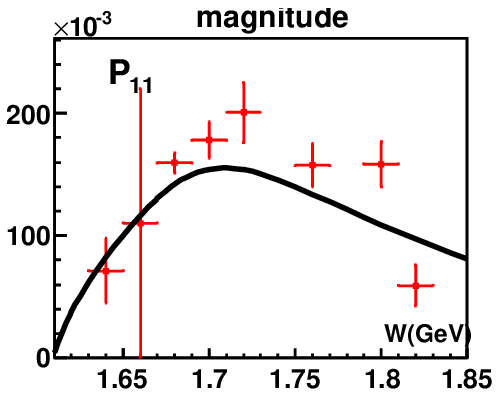}&
\hspace{-3mm}\includegraphics[width=0.185\textwidth]{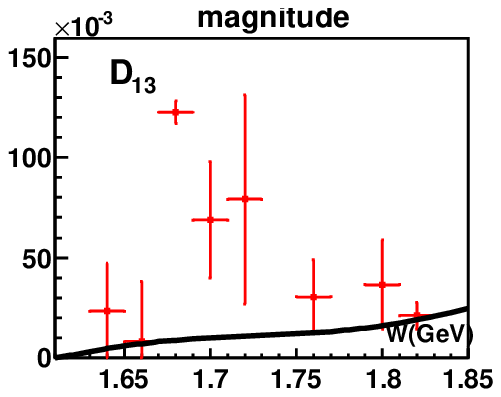}
\\
&
\hspace{-3mm}\includegraphics[width=0.185\textwidth]{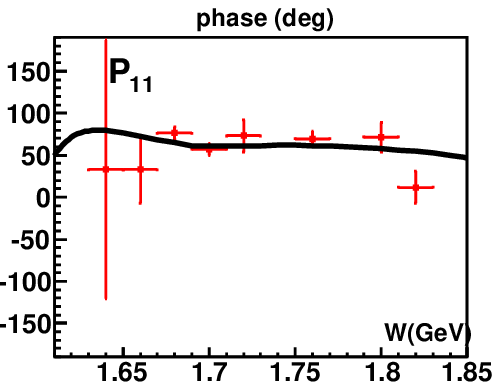}&
\hspace{-3mm}\includegraphics[width=0.185\textwidth]{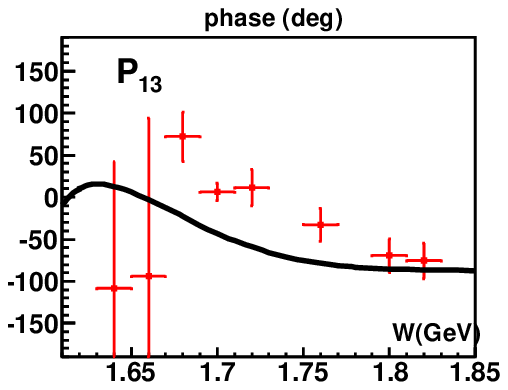}&
\hspace{-3mm}\includegraphics[width=0.185\textwidth]{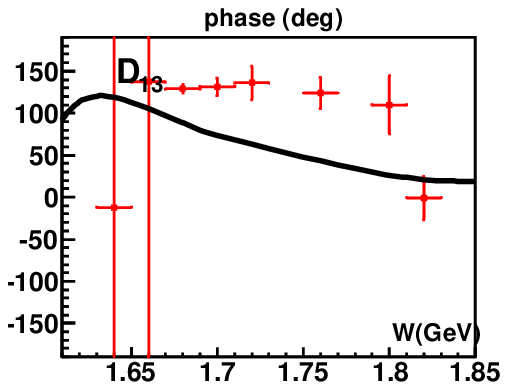}&
\hspace{-3mm}\includegraphics[width=0.185\textwidth]{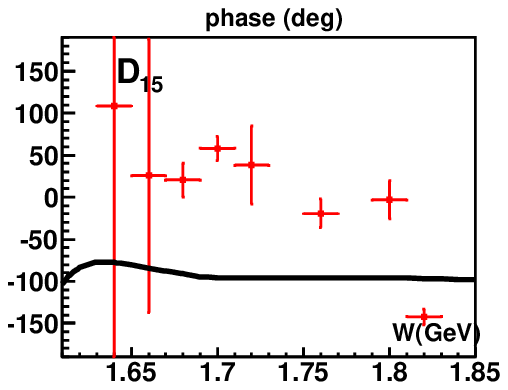}
\\
\end{tabular}
\end{center}
\caption{\label{piLambdaK2_D}Ambiguous solutions for the
decomposition of the $\pi N\to \Lambda K$  scattering amplitudes
with $S$, $P$ and $D$ waves. The solid line is the energy dependent
solution BnGa2011-02. The first solution given in the first two lines is chosen as the closest to the
energy-dependent fit. The solutions 2 to 6 given in the subsequent pairs of lines
differ from the
energy-dependent fit with increasing $\chi^2$. The ``best" solution agrees with the
BnGa energy dependent fit with $\chi^2/N=188/72$. There is a multitude of further solutions:
one may taken seven (or six, $\cdots$) energy points from the first solution and the missing points
from the second (or third, $\cdots$) solution
to obtain additional solutions. }
\end{figure*}
\begin{figure*}
\begin{center}
\begin{tabular}{ccccc}
\hspace{0mm}\includegraphics[width=0.185\textwidth]{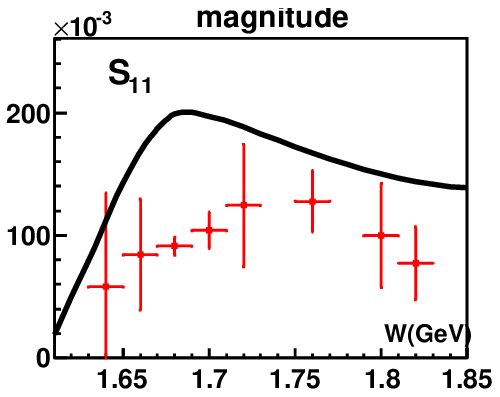}&
\hspace{-3mm}\includegraphics[width=0.185\textwidth]{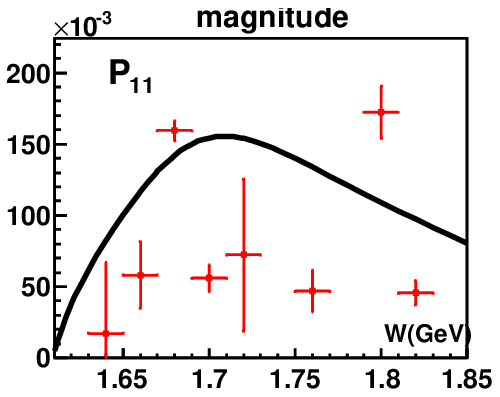}&
\hspace{-3mm}\includegraphics[width=0.185\textwidth]{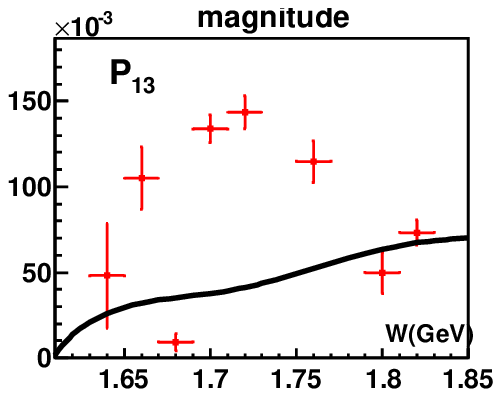}&
\hspace{-3mm}\includegraphics[width=0.185\textwidth]{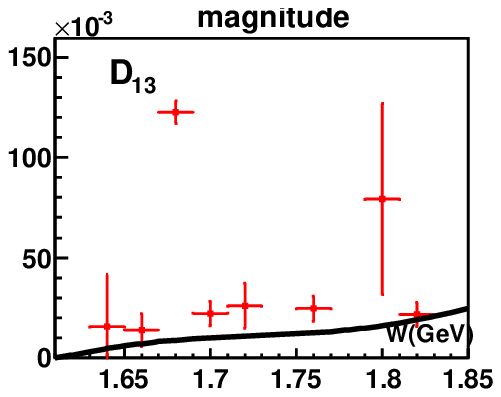}&
\hspace{-3mm}\includegraphics[width=0.185\textwidth]{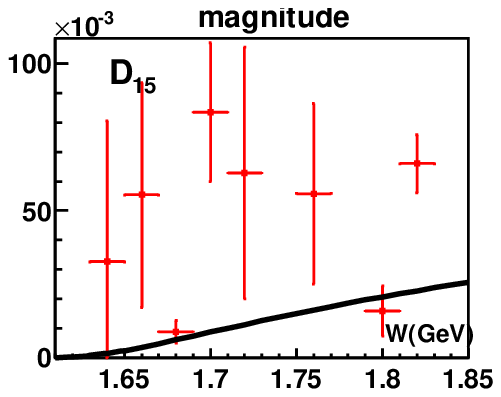}
\\
&
\hspace{-3mm}\includegraphics[width=0.185\textwidth]{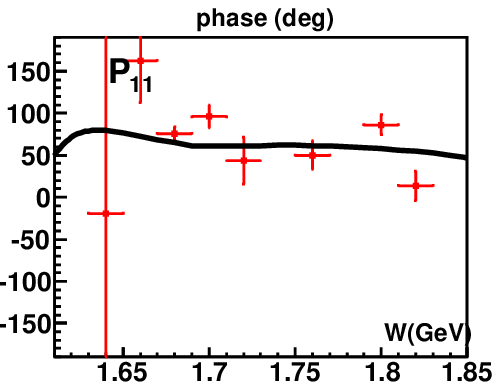}&
\hspace{-3mm}\includegraphics[width=0.185\textwidth]{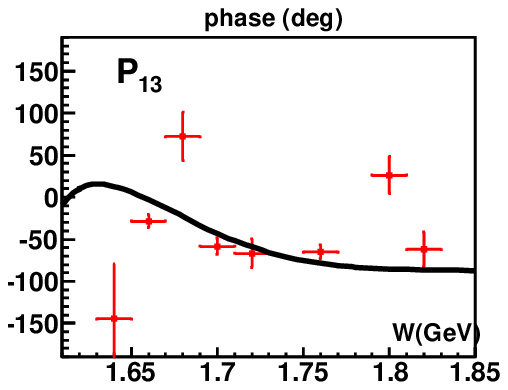}&
\hspace{-3mm}\includegraphics[width=0.185\textwidth]{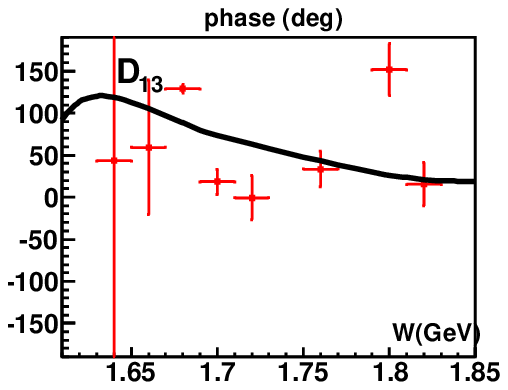}&
\hspace{-3mm}\includegraphics[width=0.185\textwidth]{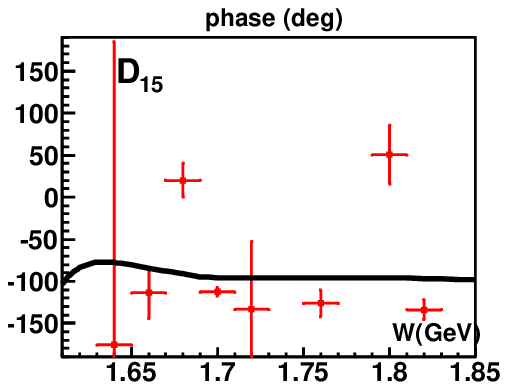}
\\
\hspace{0mm}\includegraphics[width=0.185\textwidth]{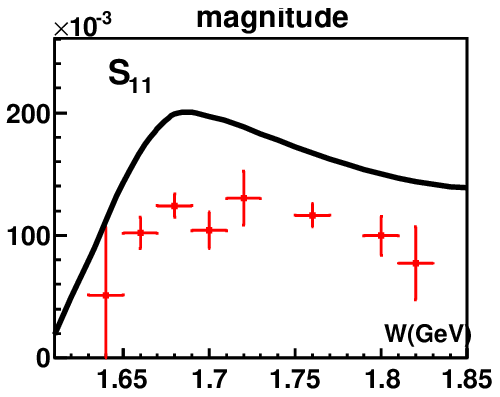}&
\hspace{-3mm}\includegraphics[width=0.185\textwidth]{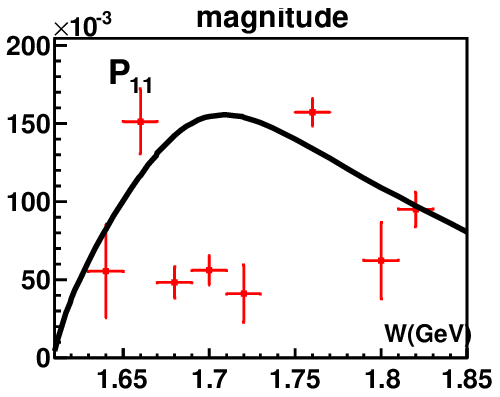}&
\hspace{-3mm}\includegraphics[width=0.185\textwidth]{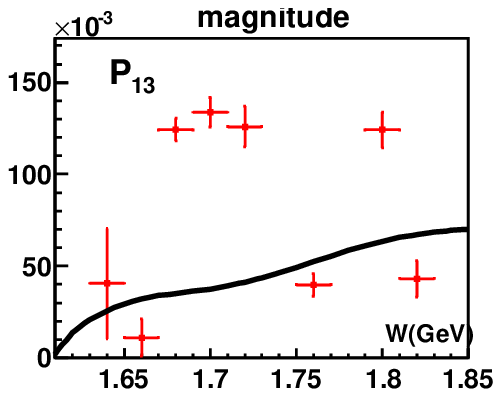}&
\hspace{-3mm}\includegraphics[width=0.185\textwidth]{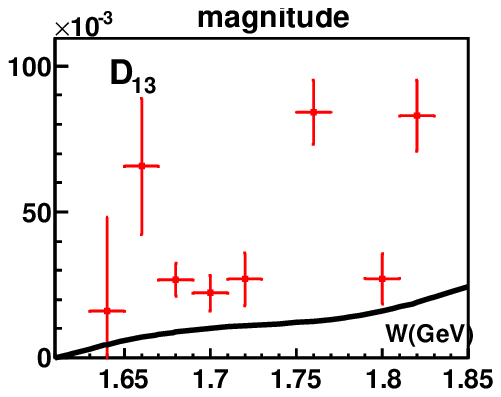}&
\hspace{-3mm}\includegraphics[width=0.185\textwidth]{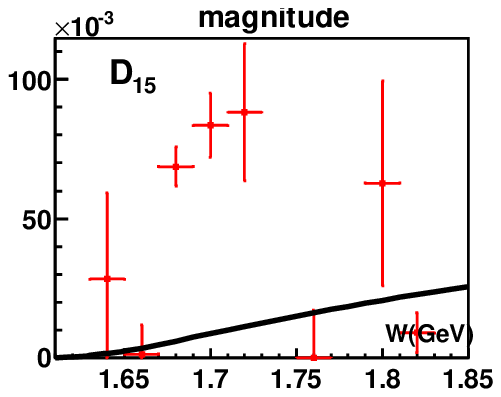}
\\
&
\hspace{-3mm}\includegraphics[width=0.185\textwidth]{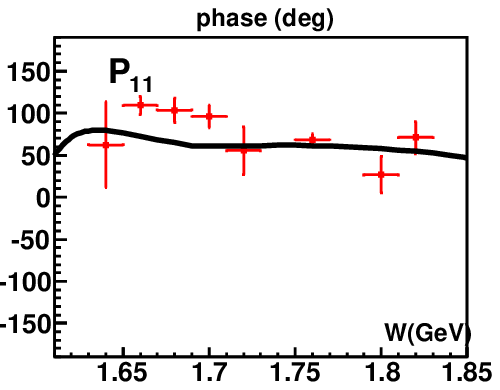}&
\hspace{-3mm}\includegraphics[width=0.185\textwidth]{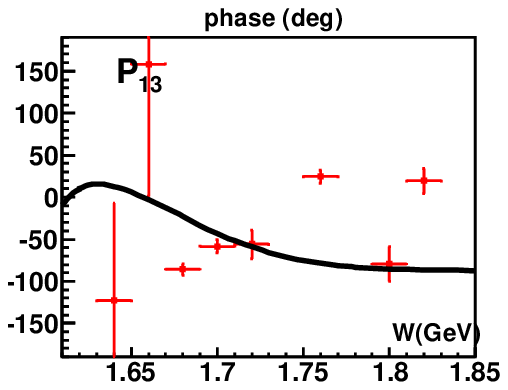}&
\hspace{-3mm}\includegraphics[width=0.185\textwidth]{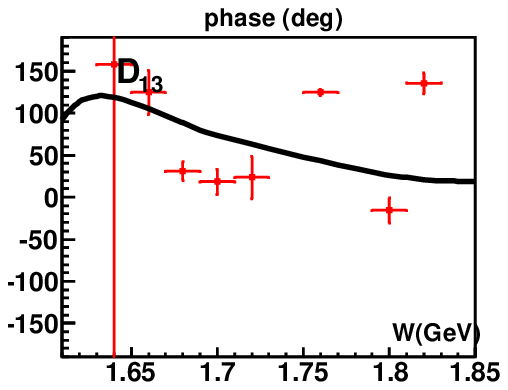}&
\hspace{-3mm}\includegraphics[width=0.185\textwidth]{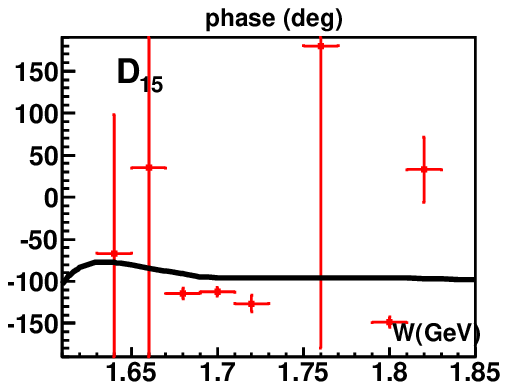}
\\
{\bf Fig. \protect\ref{piLambdaK2_D}} \ continued.&&&&
\end{tabular}

\end{center}
\end{figure*}

A better fit to the data can be achieved when $D$ waves are admitted
in addition to $S$ and $P$ waves. The fit is shown in
Fig.~\ref{piLambdaK_obeser2} as solid line, the $\chi^2$ of the fit
is given in Table~\ref{Table_D}. The quality of the fit is now
acceptable but the price one has to pay is the increase in the
number of ambiguous solutions. Numerically, we now find, for each
energy bin, six different solutions. Again, these are sorted
according to their proximity to the energy dependent solution,
starting from the upper row on Fig.~\ref{piLambdaK2_D}. The total
number of possible solutions is much larger than 6: At each energy,
there are 6 independent solutions, and they can be combined in any
order. Without the {\it a priori} knowledge of an energy independent
solution, there seems to be little chance to choose the correct
solution among the numerous ambiguous solutions which all reproduce
the data with exactly the same $\chi^2$.

The second price for the inclusion of $D$-waves are much larger
error bars of the amplitudes. The two newly added amplitudes
$D_{13}$ and $D_{15}$ are smaller by nearly one order of magnitude
when compared to the leading $S_{11}$ wave, and the energy dependent
fit overestimates their contributions.

It may be surprising that not only the amplitudes, moduli and
phases, are different in the six solutions but also their errors.
This is due to the fact that the different solutions are often close
to each other; depending on small details, the fit may identify a
clear minimum or find an effective minimum of two or more close-by minima.

\section{\boldmath Energy independent PWA in the  $1840-2270$\,MeV region}

In the region $1840-2270$ MeV region, a {\it complete} experiment
has been performed; differential cross section $d\sigma/d\Omega$,
$\Lambda$ polarization $P$, and spin-rotation angle $\beta$ were
measured \cite{Baker:1978qm,Saxon:1979xu,Bell:1983dm}. From these
data we select 7 bins of $10-20$\,MeV width which have all three
experimental observables. The complete data set eliminates the
ambiguities which we have near threshold. The only ambiguities we
could have now are related to the quality of the data.
Let us note at this point that $\beta$ has be measured only in a limited
range of angles ($\cos\theta\ge 0$). The very rapid changes or even the discontinuities of
the $\beta$ observable happens at points where $P^2\approx 1$
and where both observables, $R$ and $A$, are small. If they both
change the sign, the phase of $\beta$ varies by 180 degrees.

\subsection{\boldmath Solution with $S$,$P$ and $D$ waves}

\begin{figure*}[ph]
\begin{center}
\begin{tabular}{cccc}
\hspace{-1mm}\includegraphics[width=0.21\textwidth]{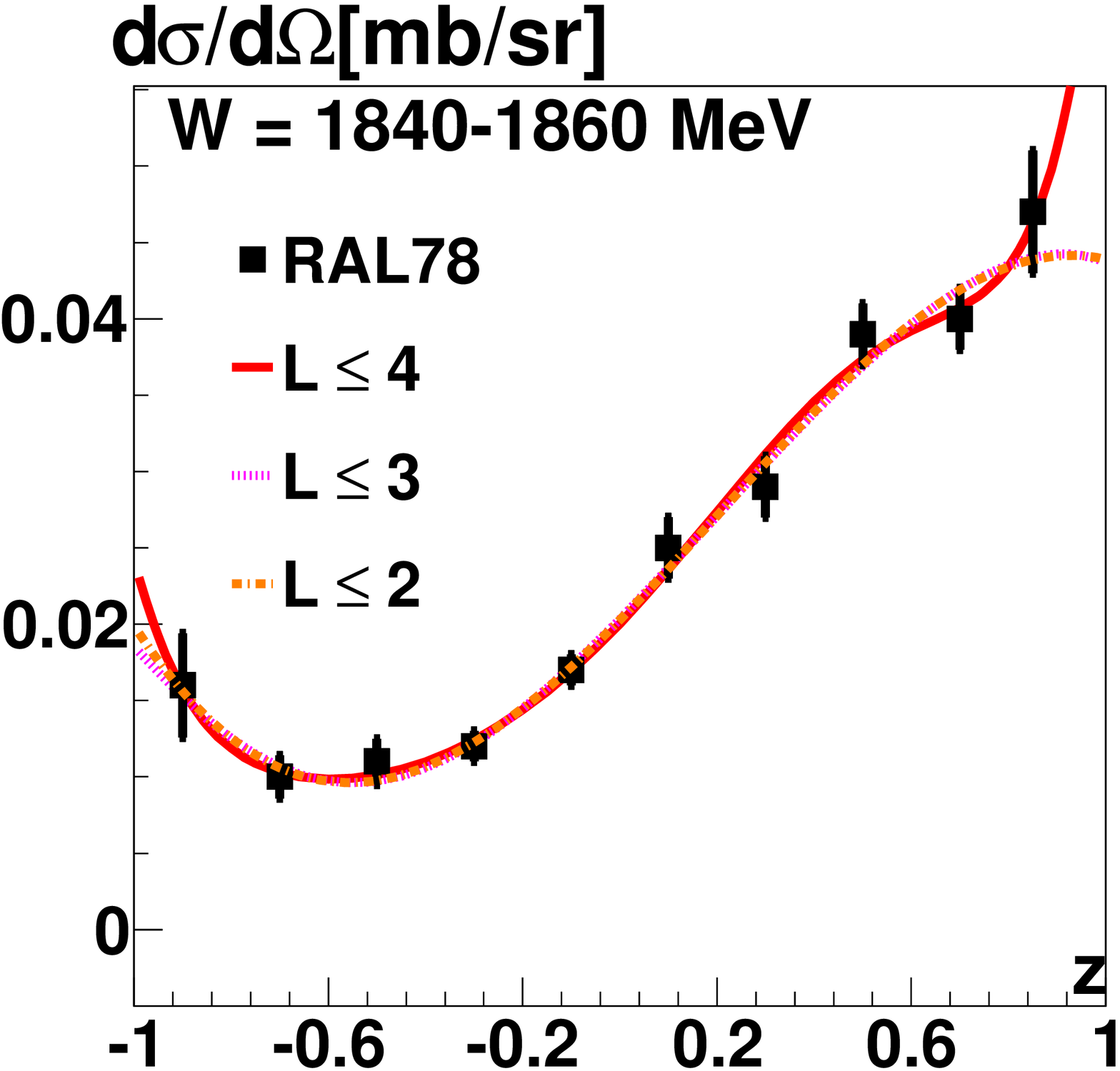}&
\hspace{-4mm}\includegraphics[width=0.21\textwidth]{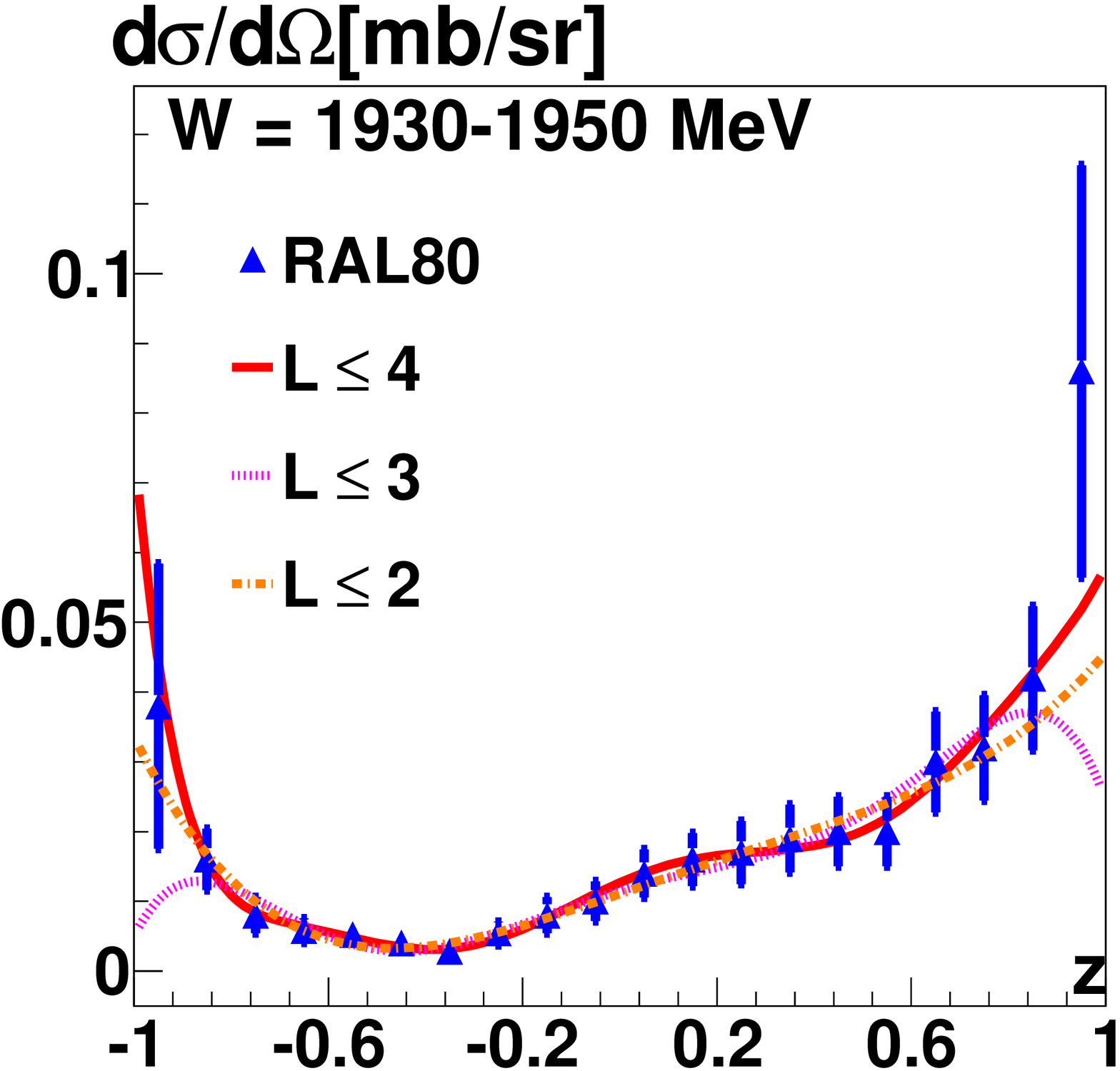}&
\hspace{-1mm}\includegraphics[width=0.21\textwidth]{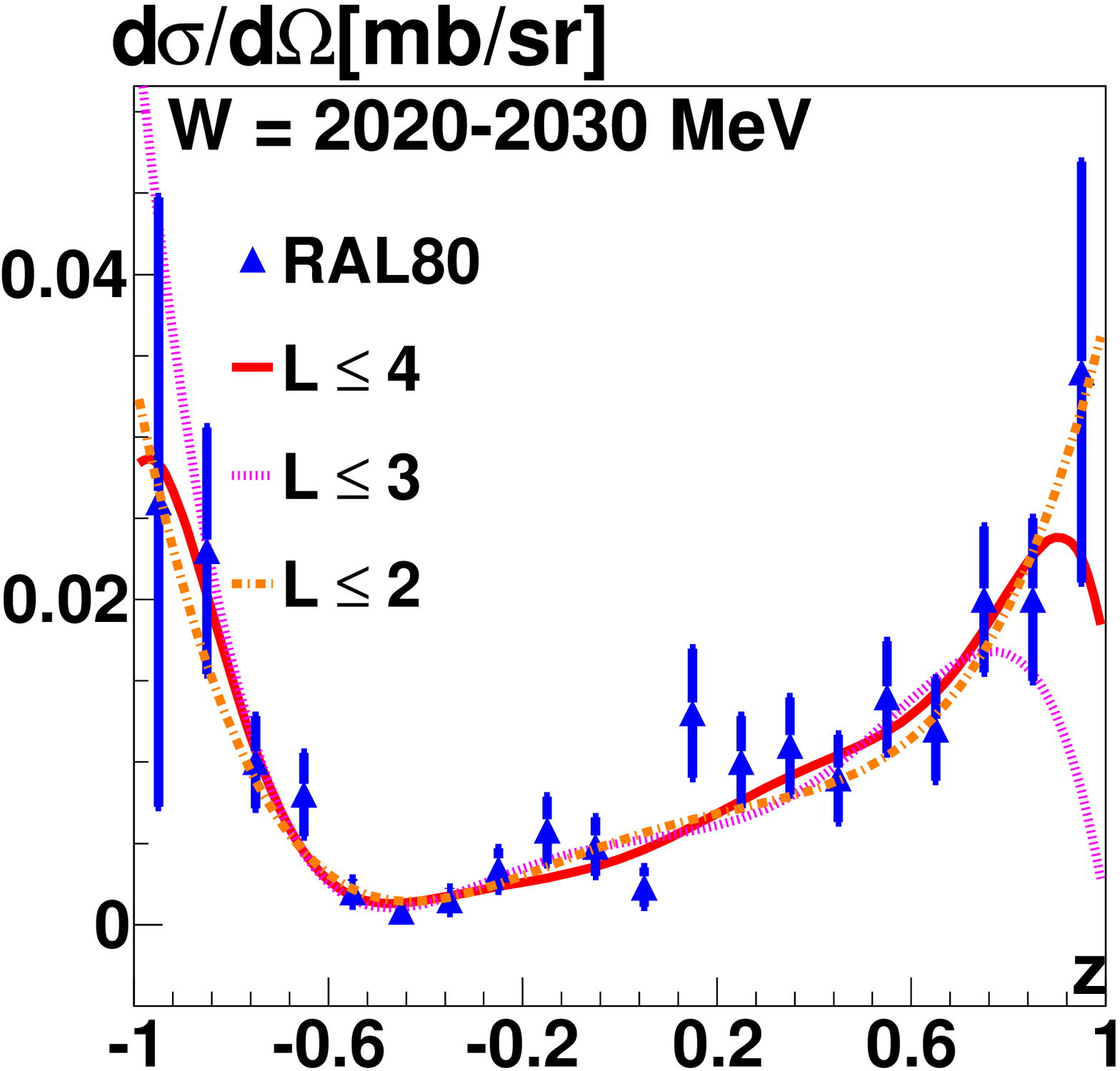}&
\hspace{-4mm}\includegraphics[width=0.21\textwidth]{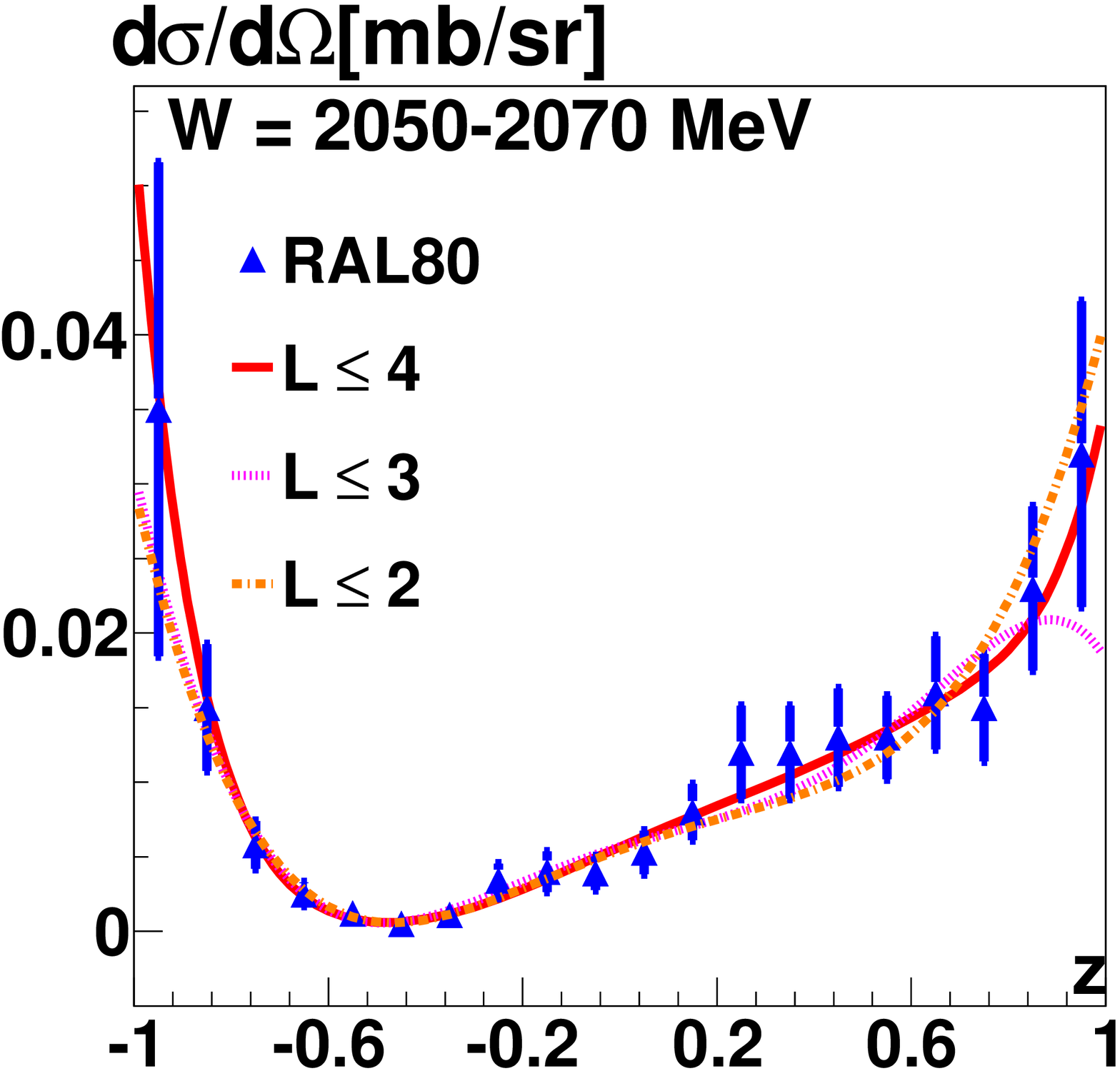}\\
\hspace{-4mm}\includegraphics[width=0.21\textwidth]{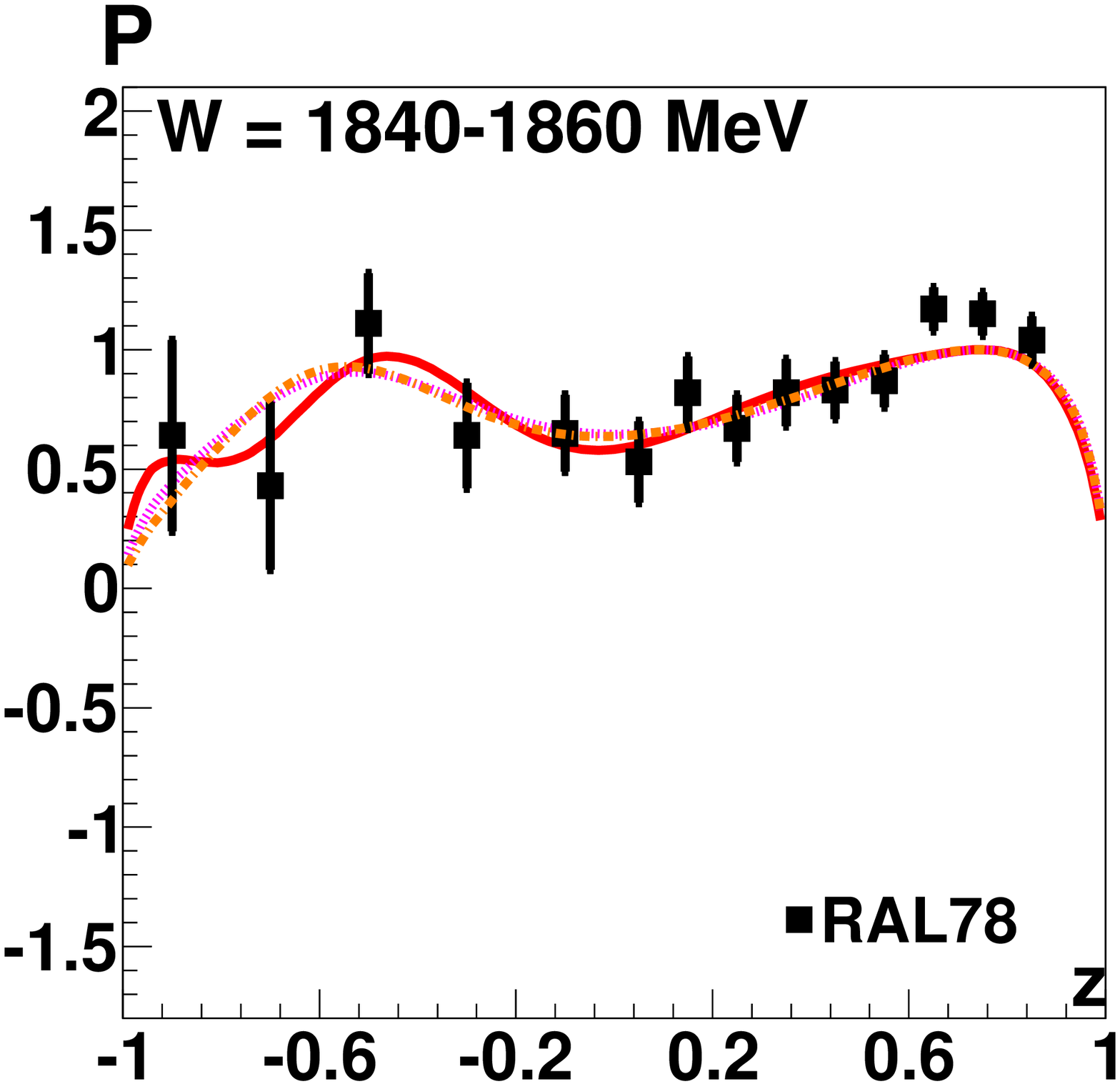}&
\hspace{-4mm}\includegraphics[width=0.21\textwidth]{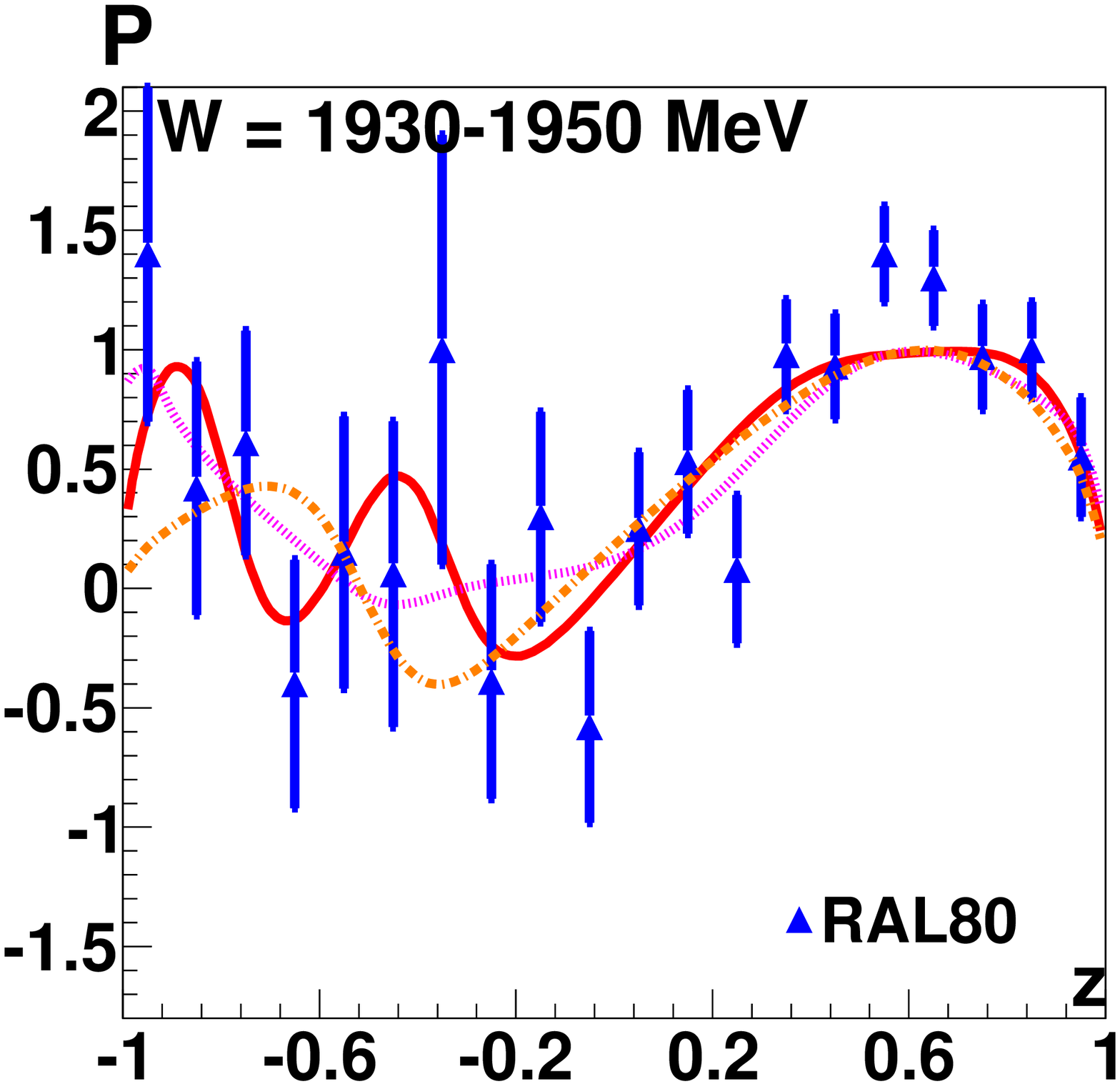}&
\hspace{-4mm}\includegraphics[width=0.21\textwidth]{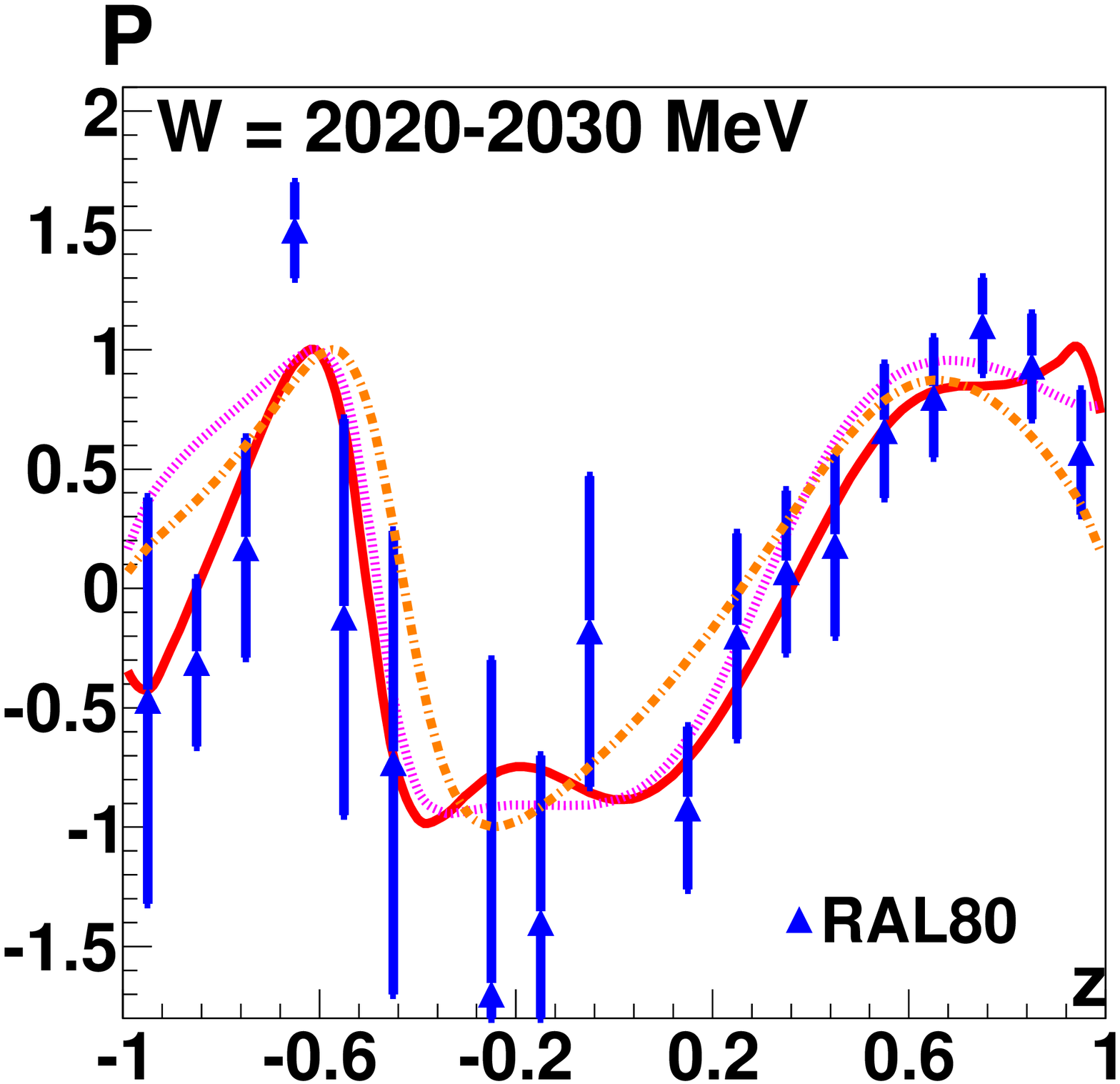}&
\hspace{-4mm}\includegraphics[width=0.21\textwidth]{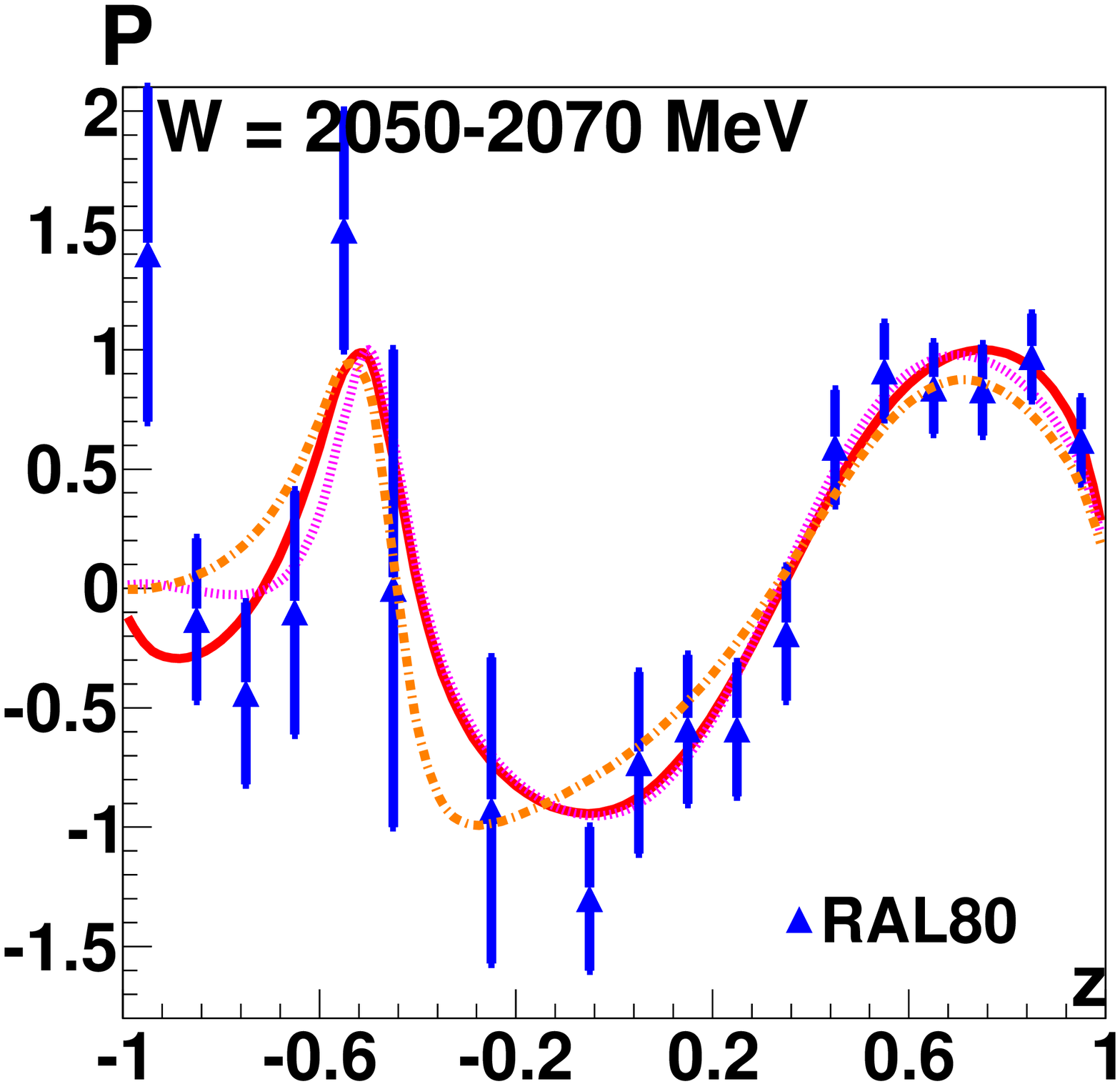}\\
\hspace{-4mm}\includegraphics[width=0.21\textwidth]{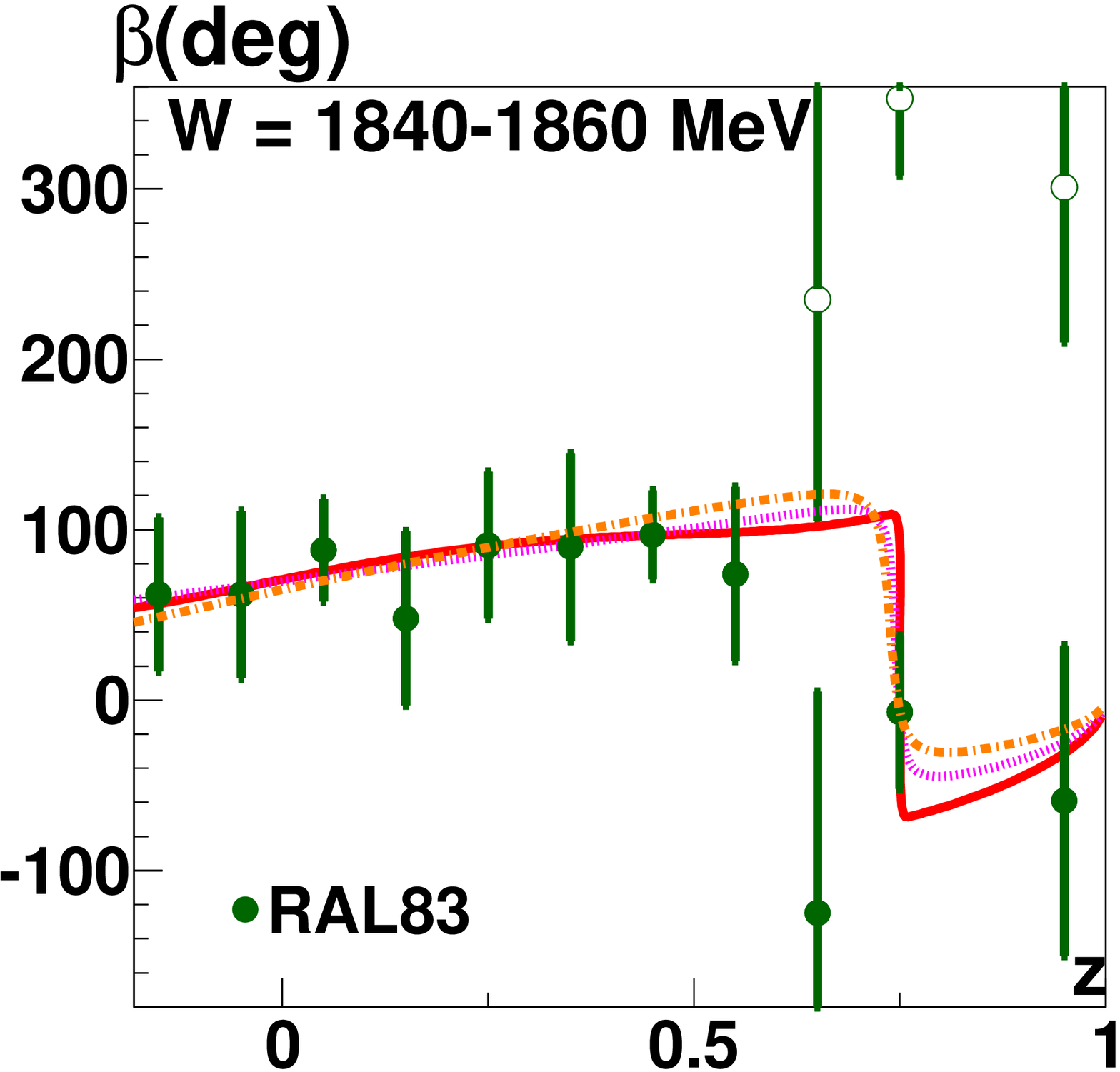}&
\hspace{-4mm}\includegraphics[width=0.21\textwidth]{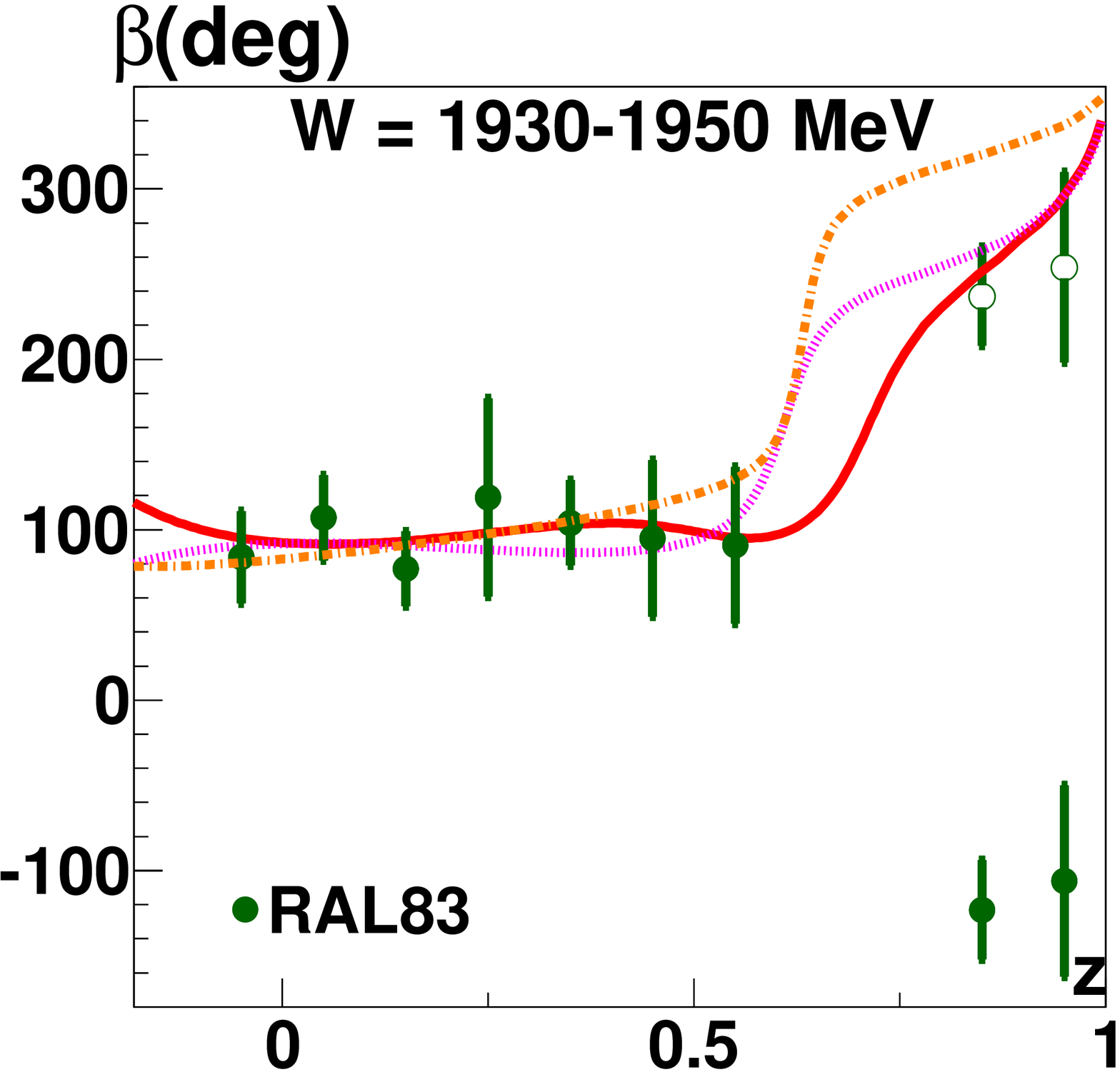}&
\hspace{-4mm}\includegraphics[width=0.21\textwidth]{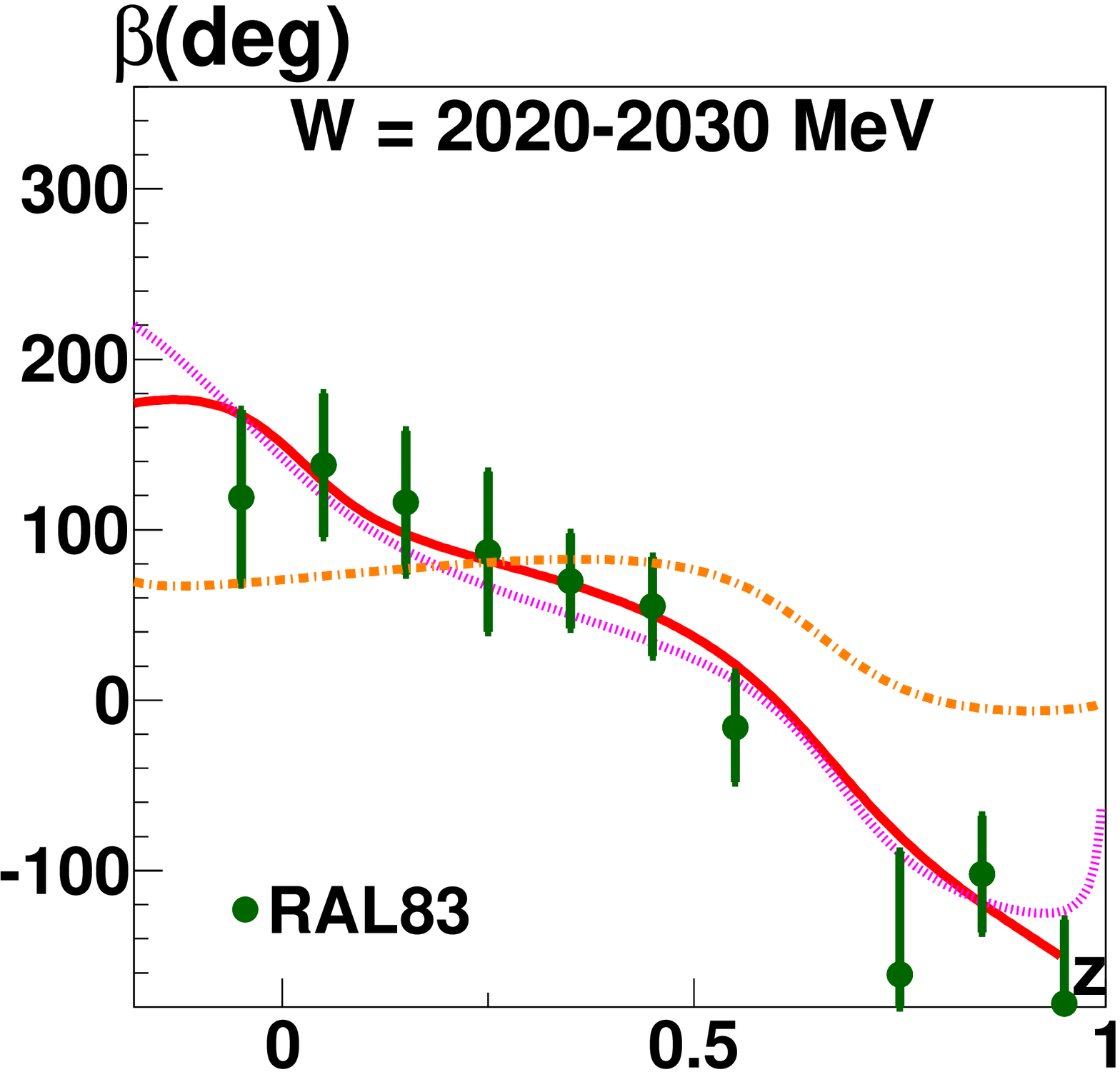}&
\hspace{-4mm}\includegraphics[width=0.21\textwidth]{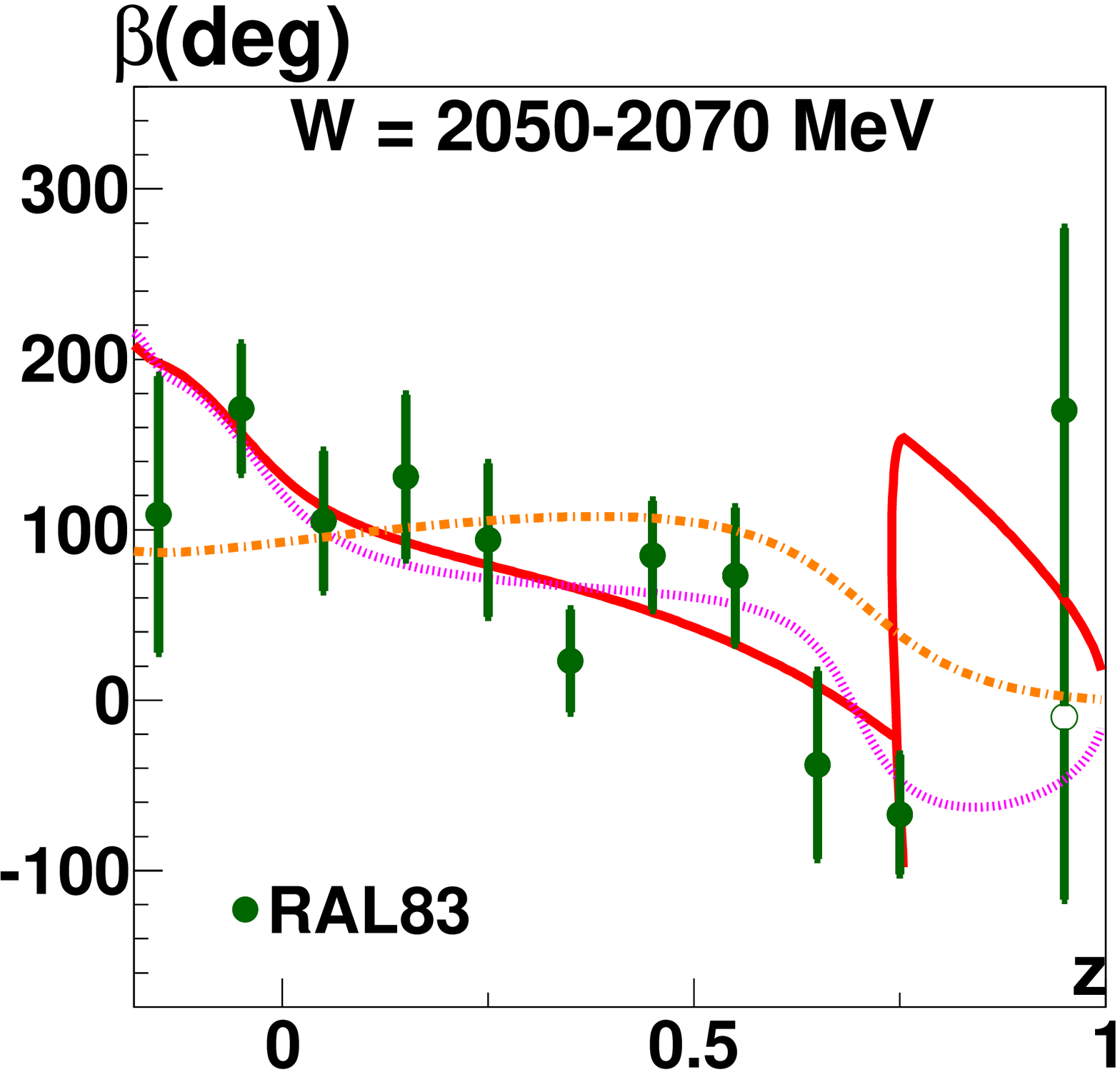}\\
\hspace{-1mm}\includegraphics[width=0.21\textwidth]{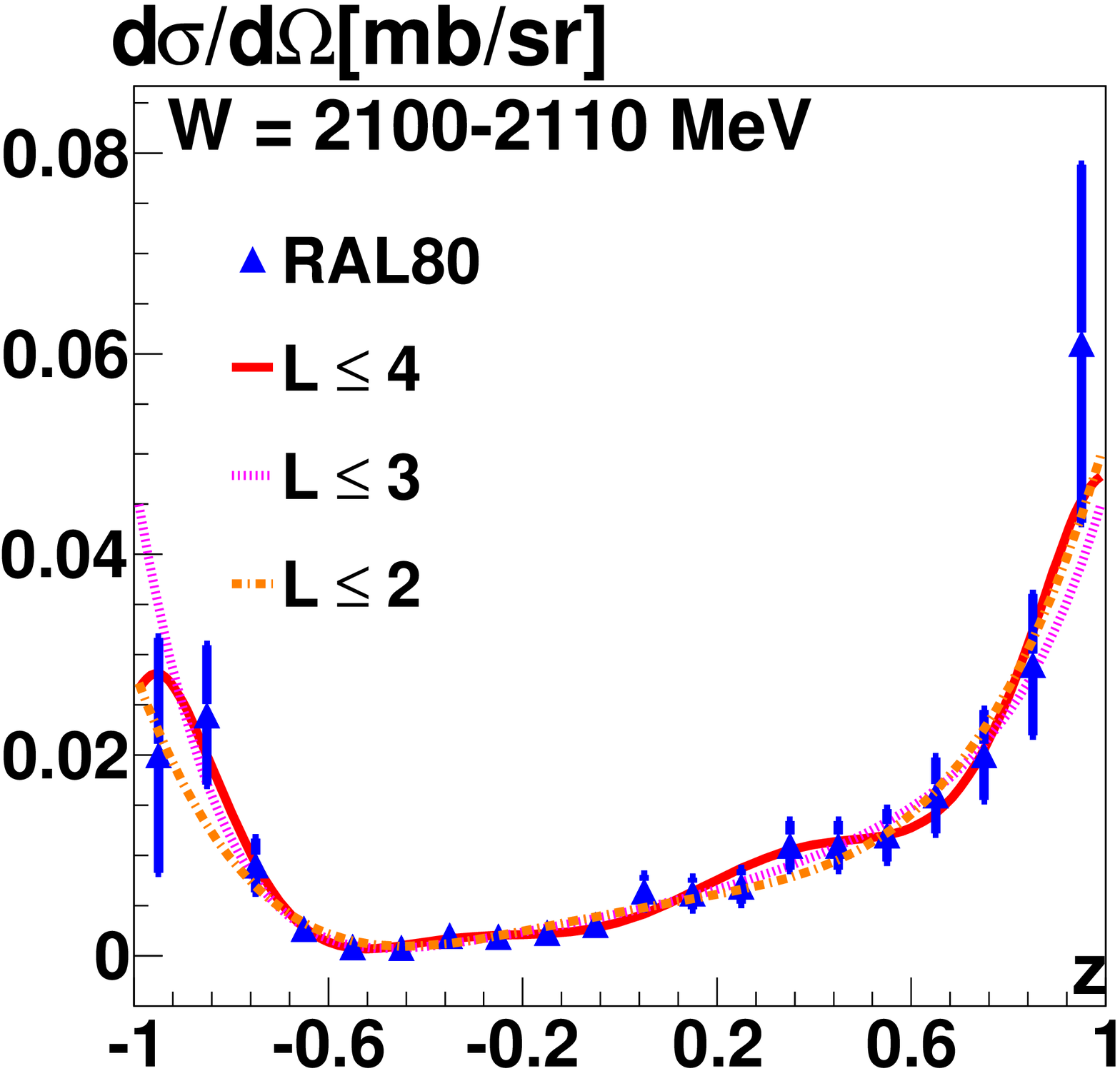}&
\hspace{-4mm}\includegraphics[width=0.21\textwidth]{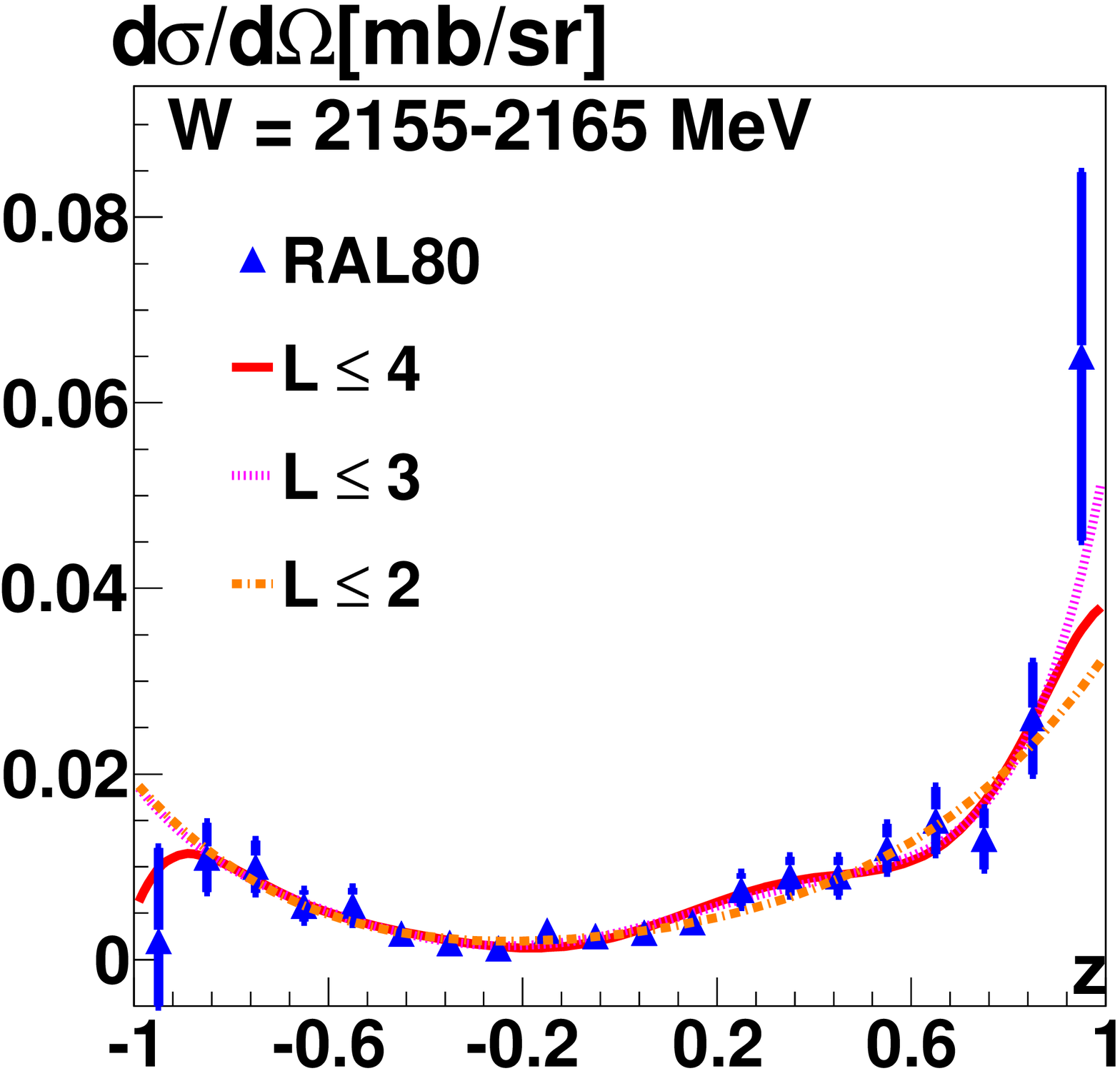}&
\hspace{-1mm}\includegraphics[width=0.21\textwidth]{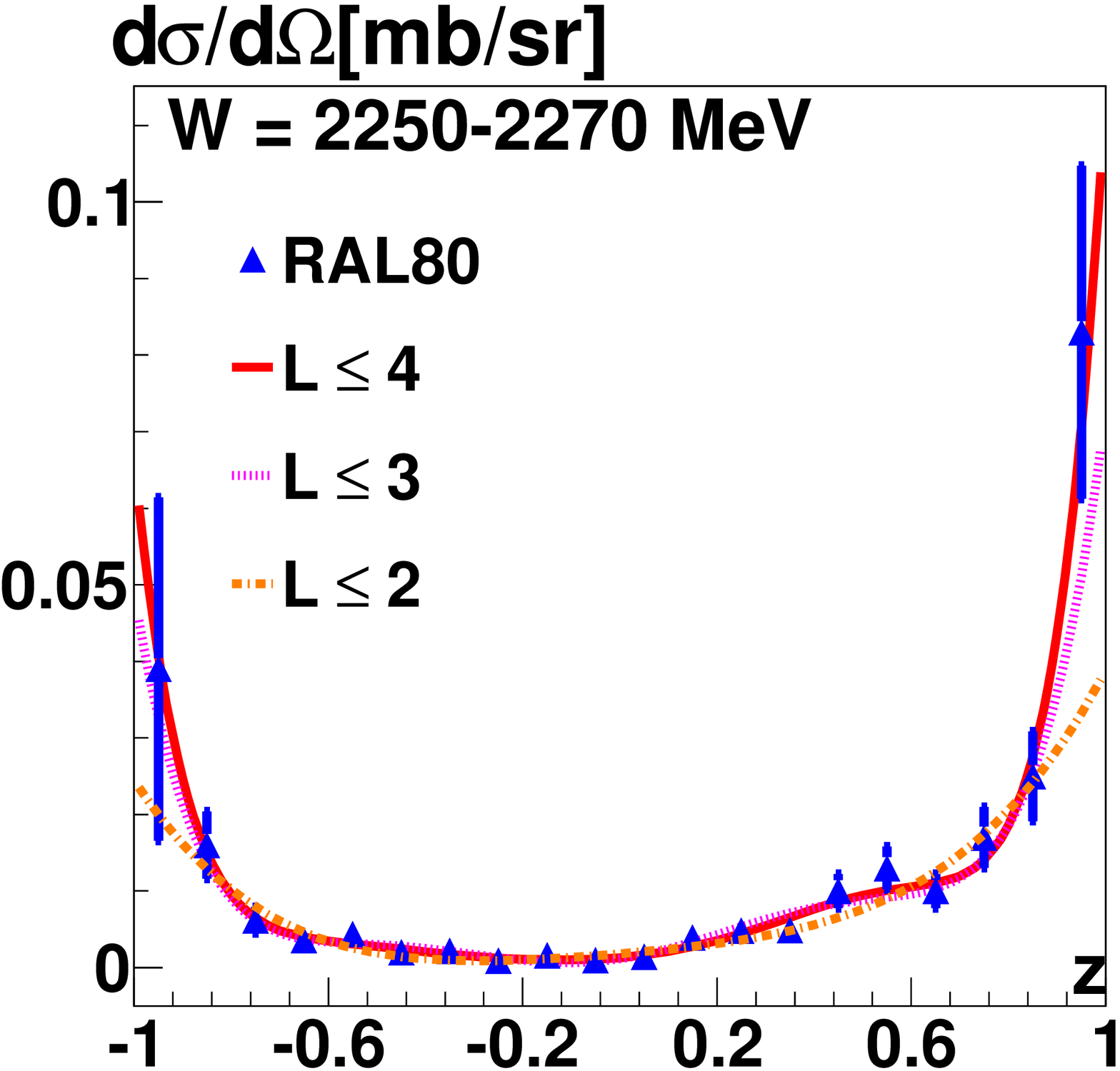}&\\
\hspace{-4mm}\includegraphics[width=0.21\textwidth]{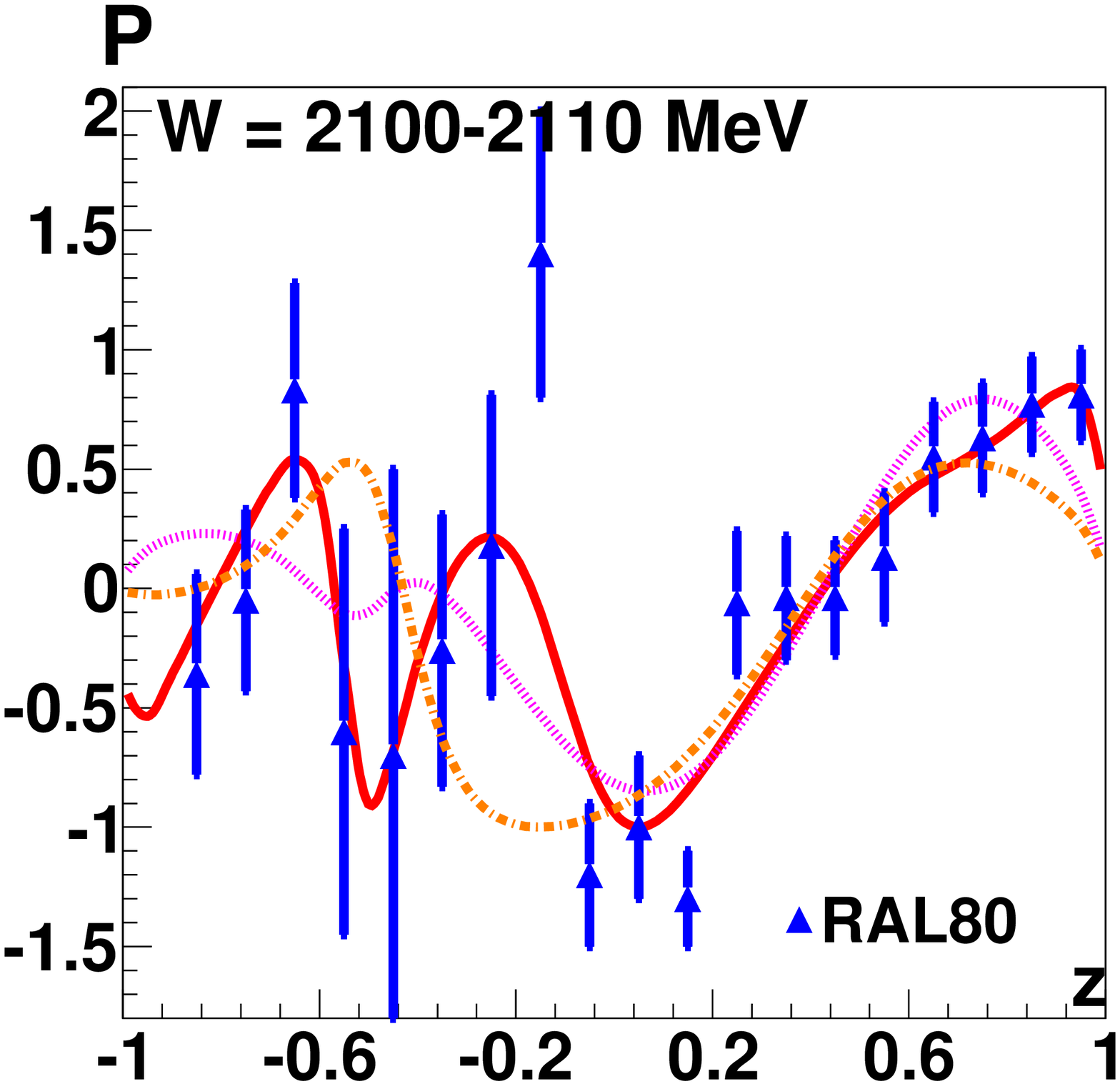}&
\hspace{-4mm}\includegraphics[width=0.21\textwidth]{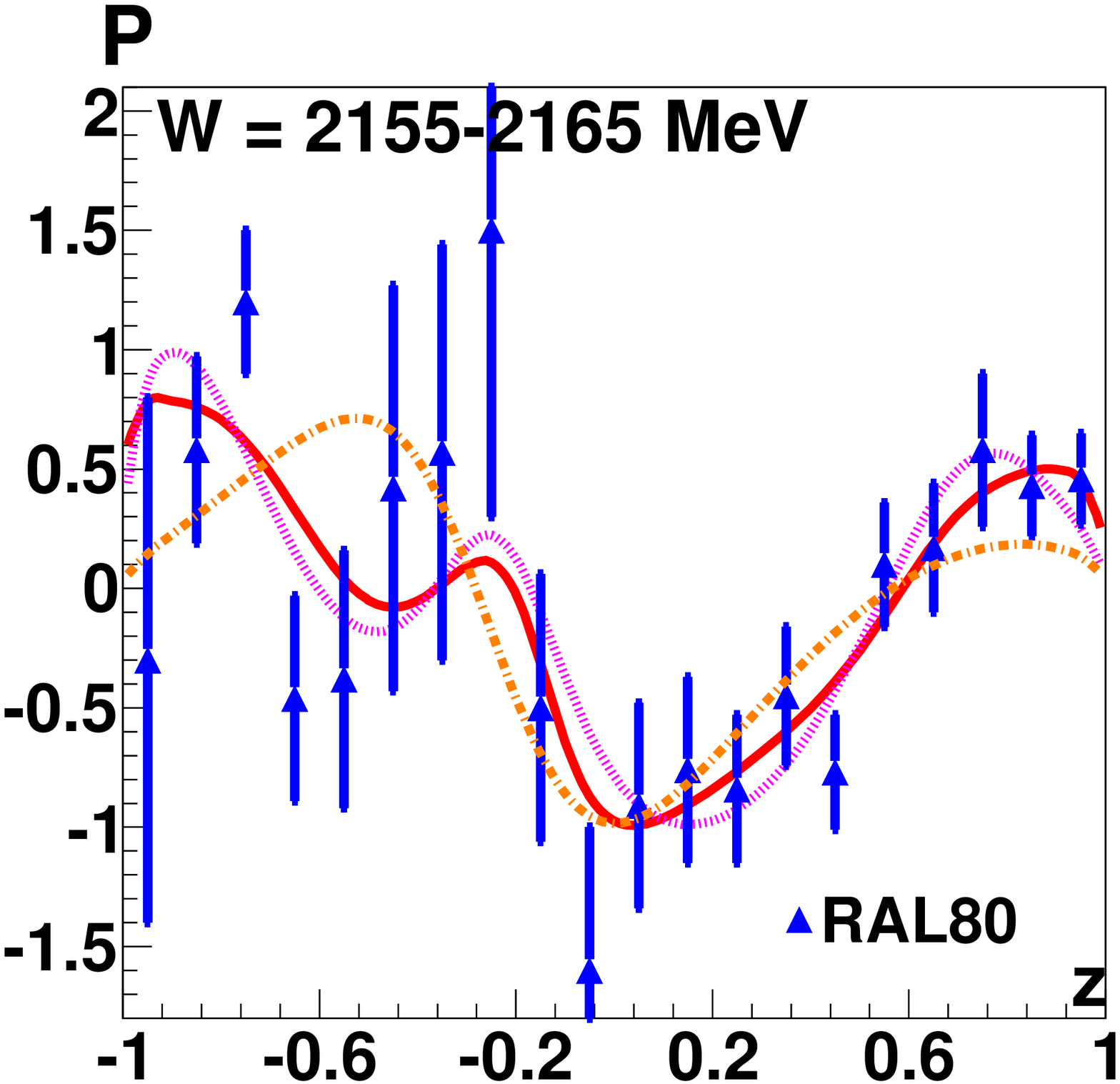}&
\hspace{-4mm}\includegraphics[width=0.21\textwidth]{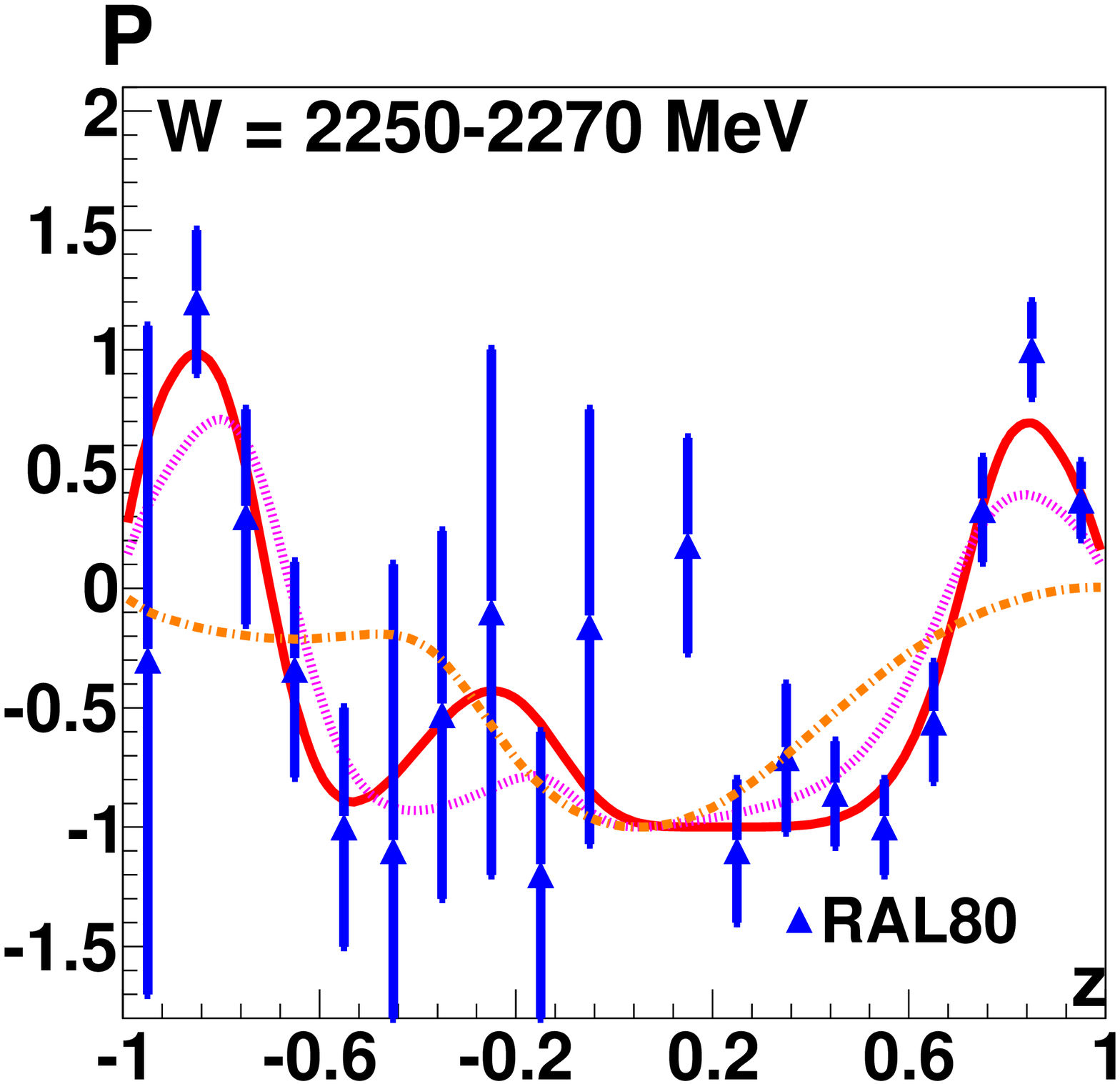}&\\
\hspace{-4mm}\includegraphics[width=0.21\textwidth]{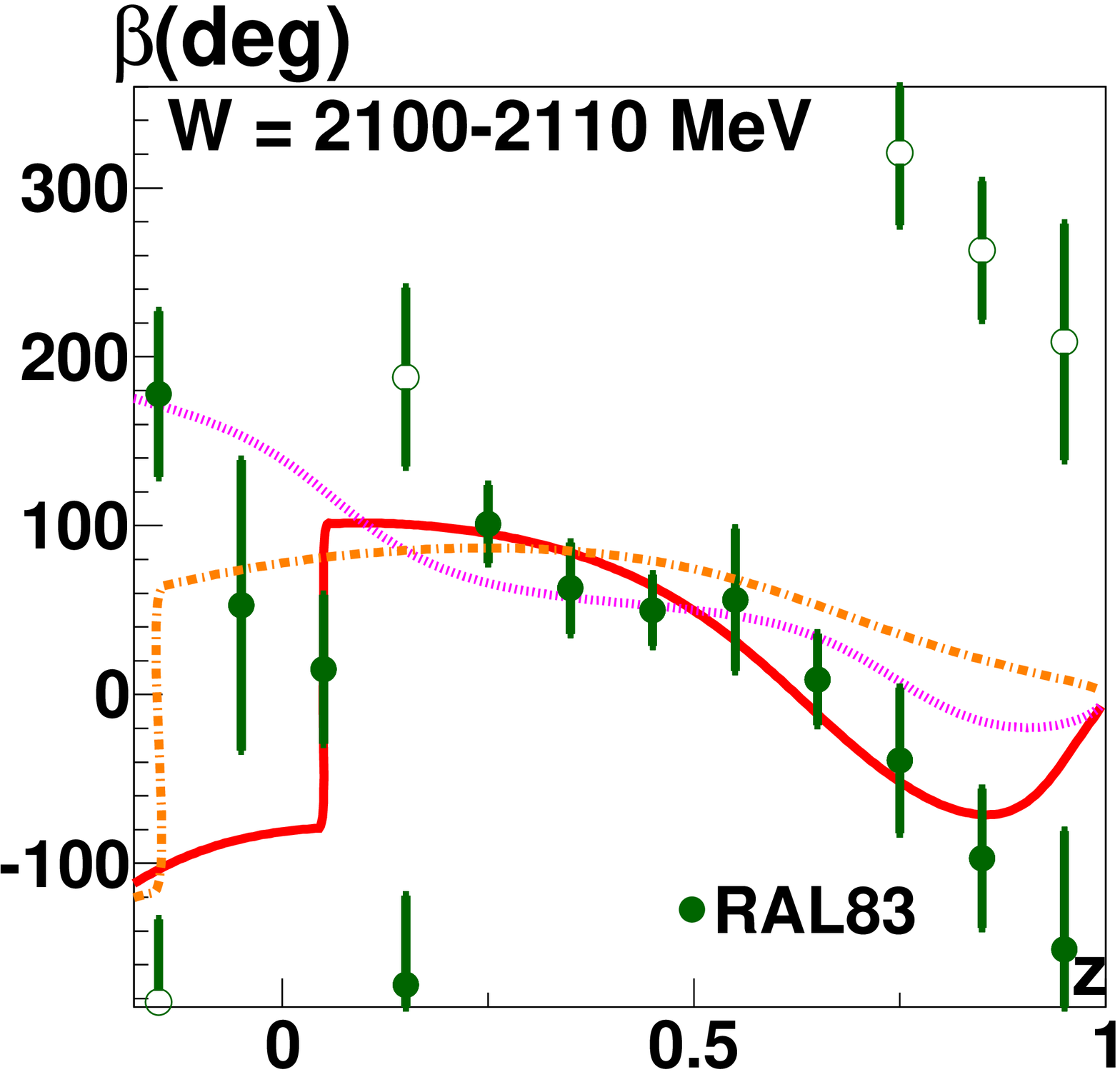}&
\hspace{-4mm}\includegraphics[width=0.21\textwidth]{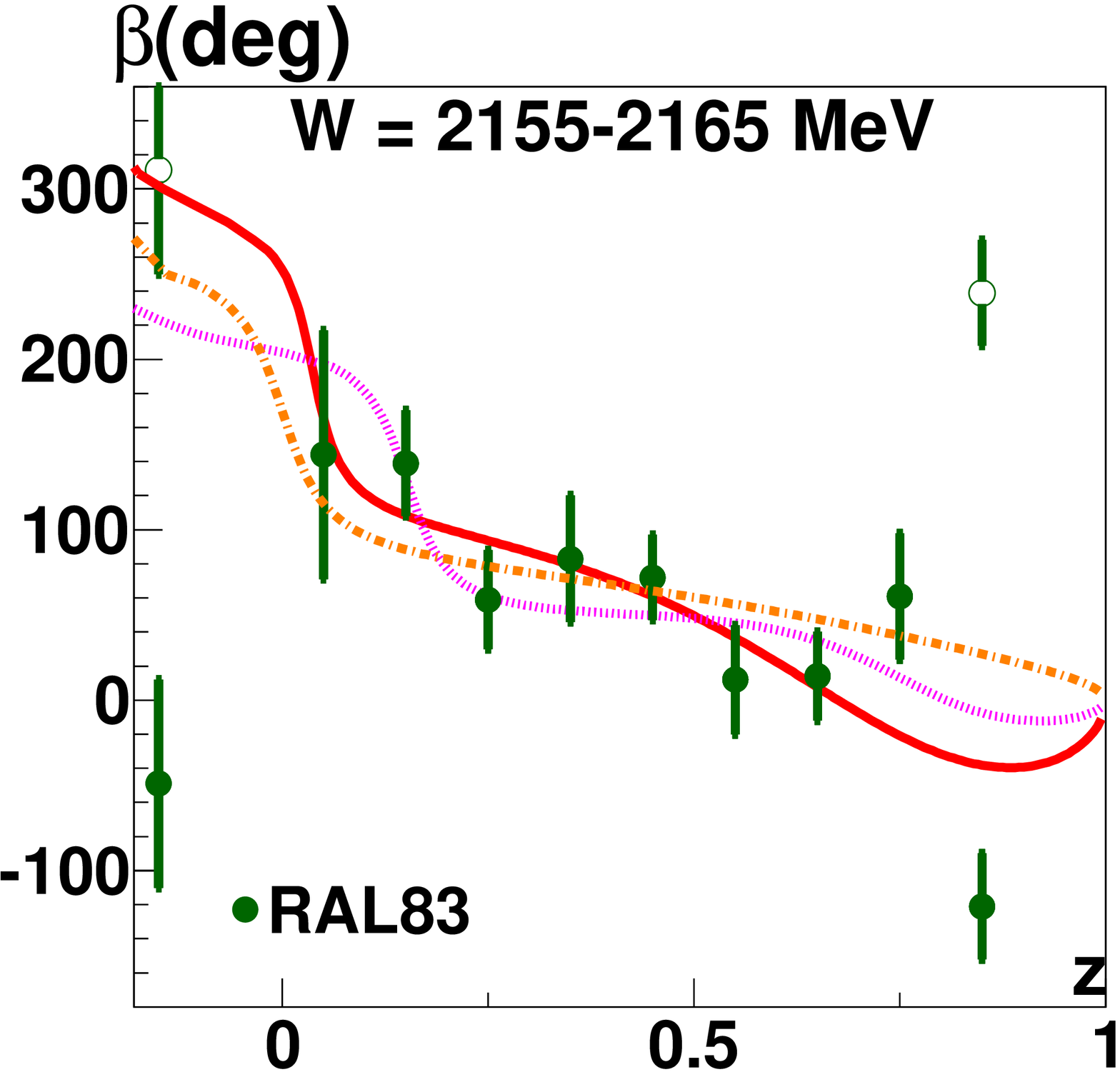}&
\hspace{-4mm}\includegraphics[width=0.21\textwidth]{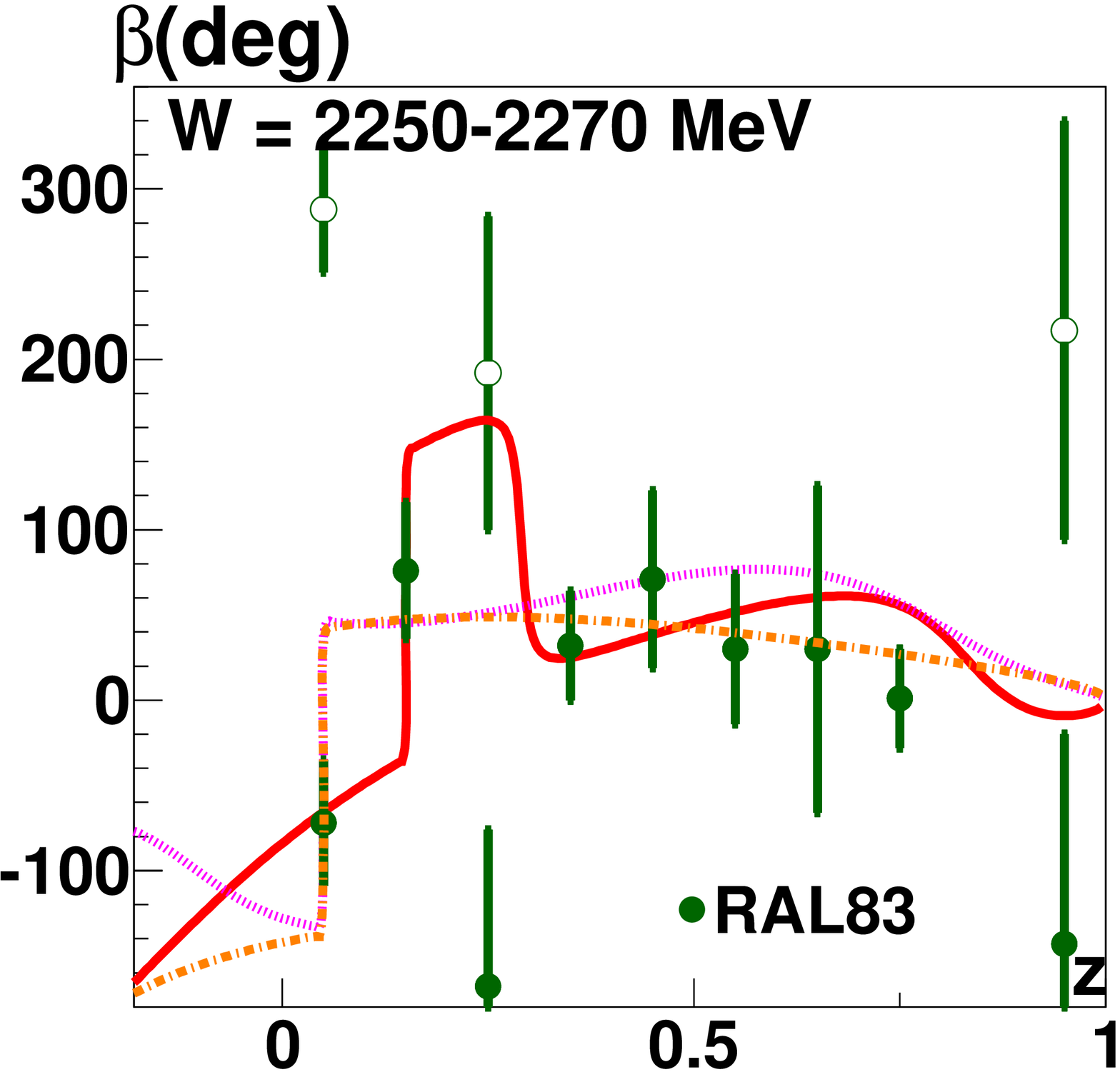}&
\\
\end{tabular}
\end{center}
\caption{\label{piLambdaK_obeser_high1} (Color online) Energy
independent fit (red lines) with different maximum angular momentum
for $\pi^- p \rightarrow K^0 \Lambda$ reaction in the region
$1840-2270$ MeV. The experimental data are from RAL78
\cite{Baker:1978qm}, RAL80 \cite{Saxon:1979xu}, and RAL83
\cite{Bell:1983dm}. Note that $\beta$ is 360-degree cyclic which
leads to additional data points shown by empty circles. The
BnGa2010-02 fit to the data is shown in figs. 1 - 4 of
\cite{Anisovich:2010an}.} \vspace{2mm}
\end{figure*}

As the first step, we fit the data with $S$, $P$ and $D$ waves. The
fit is shown in Fig.~\ref{piLambdaK_obeser_high1} as dotted curve,
the quality of the fit in terms of $\chi^2$'s is given in Table
\ref{Table_high1}. The fit to the differential cross sections is
satisfactory, partly even excellent. More problems originate from
the polarization variables: through interference, they are more
sensitive to the presence of small waves. In particular the $\beta$
parameter is badly described.

\begin{figure*}
\begin{center}
\begin{tabular}{ccccc}
\hspace{0mm}\includegraphics[width=0.185\textwidth]{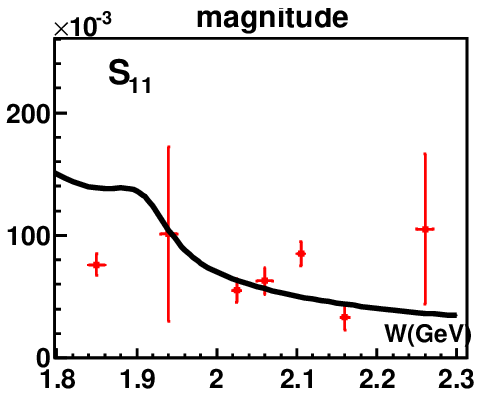}&
\hspace{-3mm}\includegraphics[width=0.185\textwidth]{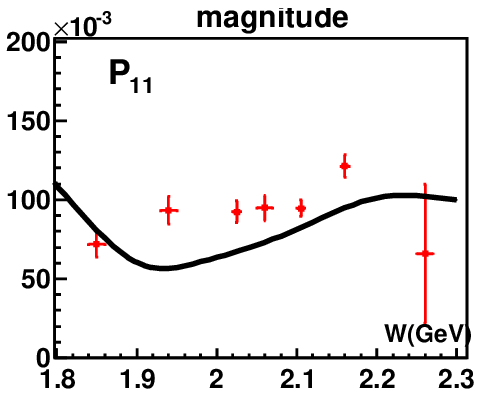}&
\hspace{-3mm}\includegraphics[width=0.185\textwidth]{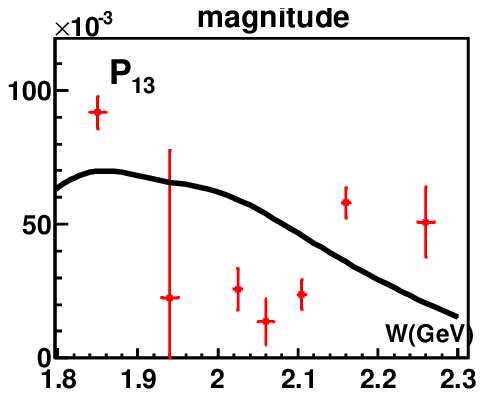}&
\hspace{-3mm}\includegraphics[width=0.185\textwidth]{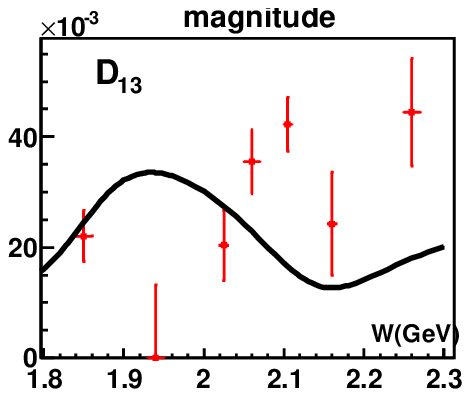}&
\hspace{-3mm}\includegraphics[width=0.185\textwidth]{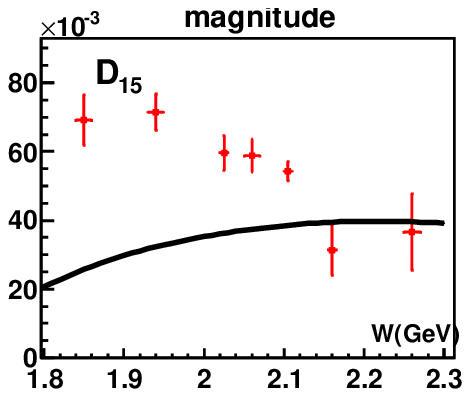}
\\
&\hspace{-3mm}\includegraphics[width=0.185\textwidth]{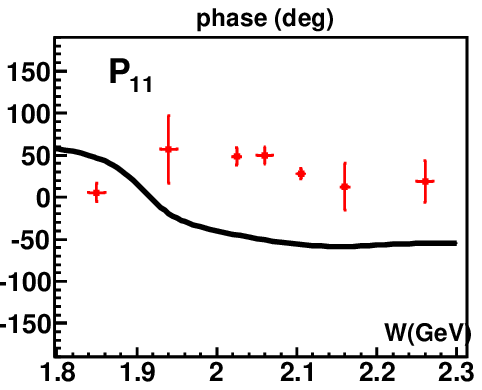}&
\hspace{-3mm}\includegraphics[width=0.185\textwidth]{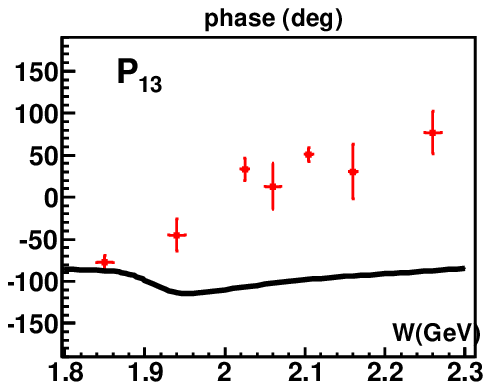}&
\hspace{-3mm}\includegraphics[width=0.185\textwidth]{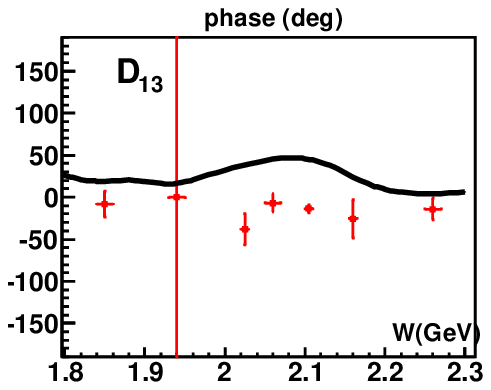}&
\hspace{-3mm}\includegraphics[width=0.185\textwidth]{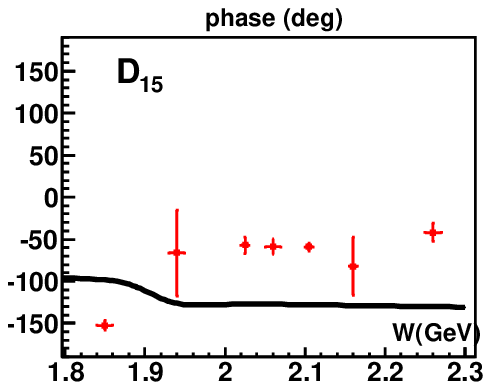}
\end{tabular}\end{center}
\caption{\label{piLambdaK2_F_HighD} Decomposition of the $\pi N\to
\Lambda K$ scattering amplitudes with $S$, $P$,  and $D$ waves. The
solid line is the energy dependent solution BnGa2011-02.
There is no good agreement between the energy dependent and independent solutions,
$\chi^2/N=1982/63$.}
\begin{center}
\begin{tabular}{ccccc}
\hspace{-3mm}\includegraphics[width=0.185\textwidth]{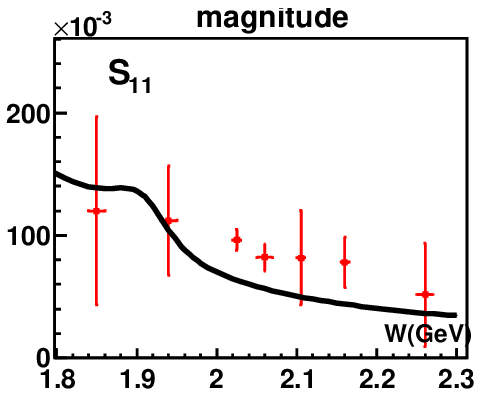}&
\hspace{-3mm}\includegraphics[width=0.185\textwidth]{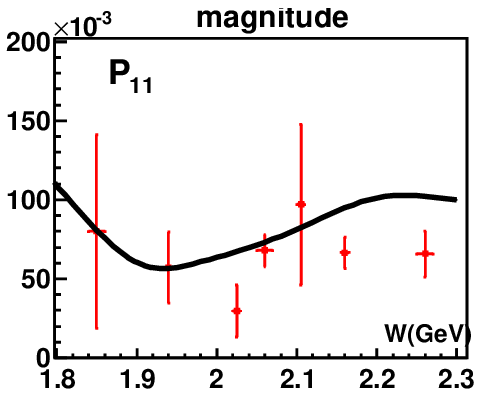}&
\hspace{-3mm}\includegraphics[width=0.185\textwidth]{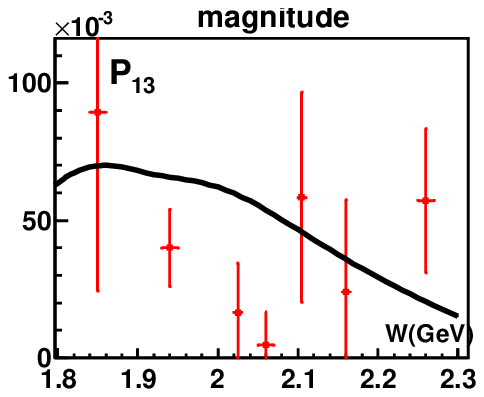}&
\hspace{-3mm}\includegraphics[width=0.185\textwidth]{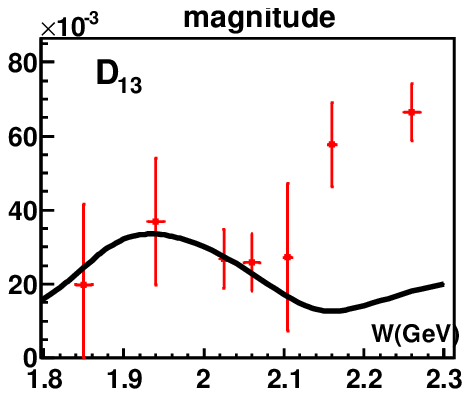}&
\hspace{-3mm}\includegraphics[width=0.185\textwidth]{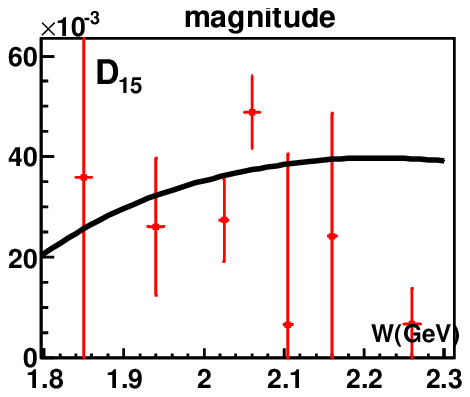}\\
&
\hspace{-3mm}\includegraphics[width=0.185\textwidth]{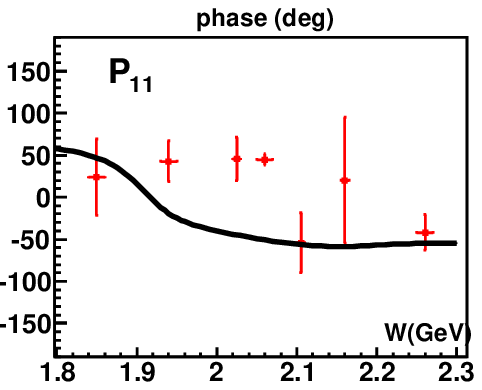}&
\hspace{-3mm}\includegraphics[width=0.185\textwidth]{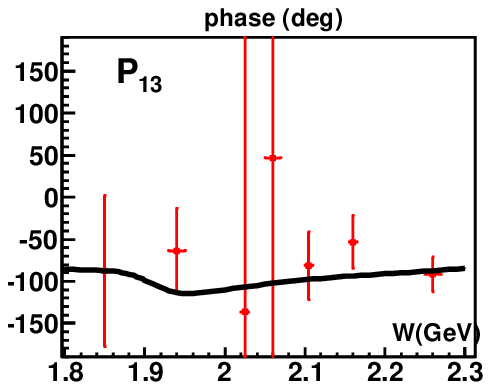}&
\hspace{-3mm}\includegraphics[width=0.185\textwidth]{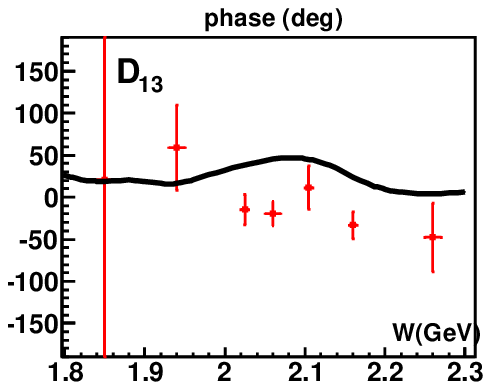}&
\hspace{-3mm}\includegraphics[width=0.185\textwidth]{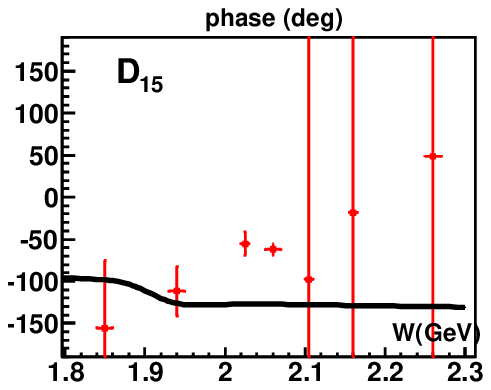}\\
\end{tabular}
\vspace{-10mm}
\end{center}
\end{figure*}

The resulting amplitudes are shown in Fig.~\ref{piLambdaK2_F_HighD}.
Magnitudes and phases of the reconstructed amplitudes resemble only
vaguely the curves representing the energy dependent fit. We
anticipate the need of higher partial waves with $L>2$.


%
%

\begin{figure}[pt]
\begin{tabular}{cc}
\hspace{0mm}\includegraphics[width=0.185\textwidth]{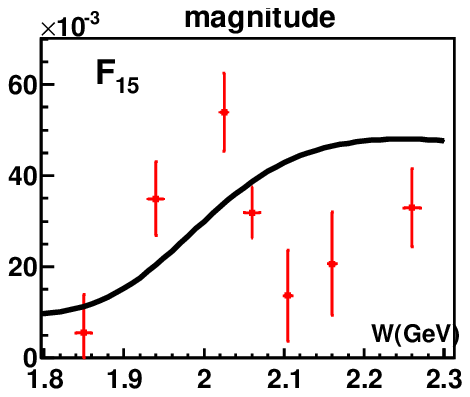}&
\hspace{-3mm}\includegraphics[width=0.185\textwidth]{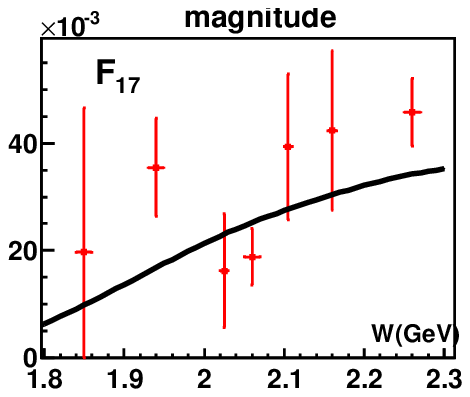}\\
\hspace{0mm}\includegraphics[width=0.185\textwidth]{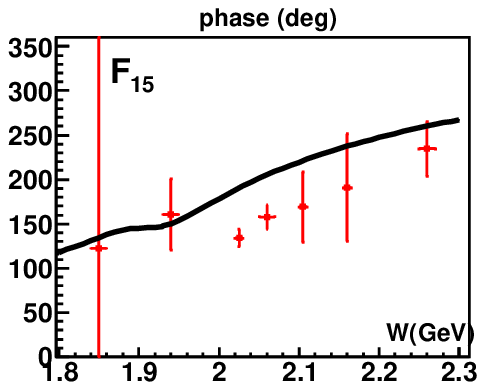}&
\hspace{-3mm}\includegraphics[width=0.185\textwidth]{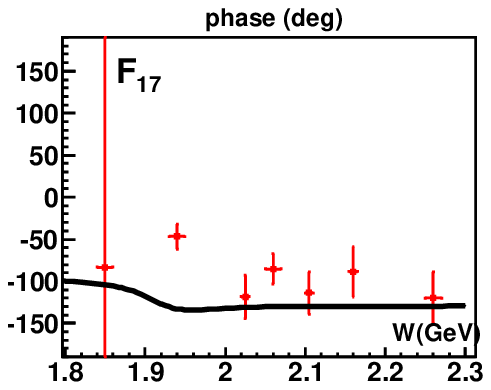}
\end{tabular}
\caption{\label{piLambdaK2_F_High1} Decomposition of the $\pi N\to
\Lambda K$ scattering amplitudes with $S$, $P$, $D$, and $F$ waves.
The solid line is the energy dependent solution BnGa2011-02.
There is still no good agreement between the energy dependent and independent solutions,
$\chi^2/N=868/91$.}
\end{figure}

\subsection{Solution with $S$,$P$, $D$ and $F$ waves}
As the next step, we include additionally $F$ waves in the energy
independent PWA. The dashed curve in
Fig.~\ref{piLambdaK_obeser_high1} represents this fit.
The scattering amplitudes in this case (with $S$,$P$, $D$, and $F$
waves included) are shown in Fig.~ \ref{piLambdaK2_F_High1}. The
amplitude errors are again obtained from solutions which differ from
the best solution by $\delta \chi^2 < 1$.

The quality of the fit is given in Table~\ref{Table_high2}. The $F$
waves improve the fit, and the $F$-wave contribution is significant.
But still, the data on the $\beta$ parameter are not yet
satisfactorily reproduced in the region above 2\,GeV. Hence we
extend the number of partial waves to include waves with $L=4$
($G$-waves).

\begin{table}
\phantom{\large dummy}
\end{table}
\begin{table}[pb]
\caption{\label{Table_high1} Quality of the energy independent fit
in the region $1840-2270$ MeV using $S$, $P$ and $D$ waves:
$\chi^2/N_{\rm data}$ and number of data points (in brackets). The
overall $\chi^2/N_{\rm data}$ is 1.52.}
\renewcommand{\arraystretch}{1.15}\begin{center}
\begin{tabular}{|c|c|c|c|}
  \hline
  Energy bin& $d\sigma/d\Omega$& $P$& $\beta$\\
  \hline
  1840-1860 & 0.45 (10) & 0.92 (14)  & 0.50 (11) \\
  1930-1950 & 0.38 (20)& 1.19 (20)& 1.53 (9) \\
  2020-2030 & 1.08 (20) & 1.71 (19) & 4.06 (10) \\
  2050-2070 & 0.42 (20) & 1.00 (18) & 3.12 (11) \\
  2100-2110 & 0.64 (20) & 2.60 (20) & 4.02 (12) \\
  2155-2165 & 0.75 (20) & 1.65 (20) & 3.24 (10) \\
  2250-2270 & 1.28 (20) & 4.46 (19) & 0.97 (9) \\
  \hline
\end{tabular}
\end{center}
\caption{\label{Table_high2} Quality of the $S$, $P$, $D$ and $F$
waves energy independent fit in the region $1840-2270$ MeV:
$\chi^2/N_{\rm data}$ and number of data points (in brackets). The
overall $\chi^2/N_{\rm data}$ is now 0.97.}
\renewcommand{\arraystretch}{1.15}\begin{center}
\begin{tabular}{|c|c|c|c|}
  \hline
  Energy bin& $d\sigma/d\Omega$& $P$& $\beta$\\
  \hline
  1840-1860 & 0.46 (10) & 0.91 (14)  & 0.41 (11) \\
  1930-1950 & 0.49 (20)& 0.96 (20)& 0.46 (9) \\
  2020-2030 & 1.16 (20) & 1.52 (19) & 0.58 (10) \\
  2050-2070 & 0.43 (20) & 0.76 (18) & 1.99 (11) \\
  2100-2110 & 0.45 (20) & 1.97 (20) & 3.55 (12) \\
  2155-2165 & 0.53 (20) & 0.93 (20) & 4.07 (10) \\
  2250-2270 & 0.73 (20) & 1.78 (19) & 1.32 (9) \\
  \hline
\end{tabular}
\end{center}
\end{table}


\begin{figure*}[pt]
\begin{center}
\begin{tabular}{ccccc}
\hspace{-3mm}\includegraphics[width=0.185\textwidth]{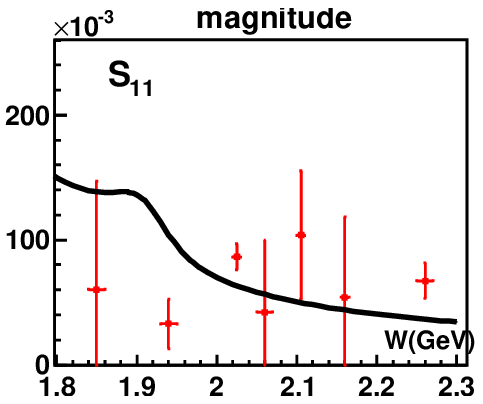}&
\hspace{-3mm}\includegraphics[width=0.185\textwidth]{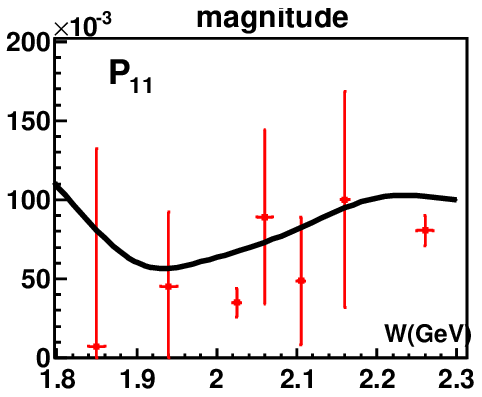}&
\hspace{-3mm}\includegraphics[width=0.185\textwidth]{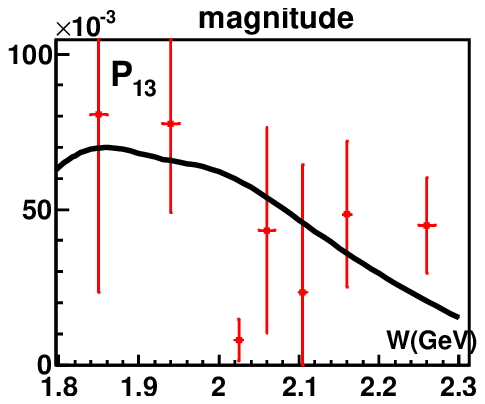}&
\hspace{-3mm}\includegraphics[width=0.185\textwidth]{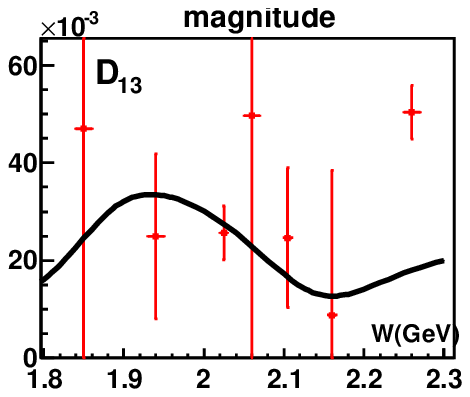}&
\hspace{-3mm}\includegraphics[width=0.185\textwidth]{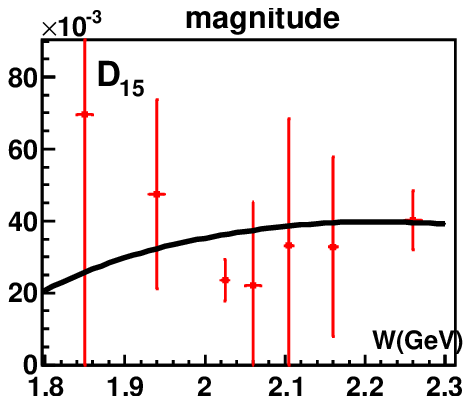}\\
&
\hspace{-3mm}\includegraphics[width=0.185\textwidth]{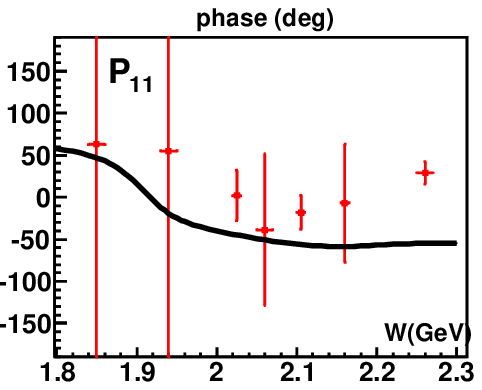}&
\hspace{-3mm}\includegraphics[width=0.185\textwidth]{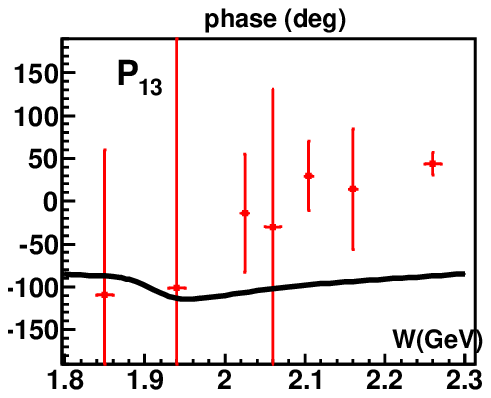}&
\hspace{-3mm}\includegraphics[width=0.185\textwidth]{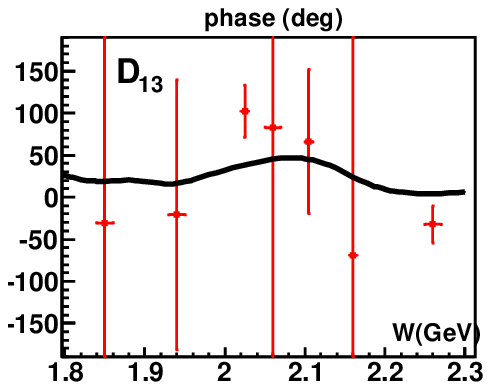}&
\hspace{-3mm}\includegraphics[width=0.185\textwidth]{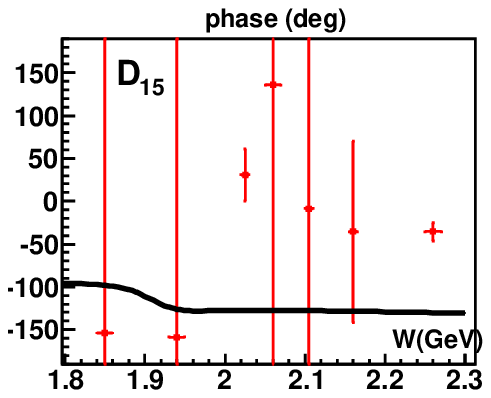}\\
\hspace{-3mm}\includegraphics[width=0.185\textwidth]{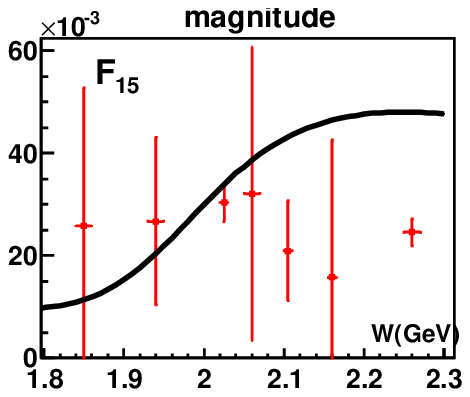}&
\hspace{-3mm}\includegraphics[width=0.185\textwidth]{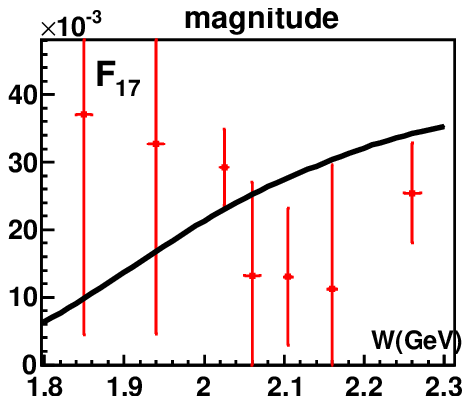}&
\hspace{-3mm}\includegraphics[width=0.185\textwidth]{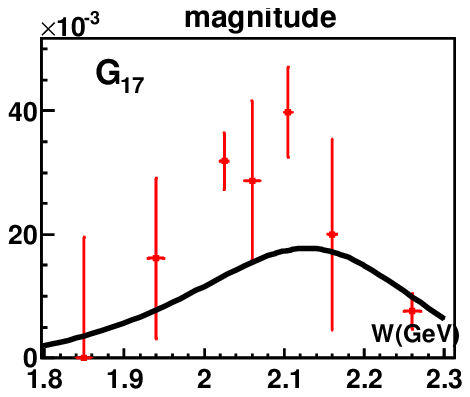}&
\hspace{-3mm}\includegraphics[width=0.185\textwidth]{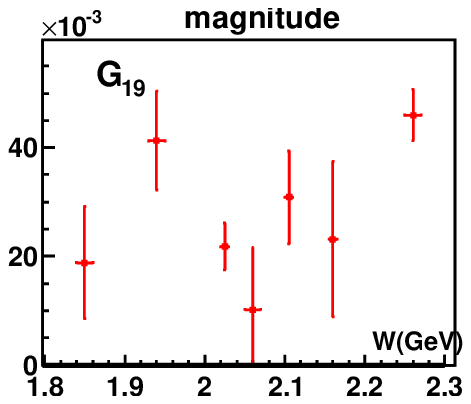}\\
\hspace{-3mm}\includegraphics[width=0.185\textwidth]{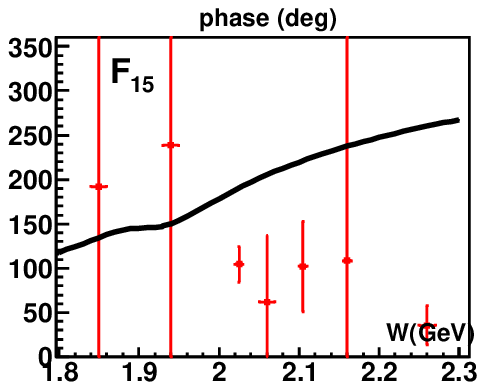}&
\hspace{-3mm}\includegraphics[width=0.185\textwidth]{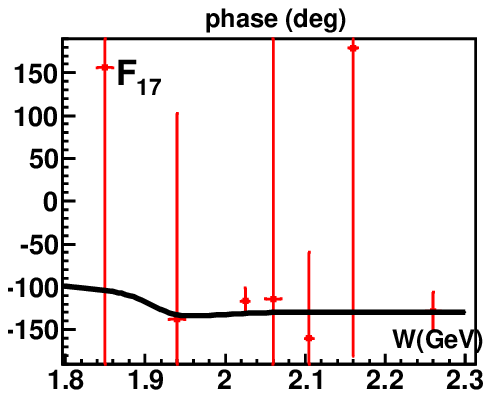}&
\hspace{-3mm}\includegraphics[width=0.185\textwidth]{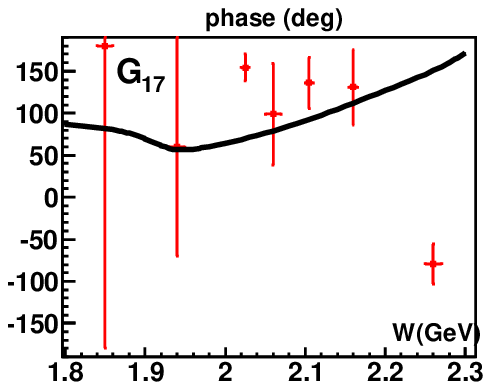}&
\hspace{-3mm}\includegraphics[width=0.185\textwidth]{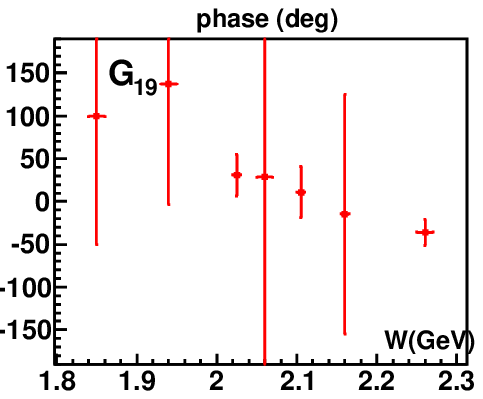}
\end{tabular}
\end{center}
\caption{\label{piLambdaK2_G_High1} Decomposition of the $\pi N\to
\Lambda K$ scattering amplitudes with $S$, $P$, $D$, $F$ and $G$
waves.  The solid line is the energy dependent solution BnGa2011-02.
The agreement between the energy dependent and independent solutions has become worse
due to an increase of random fluctuations;
$\chi^2/N=1966/105$.}
\end{figure*}
\subsection{Solution with $S$,$P$, $D$, $F$  and $G$ waves}
The $\beta$ observable is still not yet properly described, hence we
extend the list of partial waves to include the $G$-wave. Above
2\,GeV, the $\chi^2$ of the fit to the data still improves, see
Table~\ref{Table_high3} and the solid line in
Fig.~\ref{piLambdaK_obeser_high1}.

\begin{table}[ph]
\caption{\label{Table_high3} Quality of the $S$, $P$, $D$, $F$ and
$G$ waves energy independent fit in the region $1840-2270$ MeV:
$\chi^2/N_{\rm data}$ and number of data points (in brackets). The
overall $\chi^2/N_{\rm data}$ is now 0.67.}
\renewcommand{\arraystretch}{1.15}\begin{center}
\begin{tabular}{|c|c|c|c|}
  \hline
  Energy bin& $d\sigma/d\Omega$& $P$& $\beta$\\
  \hline
  1840-1860 & 0.31 (10) & 0.87 (14)  & 0.20 (11) \\
  1930-1950 & 0.14 (20)& 0.92 (20)& 0.23 (9) \\
  2020-2030 & 1.03 (20) & 1.06 (19) & 0.40 (10) \\
  2050-2070 & 0.28 (20) & 0.78 (18) & 0.74 (11) \\
  2100-2110 & 0.29 (20) & 1.22 (20) & 1.09 (12) \\
  2155-2165 & 0.61 (20) & 0.83 (20) & 1.54 (10) \\
  2250-2270 & 0.49 (20) & 0.77 (19) & 0.63 (9) \\
  \hline
\end{tabular}
\end{center}
\end{table}

The $G$ waves are small, but their inclusion clearly changes the
energy independent solution in other waves. Even in the $S_{11}$
wave energy dependent and independent solutions are now no longer
consistent. Of course, the large errors of the scattering amplitudes
are the result of the poor quality of the experimental data. While
the data seem to require even $L=4$ waves, their inclusion leads to
large uncertainties in scattering amplitudes. These are shown in
Fig.~\ref{piLambdaK2_G_High1}.

\subsection{\boldmath Error bands }

\begin{figure}
\begin{center}
\begin{tabular}{cc}
\hspace{-4mm}\includegraphics[width=0.19\textwidth,height=0.17\textwidth]
{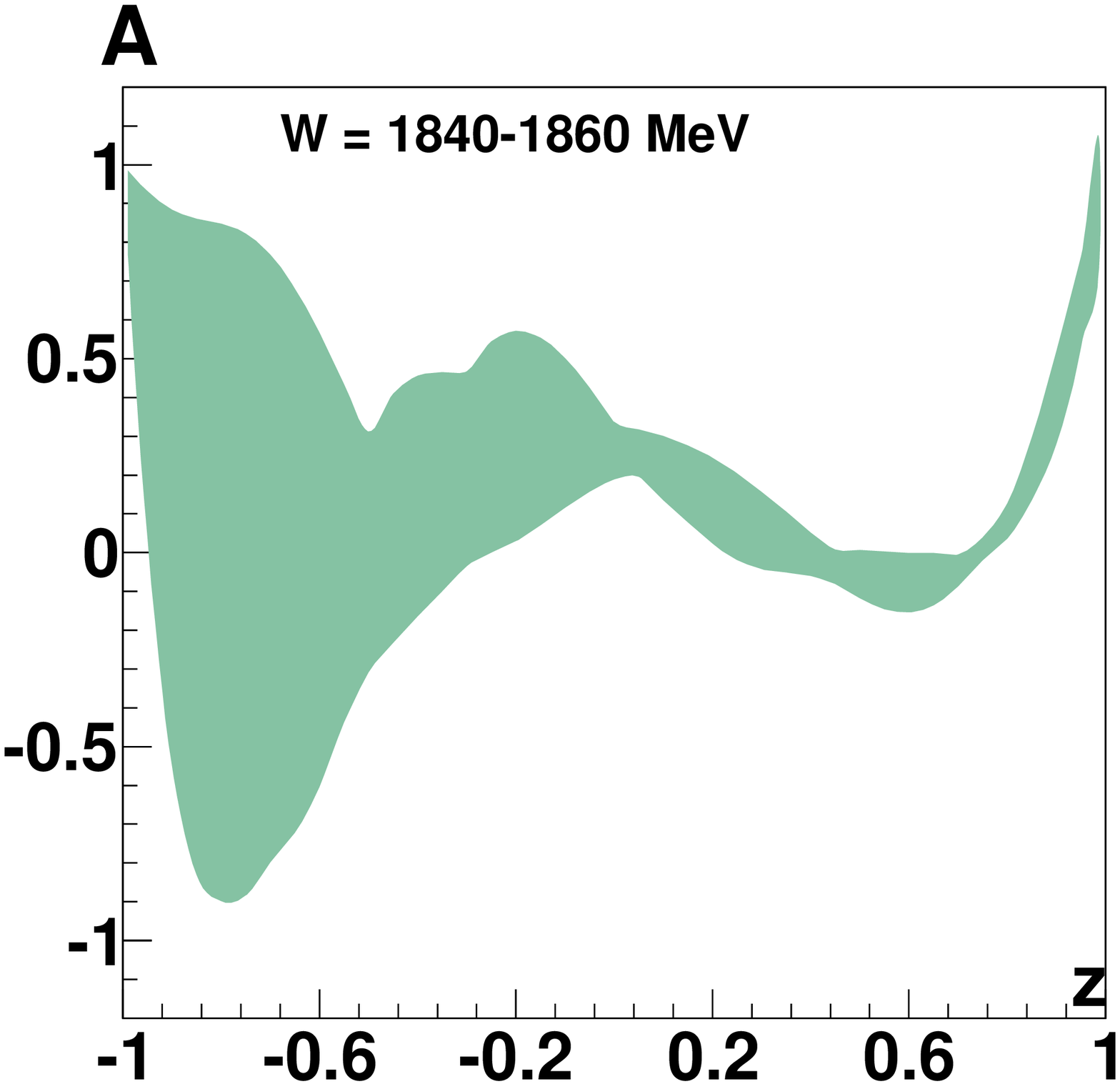}&
\hspace{-4mm}\includegraphics[width=0.19\textwidth,height=0.17\textwidth]
{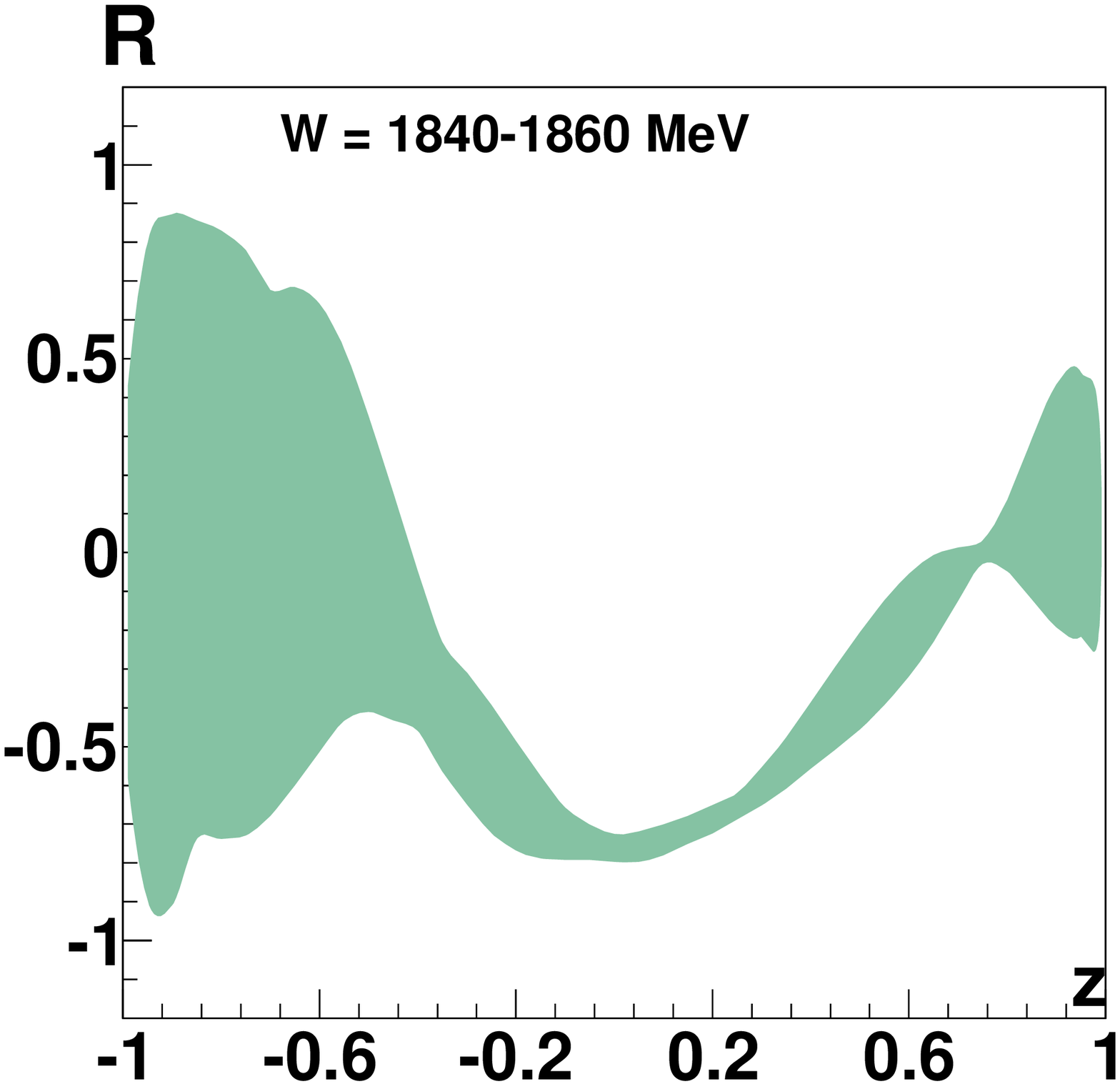}
\\
\hspace{-4mm}\includegraphics[width=0.19\textwidth,height=0.17\textwidth]
{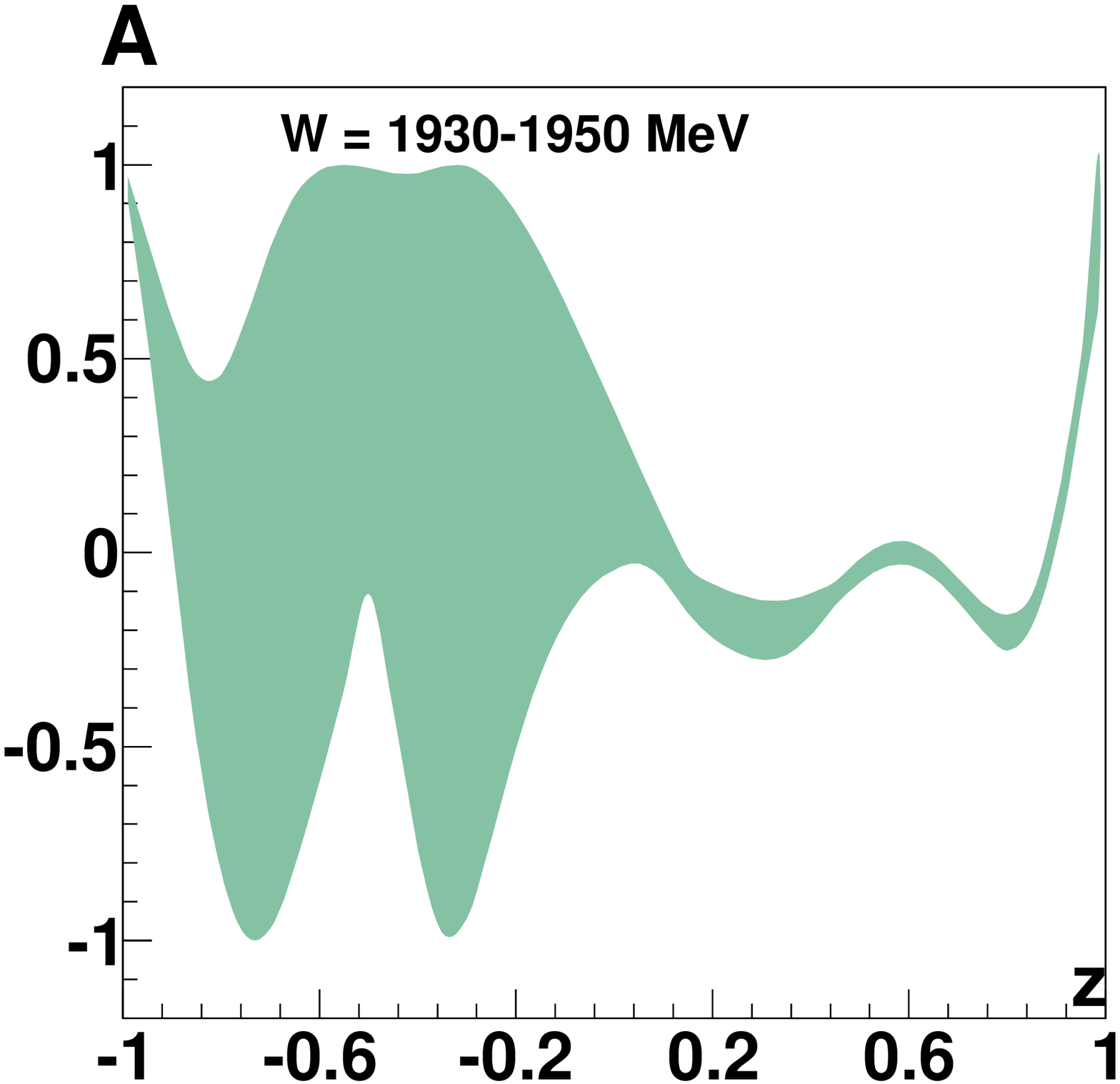}&
\hspace{-4mm}\includegraphics[width=0.19\textwidth,height=0.17\textwidth]
{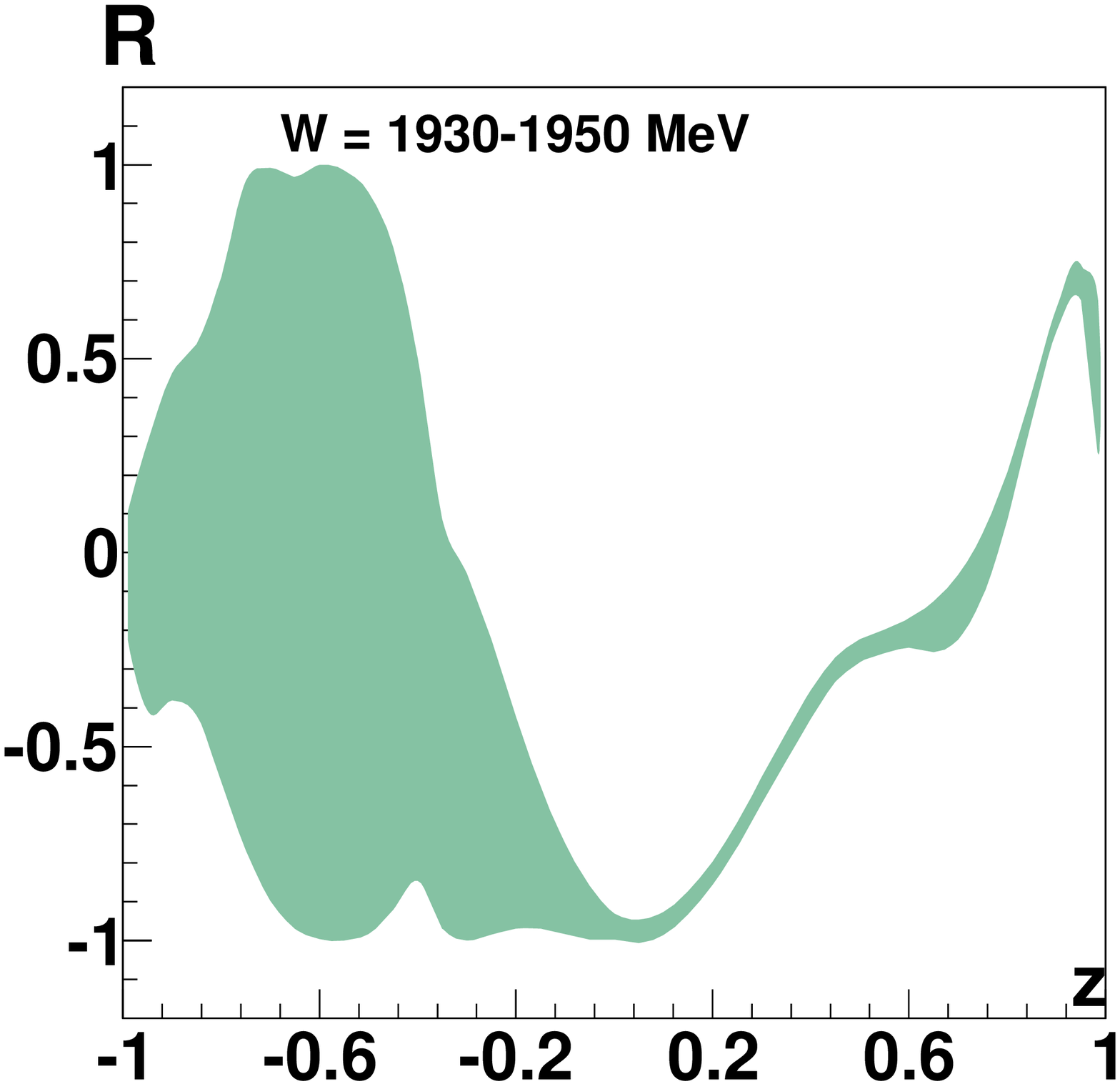}
\\
\hspace{-4mm}\includegraphics[width=0.19\textwidth,height=0.17\textwidth]
{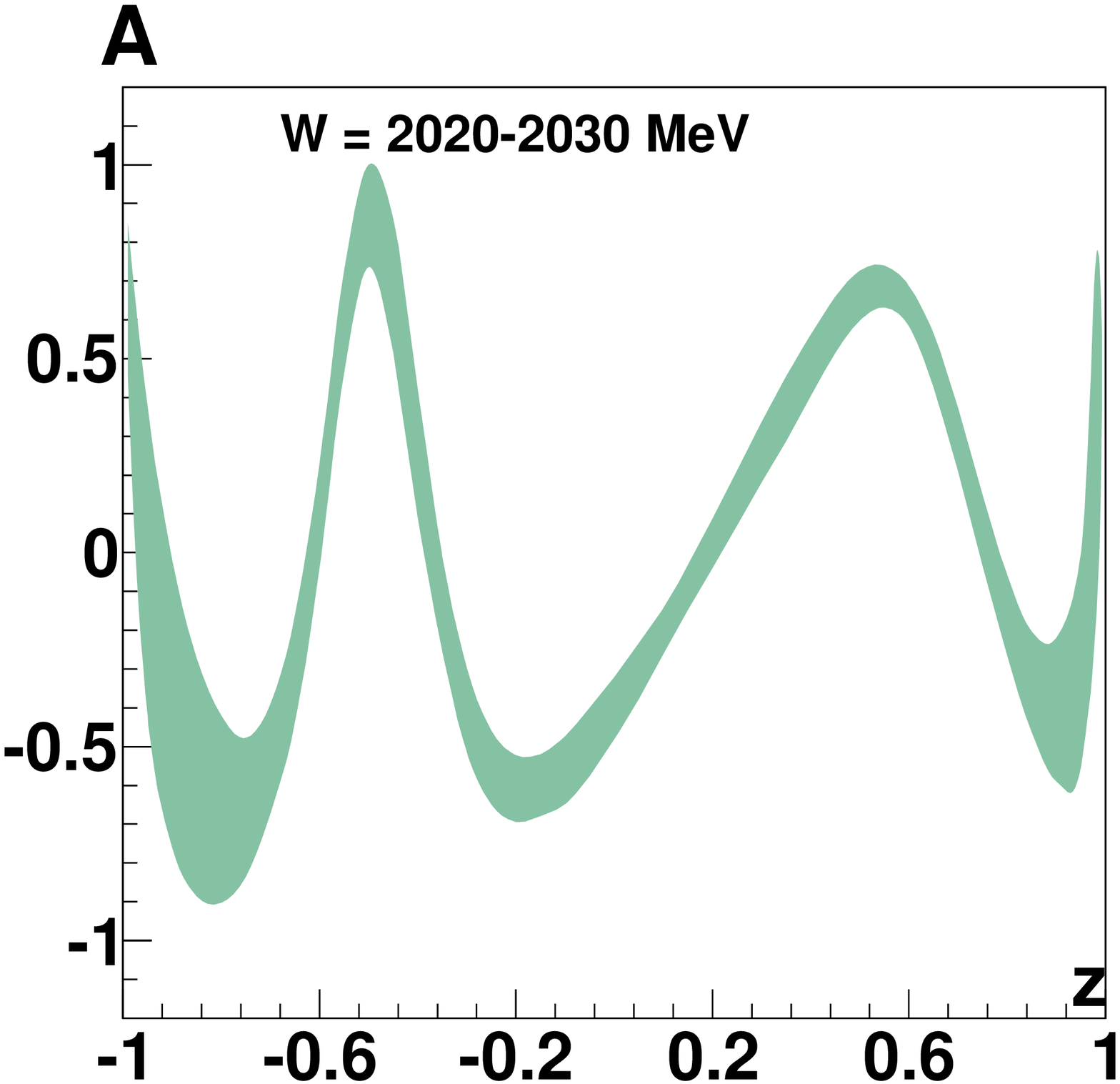}&
\hspace{-4mm}\includegraphics[width=0.19\textwidth,height=0.17\textwidth]
{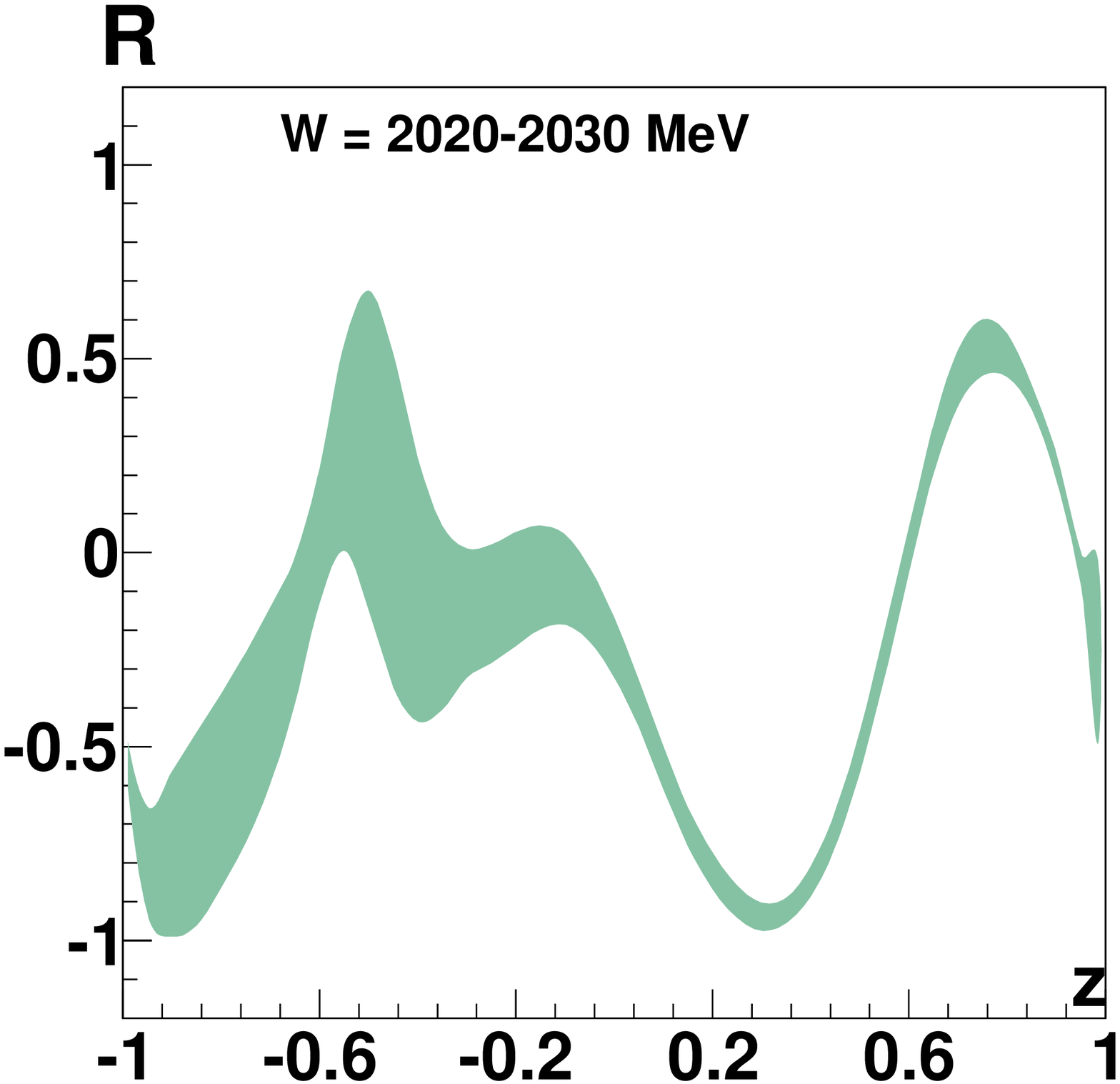}
\\
\hspace{-4mm}\includegraphics[width=0.19\textwidth,height=0.17\textwidth]
{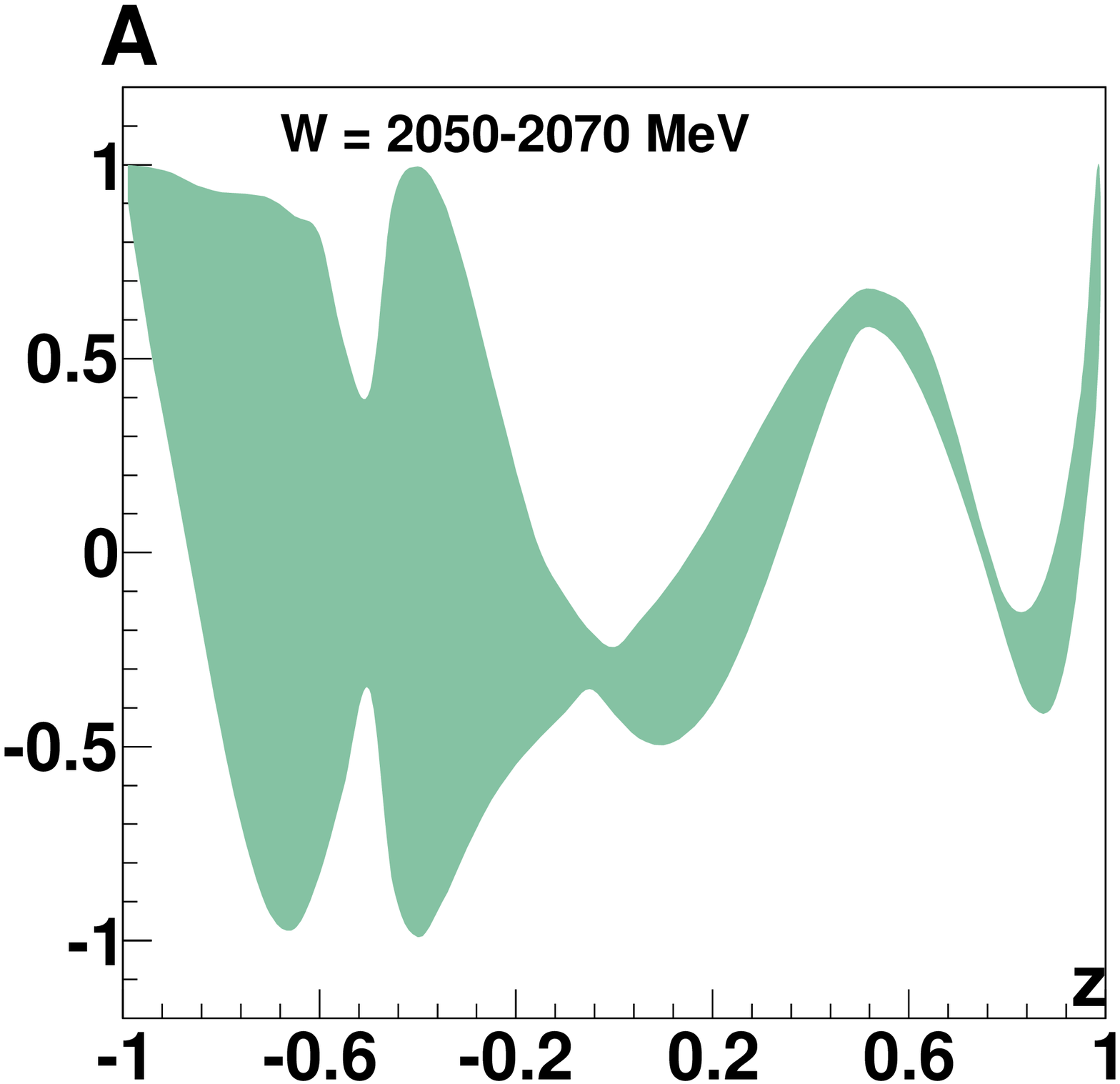}&
\hspace{-4mm}\includegraphics[width=0.19\textwidth,height=0.17\textwidth]
{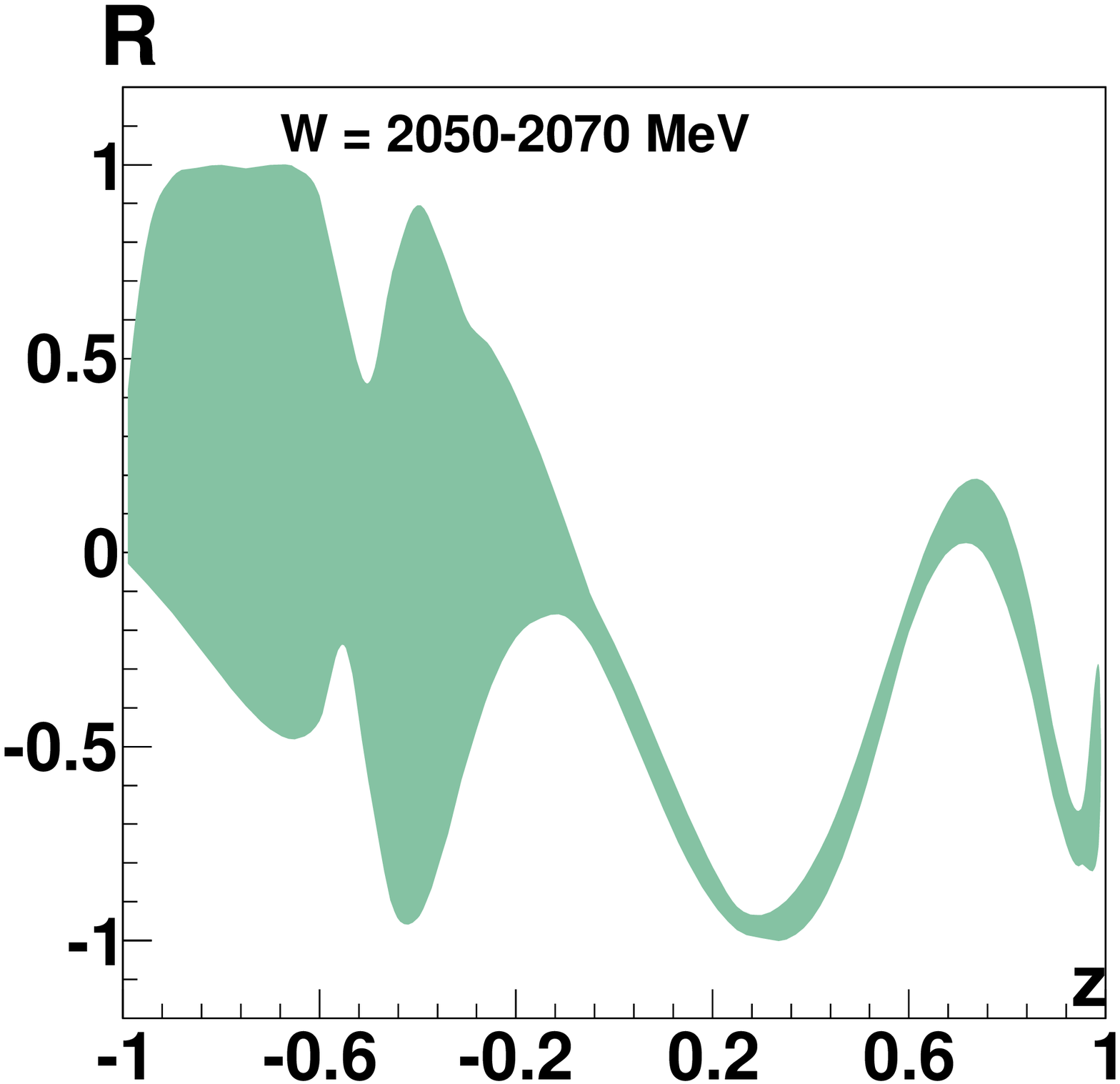}
\\
\hspace{-4mm}\includegraphics[width=0.19\textwidth,height=0.17\textwidth]
{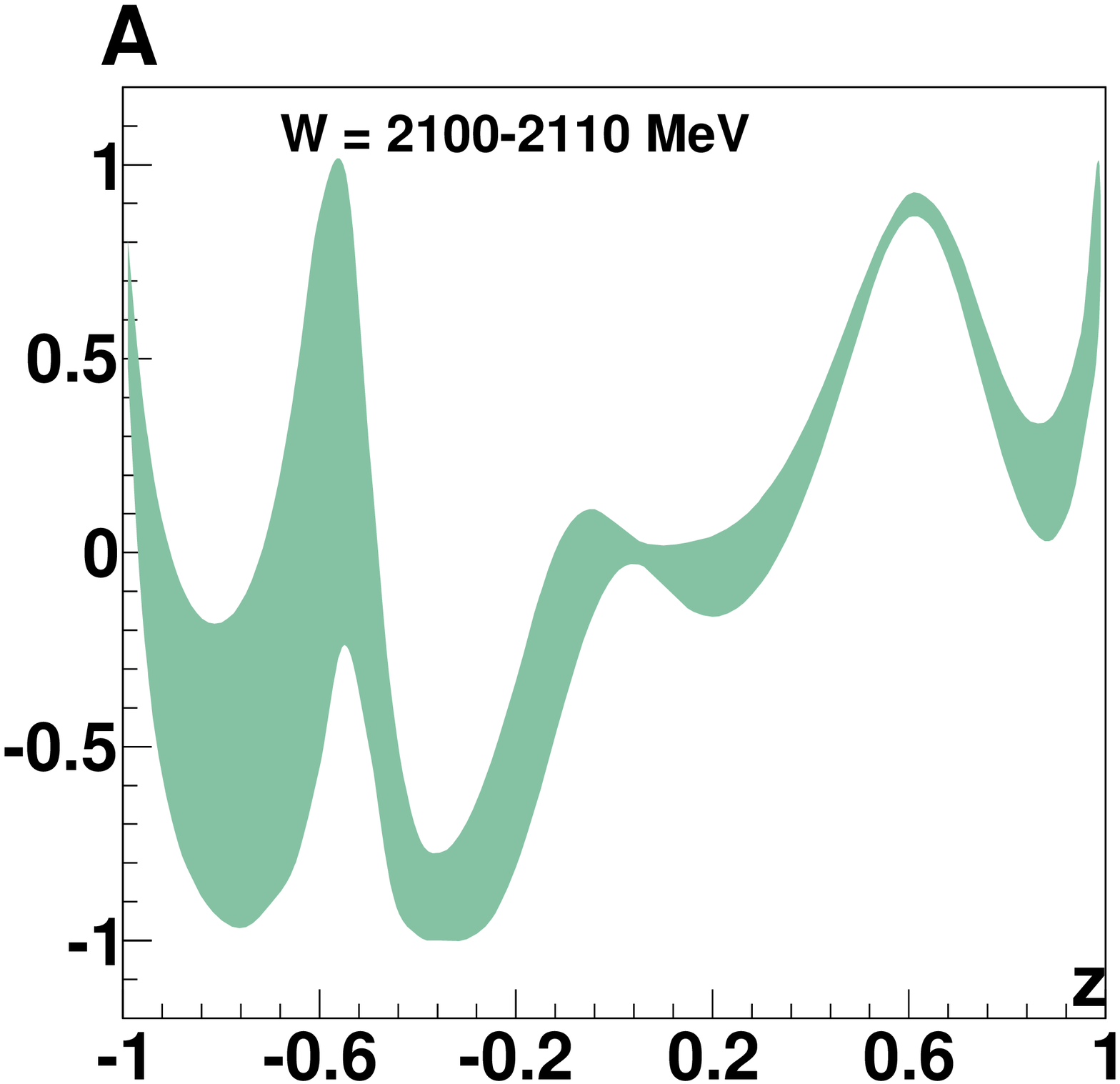}&
\hspace{-4mm}\includegraphics[width=0.19\textwidth,height=0.17\textwidth]
{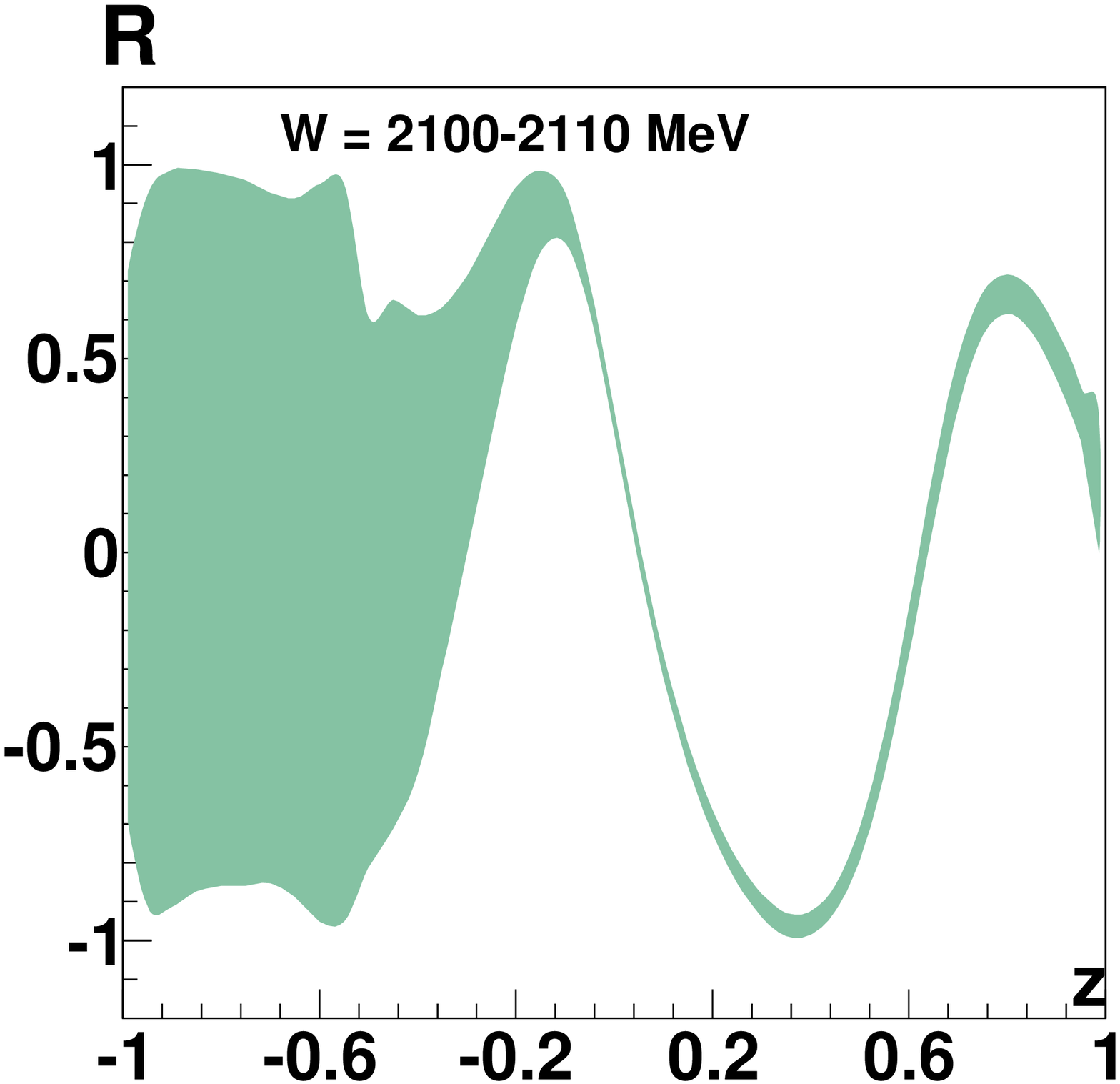}
\\
\hspace{-4mm}\includegraphics[width=0.19\textwidth,height=0.17\textwidth]
{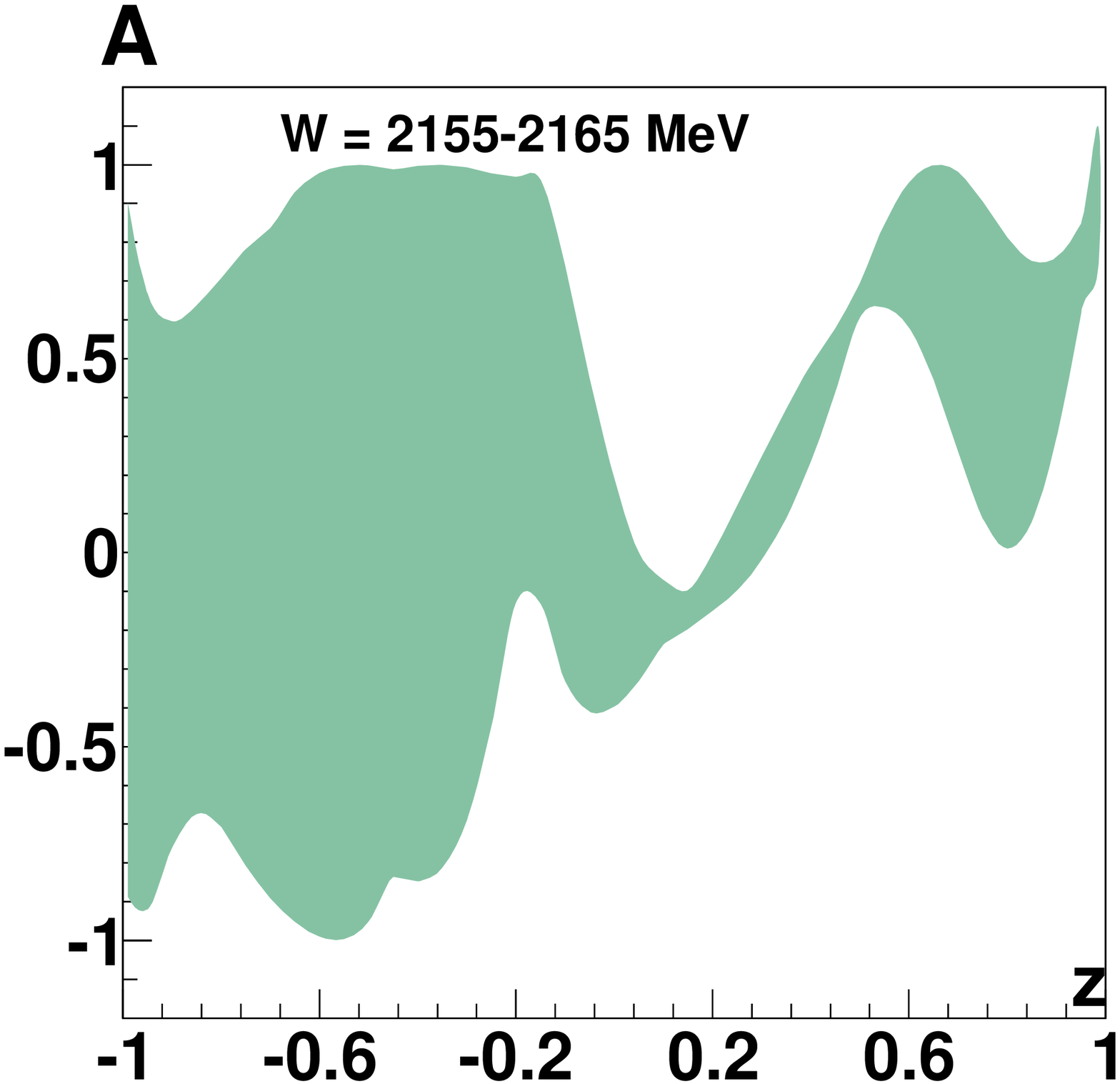}&
\hspace{-4mm}\includegraphics[width=0.19\textwidth,height=0.17\textwidth]
{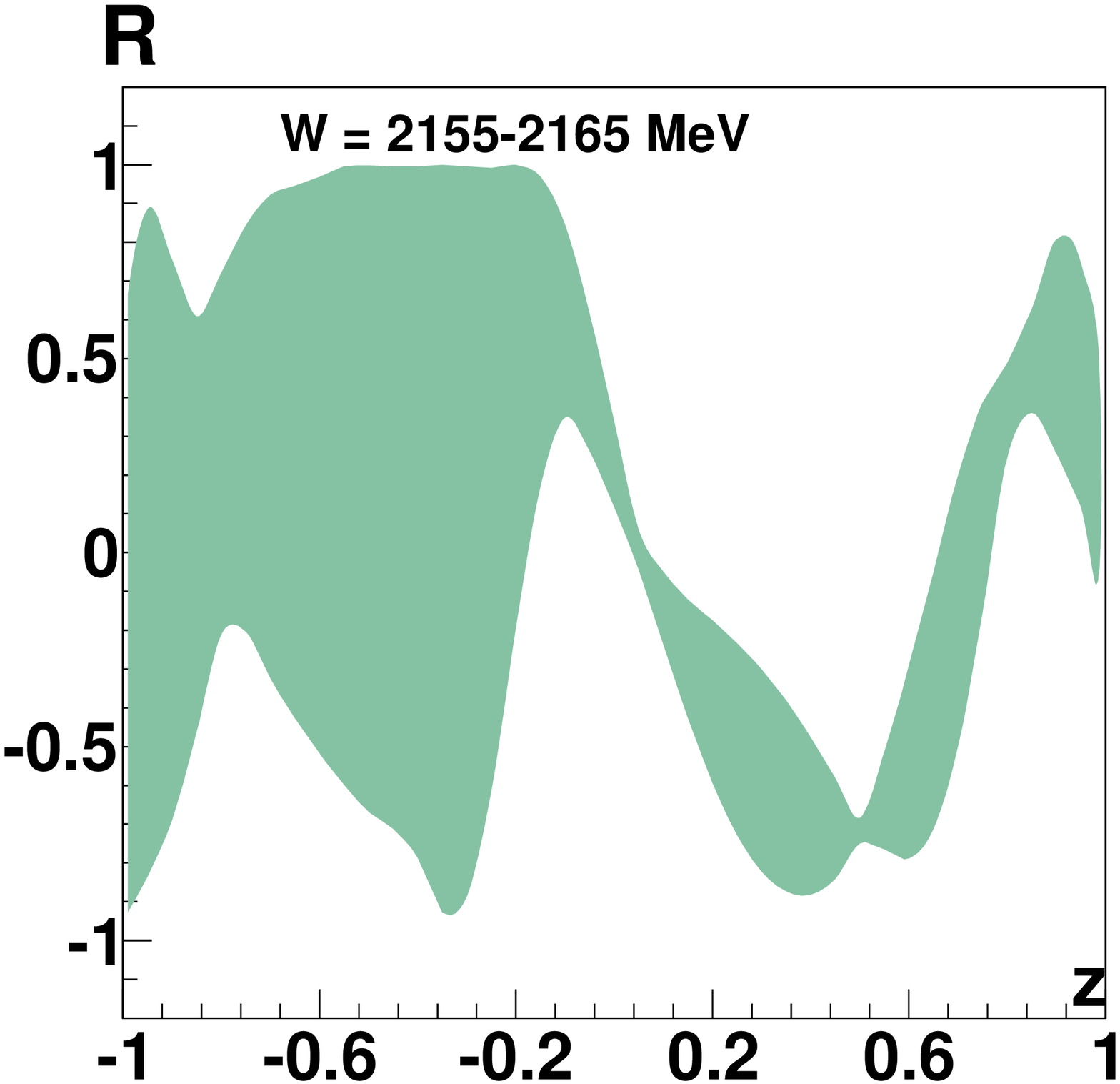}
\\
\hspace{-4mm}\includegraphics[width=0.19\textwidth,height=0.17\textwidth]
{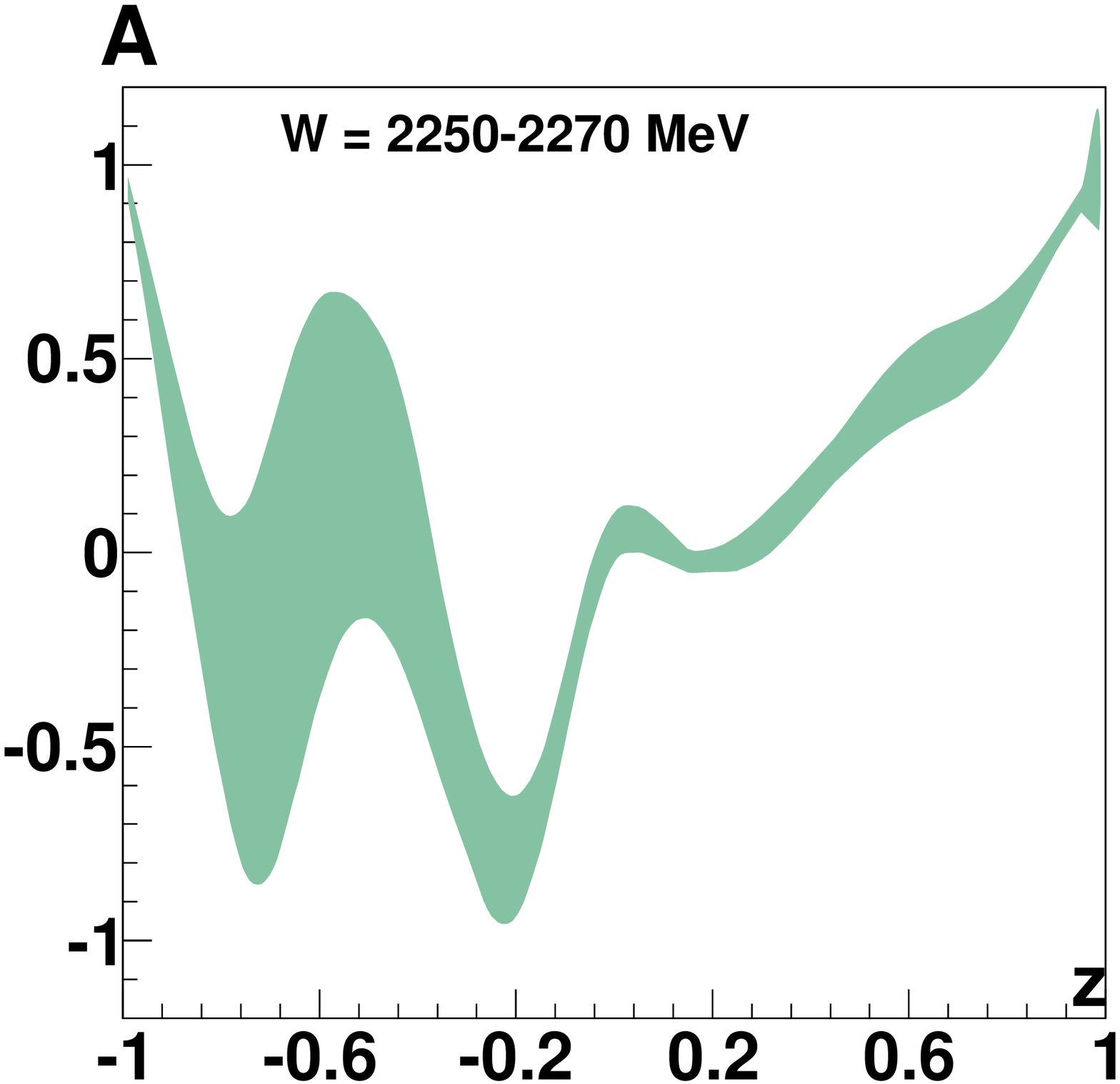}&
\hspace{-4mm}\includegraphics[width=0.19\textwidth,height=0.17\textwidth]
{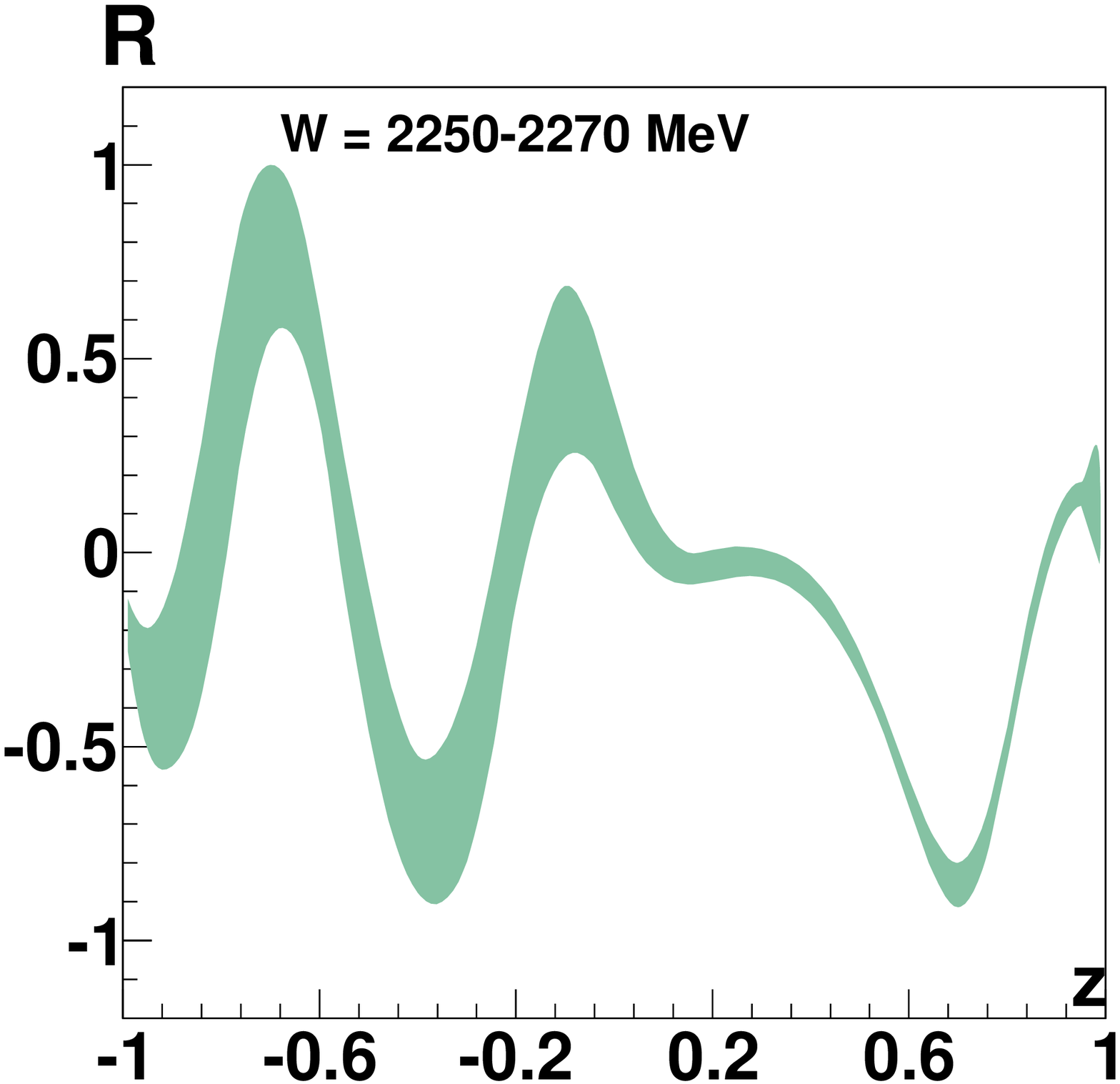}
\\
\end{tabular}
\end{center}
\caption{\label{ErrorBand} Error bands for $A$ and $R$ observables
for the best energy independent fit with $L\leq 4$ (up to the
$G$-wave).}
\end{figure}

A major source of uncertainty is the absence of the spin rotation
data in the region $z <-0.2$. To illustrate this uncertainty the
error bands for $A$ and $R$ observables were defined which include
the solutions with $\delta \chi^2 < 1$ from the best one.  They are
shown in Fig.~ \ref{ErrorBand}. We use the fits with $L\leq 4$ since
they show best the angular ranges where more precise data are
urgently needed.

\begin{figure}[pt]
\begin{center}
\begin{tabular}{cc}
\hspace{-3mm}\includegraphics[width=0.23\textwidth]{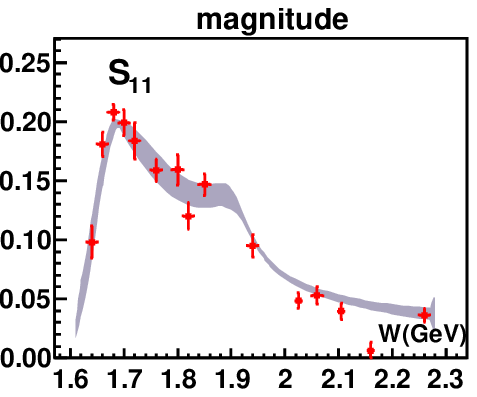}\\
\hspace{-3mm}\includegraphics[width=0.23\textwidth]{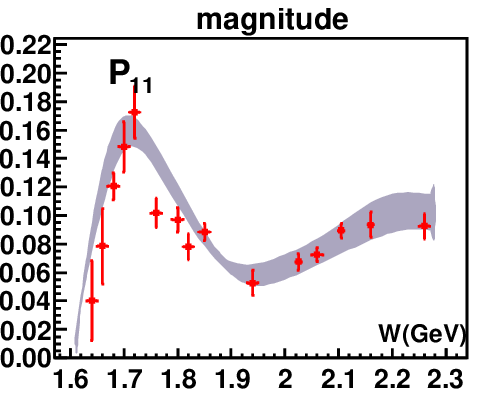}&
\hspace{-3mm}\includegraphics[width=0.23\textwidth]{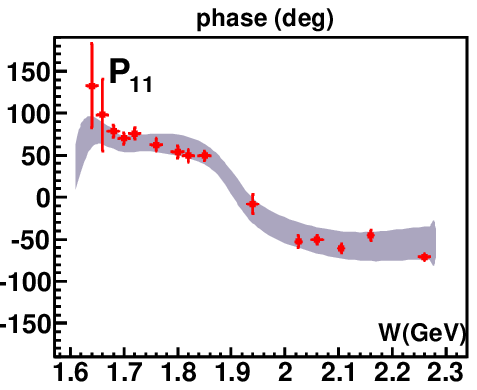}\\
\hspace{-3mm}\includegraphics[width=0.23\textwidth]{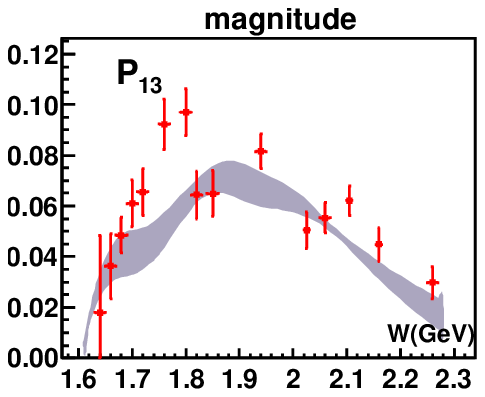}&
\hspace{-3mm}\includegraphics[width=0.23\textwidth]{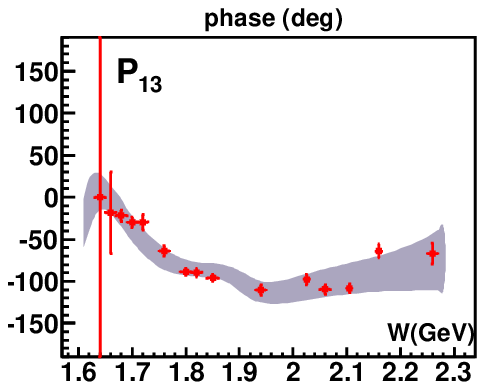}\\
\end{tabular}
\end{center}
\caption{\label{piLambdaK-freeSandP} Fit with free $S$ and $P$-waves
and $D, F$, and $G$-waves fixed to the energy-dependent solution
BnGa2011-02.}
\end{figure}
\subsection{Fits with fixed high or low partial
waves.}

Having all partial waves free in the fits leads to very large errors
and the results are of no use any longer. Apparently, the
statistical accuracy of the data is not sufficient to extract all
partial waves simultaneously. We therefore fixed the $D, F$, and $G$
waves to the energy dependent solution, and then determined the
energy independent amplitudes from a fit to the data. The result is
shown in Fig.~\ref{piLambdaK-freeSandP}. The results now look
reasonable, but suggest that the transition matrix element for the
$P_{13}$ wave in $\pi^-p\to\Lambda K^0$ could be larger than the
value found from the energy dependent fit. This matrix element is,
however, well fixed from the reactions $\gamma p\to p\pi^0$, $\gamma
p\to n\pi^+$, and $\gamma p\to \Lambda K^+$ which have much higher
statistics.  Hence we refrain from an overall refit of the data
imposing the new energy independent solution shown in
Fig.~\ref{piLambdaK2_F_High1}.

\begin{figure}[pt]
\begin{center}
\begin{tabular}{cc}
\hspace{-3mm}\includegraphics[width=0.23\textwidth]{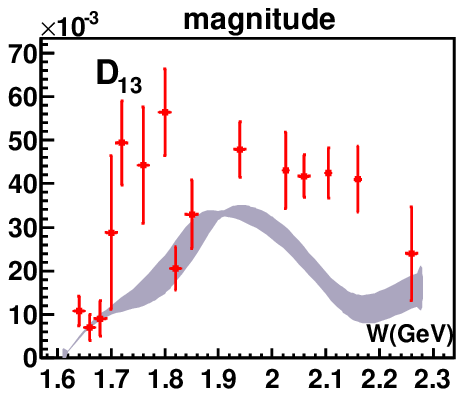}&
\hspace{-3mm}\includegraphics[width=0.23\textwidth]{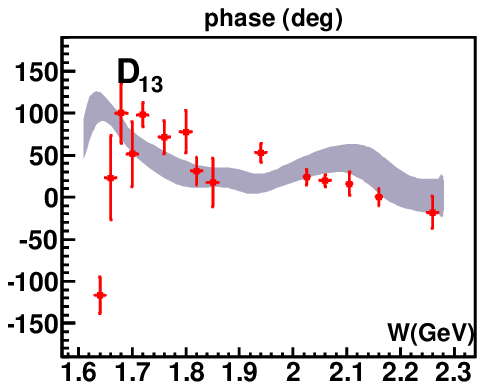}\\
\hspace{-3mm}\includegraphics[width=0.23\textwidth]{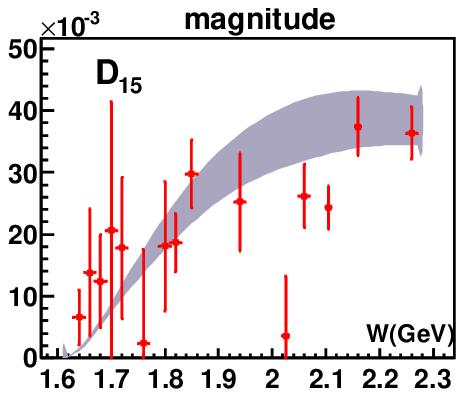}&
\hspace{-3mm}\includegraphics[width=0.23\textwidth]{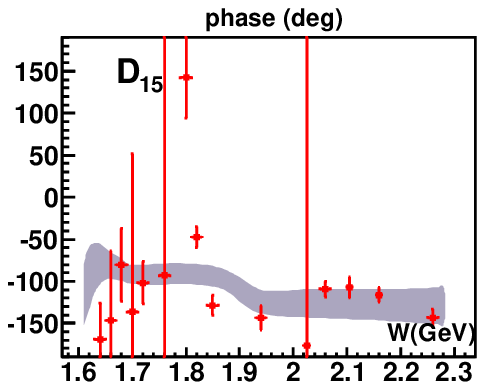}\\
\hspace{-3mm}\includegraphics[width=0.23\textwidth]{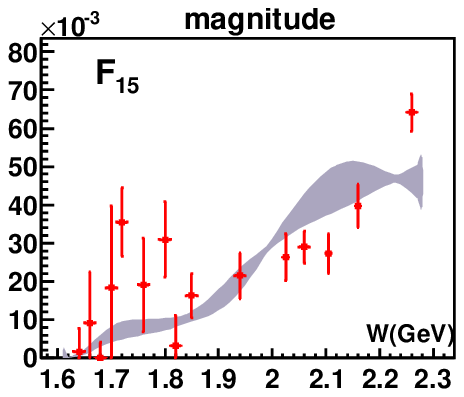}&
\hspace{-3mm}\includegraphics[width=0.23\textwidth]{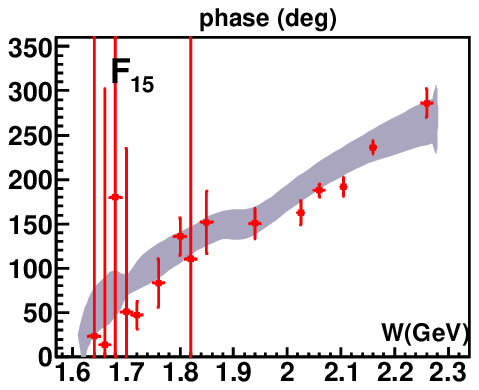}\\
\hspace{-3mm}\includegraphics[width=0.23\textwidth]{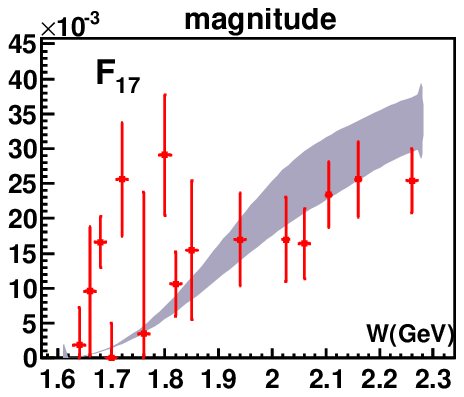}&
\hspace{-3mm}\includegraphics[width=0.23\textwidth]{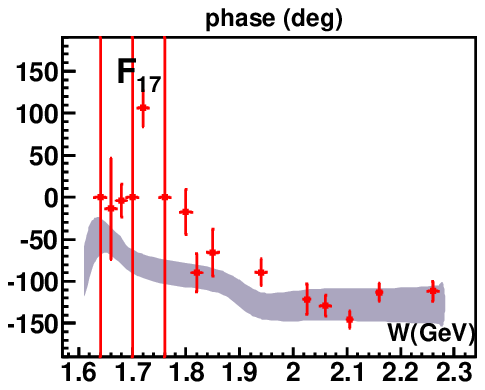}\\
\hspace{-3mm}\includegraphics[width=0.23\textwidth]{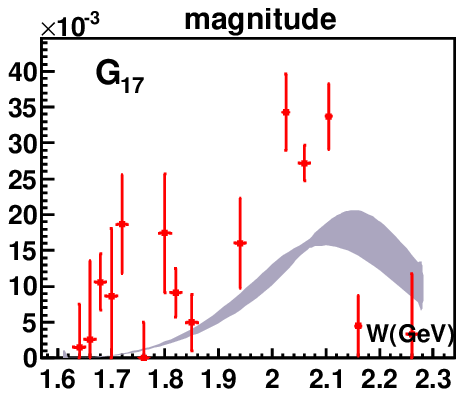}&
\hspace{-3mm}\includegraphics[width=0.23\textwidth]{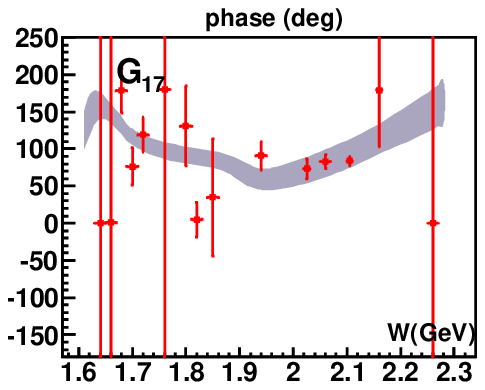}\\
\end{tabular}
\end{center}
\caption{\label{piLambdaK-freeDandFandG}Fit with free $D, F$, and
$G$-waves and $S$ and $P$-waves fixed to the energy-dependent
solution BnGa2011-02. }
\end{figure}

Alternatively, we may ask if the high partial waves can be
determined from the data, once the low partial waves are fixed. Thus
we constrained the low partial waves ($S$ and $P$-waves) to coincide
with the energy-dependent solution BnGa2011-02 while the high waves
were left free in the fit to reproduce the data on $\pi^-p\to\Lambda
K^0$. The resulting amplitudes are shown in Fig.
\ref{piLambdaK-freeDandFandG}. The results still have large error
bars but are mostly not inconsistent with the energy-dependent
solution.

\section{Summary and Conclusions}

We have studied the ambiguity problem which arises when incomplete
data are used to derive complex scattering amplitudes. We use
existing data on pion-induced production of $\Lambda$ hyperons. In
the low energy region, data for this reaction exist on the
differential cross section and $\Lambda$ polarization.

Assuming that only $S$ and $P$ waves contribute, the data can be
described with sufficient accuracy up to 1700\,MeV in invariant
mass. When higher waves are fixed to the energy dependent solution,
the reconstructed magnitudes of the $P_{11}$ and $S_{11}$ waves are
consistent with the results from the energy dependent analysis, see
Fig. \ref{piLambdaK-freeSandP}; the $P_{11}$ - $S_{11}$ phase
difference might indicate problems in the $\Lambda K$ threshold
region (see Fig. 2, left column). The $P_{13}$ wave seems to be
overestimated in the energy independent fit; the $P_{13}$ phase
motion (relative to the $S_{11}$ phase) does, however, not support
any additional feature on top of the known resonances. If $D$ waves
are admitted, the number of numerically different solution increases
to six. The ``best" solution (the two top rows in Fig. 3) is
reasonably consistent with the predicted curves even though most
waves show large error bars and an excess in magnitude of the
reconstructed amplitudes.

Above 1800\,MeV, up to 2270\,MeV in total energy, the spin rotation
angle $\beta$ has been determined and nearly complete information
exists even though not with complete solid angle coverage. The
scattering amplitudes can now be reconstructed unambiguously. If
only $S$, $P$, and $D$ are admitted, large discrepancies show up
between energy independent and energy dependent amplitudes (see
Fig.~5). Apparently, the angular distributions require at least $F$
waves, and even $G$ waves improve the quality of the fit
substantially. However, the errors of magnitude and phase of the
reconstructed amplitudes become increasingly larger.

Approximate consistency between the energy-dependent and independent
solution can be obtained by providing some {\it guidance} to the fit
by fixing the low-energy partial waves or the high-energy partial
waves, and then determining the other partial waves from a fit to
the data. This {\it guidance} leads to reasonably looking results in
particular for the low waves (see Fig.~\ref{piLambdaK-freeSandP})
but clearly, the final result on the energy-independent solution is
strongly biased by the energy-dependent solution. In the present
case, in the study of $\pi^-p\to \Lambda K^0$, the
energy-independent approach does not yield information which goes
beyond the information already provided by the energy-dependent
solution. Note that the energy dependent fit uses nearly the full
data base including photoproduction reactions while the energy
independent fit is based only on the data shown here.

\begin{figure}[pt]
\begin{center}
\begin{tabular}{cc}
\hspace{-3mm}\includegraphics[width=0.23\textwidth]{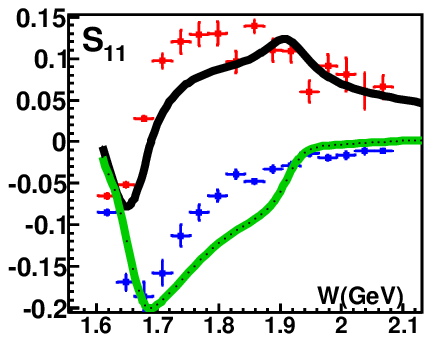}&
\hspace{-3mm}\includegraphics[width=0.23\textwidth]{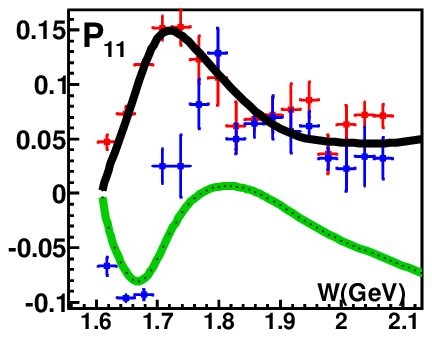}\vspace{-2mm}\\
\hspace{-3mm}\includegraphics[width=0.23\textwidth]{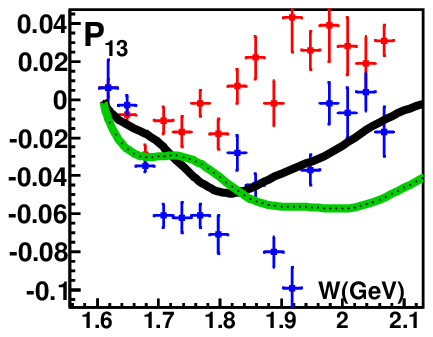}&\vspace{-2mm}\\
\hspace{-3mm}\includegraphics[width=0.23\textwidth]{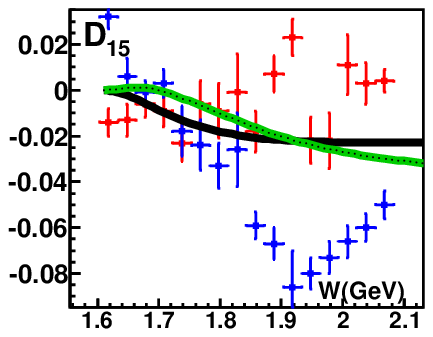}&
\hspace{-3mm}\includegraphics[width=0.23\textwidth]{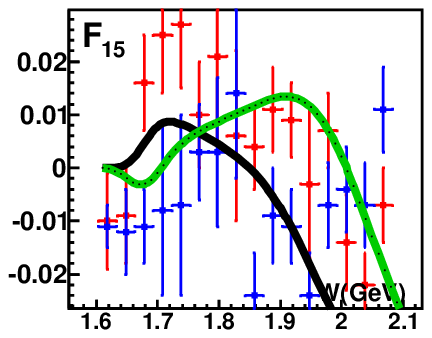}\vspace{-2mm}\\
\hspace{-3mm}\includegraphics[width=0.23\textwidth]{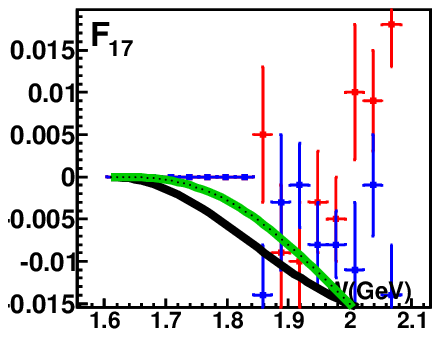}&
\hspace{-3mm}\includegraphics[width=0.23\textwidth]{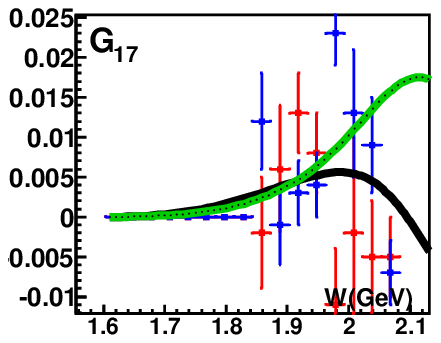}\vspace{-2mm}\\
\hspace{-3mm}\includegraphics[width=0.23\textwidth]{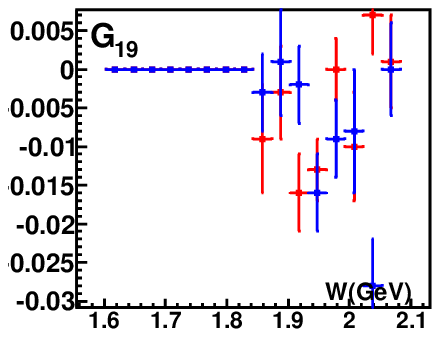}&
\hspace{-3mm}\includegraphics[width=0.23\textwidth]{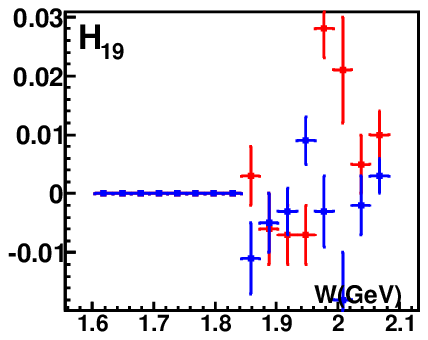}\vspace{-1mm}
\end{tabular}
\end{center}
\caption{\label{S-and-M} (Color online) Real (red) and imaginary
(blue) part of the $\pi^-p\to\Lambda K^0$ transition amplitude from
\cite{Shrestha:2012va}. The $D_{13}$ amplitude is set to zero. The
energy dependent BnGa2013 fit is shown by curves, real part (black),
imaginary part (green, grey).}
\end{figure}

The final amplitude depends critically on the procedure. In
Fig.~\ref{S-and-M}, the BnGa energy dependent fit - compatible with
the single-energy amplitudes - is compared to the energy-independent
amplitudes from \cite{Shrestha:2012va}. Large discrepancies are
observed. In the latter analysis, the initial energy-dependent fit
failed to reproduce the observables satisfactorily, and an iterative
procedure was used to derive the amplitudes at single energies. In a
first step, it was noticed that the $S_{11}$ amplitude was fitted
well with the energy-dependent fit. Hence this wave was held fixed
and only the other partial-wave amplitudes were varied. This step
lead to a $P_{11}$ amplitude which could be fitted well in the
energy-dependent fit. As a next step, both the $S_{11}$ and $P_{11}$
amplitudes were held fixed at their energy-dependent values while
the other amplitudes were varied. The energy-dependent fits
indicated that $D_{13}$ amplitude is small; it was hence set to
zero. It is unproven that this procedure converges to a ``correct"
solution. Both, the BnGa amplitudes and the amplitudes from
\cite{Shrestha:2012va} are compatible with all observables.
Identical data have been used. A priori, there is no objective
reason to trust one result better than the other one. We have to
conclude that the ``guidance" offered during the fits has a
significant impact on the final results. One should have this in
mind when the real and imaginary part of scattering amplitudes from
single channel analysis are used for further analysis. The
amplitudes look like data point with error bars. They are not. The
``data points" are the results of complex procedure which converges
only after personal judgement.

At present, the aim of this development is the reconstruction of
amplitudes from photoproduction data. This is an even more demanding
task but the data will have (and need to have) much smaller
statistical and systematic errors. If this can be achieved with one
unique solution, the data will return amplitudes from which the
spectrum of nucleon and $\Delta$ resonances can be deduced
unambiguously. At least, the energy-independent analysis will be a
valuable test of energy dependent solutions.

\section*{Acknowledgements}

We acknowledge financial support from the Deutsche
Forschungsgemeinschaft within the Sonderforschungsbereich SFB/TR16
(DFG). This work is also supported by Russian Foundation for Basic
Research 13-02-00425 À.


\begin{thebibliography}{99}
\bibitem{Arenhovel:1998vj}
  H.~Arenh\"ovel, W.~Leidemann and E.~L.~Tomusiak,
  Nucl.\ Phys.\ A {\bf 641}, 517 (1998).
  \bibitem{Hohler:1979yr}
  G.~H\"ohler, F.~Kaiser, R.~Koch and E.~Pietarinen,
  ``Handbook Of Pion Nucleon Scattering,''
Published by Fachinform. Zentr. Karlsruhe 1979, 440 P. (Physics
Data, No.12-1 (1979)).

\bibitem{Hohler:1993xq}
  G.~H\"ohler,
$\pi N$ Newslett. {\bf 9}, 108 (1993).

\bibitem{Cutkosky:1980rh}
  R.~E.~Cutkosky {\it et al.},  ``Pion - Nucleon Partial Wave Analysis,'' 4th Int. Conf. on Baryon Resonances,
  Toronto, Canada, Jul 14-16, 1980. Published in Baryon
1980:19 (QCD161:C45:1980)

 \bibitem{Arndt:2006bf}
  R.~A.~Arndt, W.~J.~Briscoe, I.~I.~Strakovsky and R.~L.~Workman,
  Phys.\ Rev.\  C {\bf 74}, 045205 (2006).

 \bibitem{Shrestha:2012va}
  M.~Shrestha and D.~M.~M.~Manley,
  Phys.\ Rev.\ C {\bf 86}, 045204 (2012).

 \bibitem{Shrestha:2012ep}
  M.~Shrestha and D.~M.~Manley,
  Phys.\ Rev.\ C {\bf 86} (2012) 055203.

\bibitem{Knasel:1975rr}
  T.~M.~Knasel {\it et al.},
  Phys.\ Rev.\  D {\bf 11}, 1 (1975).

\bibitem{Baker:1978qm}
  R.~D.~Baker {\it et al.},
  Nucl.\ Phys.\  B {\bf 141}, 29 (1978).

  \bibitem{Saxon:1979xu}
  D.~H.~Saxon {\it et al.},
  Nucl.\ Phys.\  B {\bf 162}, 522 (1980).

\bibitem{Bell:1983dm}
  K.~W.~Bell {\it et al.},
  Nucl.\ Phys.\  B {\bf 222}, 389 (1983).

\bibitem{Gersten:1969ae}
  A.~Gersten,
  Nucl.\ Phys.\ B {\bf 12}, 537 (1969).

\bibitem{Barrelet:1971pw}
  E.~Barrelet,
  Nuovo Cim.\ A {\bf 8} (1972) 331.

\bibitem{Baker:1976xp}
  R.~D.~Baker,
  ``Barrelet Zeros in Partial Wave Analysis,''
  RL-76-013.

\bibitem{Anisovich:2010an}
  A.~V.~Anisovich, E.~Klempt, V.~A.~Nikonov, A.~V.~Sarantsev and U.~Thoma,
  Eur.\ Phys.\ J.\ A {\bf 47}, 27 (2011).

\bibitem{Anisovich:2011fc}
  A.~V.~Anisovich, R.~Beck, E.~Klempt, V.~A.~Nikonov, A.~V.~Sarantsev and U.~Thoma,
  Eur.\ Phys.\ J.\ A {\bf 48}, 15 (2012).


\end{thebibliography}
\end{document}